\documentclass{JHEP}
\usepackage{epsfig}

\newcommand{\be}{\begin{equation}}
\newcommand{\ee}{\end{equation}}
\newcommand{\bea}{\begin{eqnarray}}
\newcommand{\eea}{\end{eqnarray}}

\preprint{MIT-CTP-2959\\ NSF-ITP-00-20\\ {\tt hep-th/0004092}}

\title{Mirror symmetry by O3-planes}

\author{Bo Feng
\\
Center for Theoretical Physics,
\\ Massachusetts Institute of Technology\\ Cambridge MA 02139\\
\email{fengb@ctpyellow.mit.edu}
}
\author{Amihay Hanany
\\
Center for Theoretical Physics,
\\ Massachusetts Institute of Technology\\ Cambridge MA 02139\\
\email{hanany@mit.edu}
}

\abstract{ We construct the three dimensional
mirror theory of $SO(2k) $ and $SO(2k+1)$ gauge groups by using  O3-planes.
An essential ingredient in constructing the mirror is the splitting of 
a physical brane
 (NS-brane or
D5-brane) on O3-planes. In particular, matching the
dimensions of moduli spaces of mirror pair (for example, the $SO(2k+1)$ and
its mirror) there is a D3-brane
creation or annihilation accompanying  the splitting. This novel dynamical
process gives a nontrivial prediction for strongly coupled field theories,
which will be very interesting to check by Seiberg-Witten curves.
Furthermore, 
applying the same idea, we revisit the mirror theory of $Sp(k) $
gauge group and find new mirrors which differ from  previously
known results. Our new result for $Sp(k)$ gives another example 
to a previously observed fact,
which  shows that different theories can be mirror to the same theory.
We also discussed the phenomena such as {\sl ``hidden FI-parameters''}
when the number of flavors and the rank of the gauge group satisfy 
certain relations, {\sl ``incomplete Higgsing''} for the mirror of
$SO(2k+1)$ and the {\sl ``hidden global symmetry''}. 
After discussing the mirror for a single $Sp$ or $SO$ gauge group, we extend
the study to a product of two gauge groups in two different models, namely
the elliptic  and the non-elliptic models.} 


\begin{document}


\section{Introduction}
In \cite{Int}, Intriligator and Seiberg found a new duality, the
so-called ``mirror symmetry'',  between two different $N=4$ gauge
theories in three dimensions. There exists such a mirror duality
in three dimensions due to several  special properties. First, the
$N=4$ theory has a global R-symmetry $SO(4) $ which can be
rewritten  as $SU(2) _L \times SU(2) _R$, i.e., as the direct product
of two independent $SU(2) $ factors. This is one crucial property
for mirror symmetry because one action of the mirror duality is  to
simply interchange these two $SU(2) $ factors\footnote{ When we
discuss the mirror duality of $N=2$ theory in three dimensions, we
must enhance the  explicit $U(1) $ global R-symmetry to two $U(1)
$'s, i.e., $U(1) \times U(1) $. Otherwise there is no good way to
define the mirror theory. For details see \cite{Boer3,Aha}.}. Under
the global R-symmetry, the vector multiplet is in the adjoint of
$SU(2) _L$ and is invariant under $SU(2) _R$ while the
hypermultiplet is in the adjoint of $SU(2) _R$ and is invariant under
$SU(2) _L$ (notice that both multiplets have four scalars if we
dualize the gauge field $A_\mu$ in three dimensions to a scalar).
Furthermore, the mass parameter transforms as $(3,1) $ of $SU(2)
_L \times SU(2) _R$ and the FI-parameter as $(1,3) $. So after
mirror duality, the Coulomb branch and mass parameter of one
theory change to the Higgs branch and FI-parameter of the other
 and vice versa. Such mapping has an immediate  application:
because the Higgs branch is not renormalized by quantum effects
\cite{Arg}, we can get the exact result about the Coulomb branch
of one theory which is corrected by quantum effects by studying
the Higgs branch of the mirror theory which can be studied at the 
classical level. Because of this and other good applications of
mirror duality (for details, see \cite{Int}),
 a lot of work \cite{Boer1,Boer2,Han1,Por,Kap,Han2,Aga,Kap2,Dorey}
 has been done in this topic to try and find new mirror pairs.

\par
There are several ways to construct the mirror pairs. The first
way is to use the arguments coming from field theory
\cite{Int,Boer1}. This method gives a lot of details how  fields
and parameters map to each other under the mirror duality.
However, this method requires a lot of results which are not easy
to get in field theory, so it is hard to use it to construct
general mirror pairs. The second way is to use  M-theory to
construct the mirror pairs as done by Porrati and Zaffaroni in
\cite{Por}. The third way is to use the geometric
realization in \cite{Hori}. The fourth way, which is also the
most popular way in the construction of mirror pairs, is given in
\cite{Han1} by using brane setups. The brane setup has the good
property of making many quantities in field theory more visible.
For example, the R-symmetry $SU(2) _L \times SU(2) _R$ corresponds
to  rotations in planes $X^{345}$ and $X^{789}$. The Coulomb
branch and Higgs branch become the positions of D3-branes in
NS-branes and D5-branes. The mass parameter and FI-parameter also
have similar geometric correspondences.
 These geometric pictures
give us some intuition to understand the problem better (for more
applications of brane setups, see  review \cite{Giv}). The key
observation in \cite{Han1} is that the mirror duality is just the
S-duality in string theory. Using the known property of S-dual
transformation of various kinds of  branes
\cite{Han1,Boer2,Kap,Han2} we can easily find the mirror pairs. In
this paper we will follow the last method.

\par
Because we will use the brane setup to find the mirror theory, let
us talk more about the general idea \cite{Han1} of the brane
construction. Given a gauge theory with gauge group and some
matter contents, first we try to find a proper brane setup which
represents the gauge theory (usually it is the Coulomb branch
given explicitly in the brane setup). After that, we move to
the  Higgs branch\footnote{Usually, we can break all gauge symmetries
 by Higgs mechanism. However, in some cases after  Higgsing there
are still some massless gauge fields. We call the latter case
{\sl ``incomplete Higgsing}} of the theory by splitting the D3-branes
between NS-branes and D5-branes. Then we make the S-duality
transformation (mirror transformation)  which changes the NS-brane
to D5-brane, D5-brane to NS-brane and D3-brane to itself,
while perform the electric-magnetic duality in the world volume theory 
of D3-branes. 
 When the
brane setup involves an orientifold  or $ON$ plane, we need to
know the S-duality rule for them too. Finally, we read out the
corresponding gauge theory given by the S-dual brane setup---it is
the mirror theory which we want to find.

\par
In   applications, it
is  straight forward to use the above procedure to give the
mirror theory  of $U(n) $ gauge theory with some flavors or the
product of $U(n) $'s  with some bifundamentals because the brane
setup of those theories involve only  NS5-branes, D5-branes and
D3-branes and  we know how to deal with them. However, when we try
to find the mirror for a gauge group $Sp(k) $ or $SO(n) $, we must
use an orientifold plane in the brane setup. Now a problem arises
because sometimes we do not know how to read out the gauge theory
of the S-dual brane setup of these orientifolds. The orientifolds
which are involved in the construction can be divided into two
types: the orientifold three plane (O3-plane)  and the orientifold
five plane (O5-plane). Sen has given an answer about the gauge
theory under the $ON$-projection, which is the S-dual  of the
$O5^-$ plane plus a physical D5-brane, in \cite{Sen}. Using this
result, we can get the mirror theory for $Sp(k) $  \cite{Kap,Han2}
by using the orientifold five plane in the initial brane setup.
For $SO(k) $, if we insist on using the orientifold five plane
again in the brane setup, we must know what is the gauge theory
under the $ON^+$ projection which is the S-dual of $O5^+$ plane.
 It is still
an open problem to read it out.

\par
In the above paragraph, we mention that there is a difficulty to use 
orientifold five-plane to construct the mirror theory of $SO(n)$ 
gauge group.
However, for constructing the $Sp(k) $ or $SO(n) $
gauge theory we can use an O3-plane instead of the O5-plane.
Because under S-duality the O3-plane changes into another
O3-plane, we know how to read out the gauge theory (unlike the
O5-plane which becomes $ON$ plane under  S-duality). Motivated
by this observation, in this paper we use  O3-planes
 to investigate the mirror theory of $SO(n) $ and $Sp(k) $
gauge groups. In particular, we get the mirror
theory for $SO(n)$ gauge group
 which is a completely new result. Furthermore, our proposal for
the construction of the mirror theory predicts a nontrivial strong 
coupling limit
of field theories with eight supercharges.

\par
The contents of the paper are  as follows. In section 2, we discuss
some basic facts on Op-planes  which  will set the stage for 
calculating the mirrors. These include the four
kinds of O3-planes and the $s$-configuration involving 1/2NS-brane
and 1/2D5-brane. In section 3 we discuss  the splitting of 
physical D5-branes on O3-planes. It is a crucial ingredient in
our construction of mirror theory. By S-duality, we get the rules
for how a physical NS-brane can split into two 1/2NS-branes or conversely
 how two 1/2NS-branes can combine into a physical NS-brane. 
The latter  predicts a  nontrivial transition of 
strongly coupled field theories.  
After these  preparations, we give the mirror theory of a single 
gauge group with some flavors: $Sp(k) $ in section 4, $Sp'(k)$ in 
section\footnote{There are two ways to get $Sp(k) $ gauge group: by
$O3^+$-plane or $\widetilde{O3^+}$-plane. We denote 
the theory given by $O3^+$-plane as $Sp(k) $ and the theory 
given by $\widetilde{O3^+}$-plane as $Sp'(k) $.} 5,
$SO(2k) $ in section 6 and $SO(2k+1)$ in section 7. 
In sections eight and nine we generalize the mirror
construction to products of two gauge groups: $Sp(k) \times SO(2m)
$ in section 8 and $Sp'(k) \times SO(2m+1)$ in section 9. 
Finally, we give  conclusions in section 10.

\section{Some facts  concerning  O3-planes}
In this section, we summarize some facts about the O3-plane which will be
useful for the mirror construction later.

\subsection{The four kinds of O3-planes}
There are four kinds of O3-planes which we will meet in this paper
(for a more detailed discussion, see \cite{Wit1}):
$O3^+,O3^-,\widetilde{O3^+},\widetilde{O3^-}$.  However,
before entering the specific discussion of $O3$-planes let us
start from general $Op$-planes. When $p \leq 5$, there exist four
kinds of orientifolds
$Op^+,Op^-,\widetilde{Op^+},\widetilde{Op^-}$.  Among these four 
  we are very familiar with
$Op^+,Op^-,\widetilde{Op^-}$. They can be  described
perturbatively as the fixed planes of the orientifold projection
$\Omega$ which acts on the world sheet as well as the Chan-Paton
factors. By different choices of the action $\Omega$ on the
Chan-Paton factors we get two kinds of projections which we denote
as  $\pm$ projection. In the $+$ case, we can put only an even
number of 1/2Dp-branes and the corresponding plane is the $Op^+$
plane. In the $-$ case, we can put an even or odd number of
1/2Dp-branes and the corresponding plane is $Op^-$ for even number
of 1/2Dp-branes and $\widetilde{Op^-}$ for odd number of
1/2Dp-branes. For $\widetilde{Op^-}$, because there is an odd
number of 1/2Dp-branes, one 1/2Dp-brane must be stuck on the
orientifold plane so that sometimes we consider the
$\widetilde{Op^-}$ as the bound state of the $Op^-$ and the
1/2Dp-brane (for more detailed discussion, the reader is referred
to \cite{Joe}). The $\widetilde{Op^+}$ is more complicated
 and is discussed in detail by Witten
in \cite{Wit1}. In that paper, Witten observes  $O3$-planes from a
more unified point of view, namely discrete torsion (he
 deals with  $O3$-planes. However the discussion
can be easily generalized to other $Op$-planes). We can
distinguish $Op$-planes by two $Z_2$ charges $(b,c) $ with the
definition $b=\int_{RP^2} B_{NS}$ and $c=\int_{RP^{5-p}} C^{5-p}$
(the $(b,c)$ is defined under modular two and the discussion presented here comes from lecture \cite{Han3}
already given by one of the authors at ITP, Santa Barbara; see
also \cite{Han4}). The second charge $c$ exists only for $p\leq
5$. For $p>5$, it can not be defined and we are left only with two
types of $Op$-planes (it is a little mysterious that
$\widetilde{Op^-}$ does not exist for $p>5$, some arguments can be
found in \cite{Han3,Han4}). We summarize the properties of these
four $Op$-planes  according the discrete torsions $(b,c) $ in Table
\ref{tab:fourplane} (where  S-duality is applied only to $p=3$).
\begin{table}[h]
\label{tab:fourplane}
\caption{The summary of the properties of the four $Op$-planes. The charge is
in  units of physical Dp-brane.}
\begin{tabular}{|c|c|c|c|c|}  \hline
$(b,c) $   &   notation    &  charge   &   Gauge group & (b,c)
after S-duality($p=3$ only)  \\   \hline $(0,0) $   &     $ O^-$ &
$ -2^{p-5}$ &   $ SO(2n) $ & (0,0)  $O^-$
      \\   \hline
$(0,1) $   & $\widetilde{O^-}$& $\frac{1}{2}-2^{p-5}$ & $SO(2n+1)
$ & (1,0)  $O^+$ \\   \hline $(1,0) $   &     $ O^+$      & $
2^{p-5}$ &   $ Sp(n) $  &(0,1)  $\widetilde{O^-}$     \\   \hline
$(1,1) $   & $\widetilde{O^+}$      &  $2^{p-5}$ &   $ Sp'(n)$ 
& (1,1)  $\widetilde{O^+}$ \\   \hline
\end{tabular}
\end{table}

\par
These four kinds of $O$-planes are not unrelated to each other and
in fact change to each other when they pass through the
1/2NS-brane or 1/2D-brane \cite{Eva,Han2,Han4}. The change is
shown in Figure  \ref{fig:four_plane}: when
$Op^-$($\widetilde{Op^-}$)  passes through the 1/2NS-brane, it
changes to $Op^+$($\widetilde{Op^+}$)  and vice versa; when
$Op^-$($Op^+$)
 passes through
the 1/2D(p+2) -brane, it changes to
$\widetilde{Op^-}$($\widetilde{Op^+}$)  and vice versa. 

\EPSFIGURE[h]{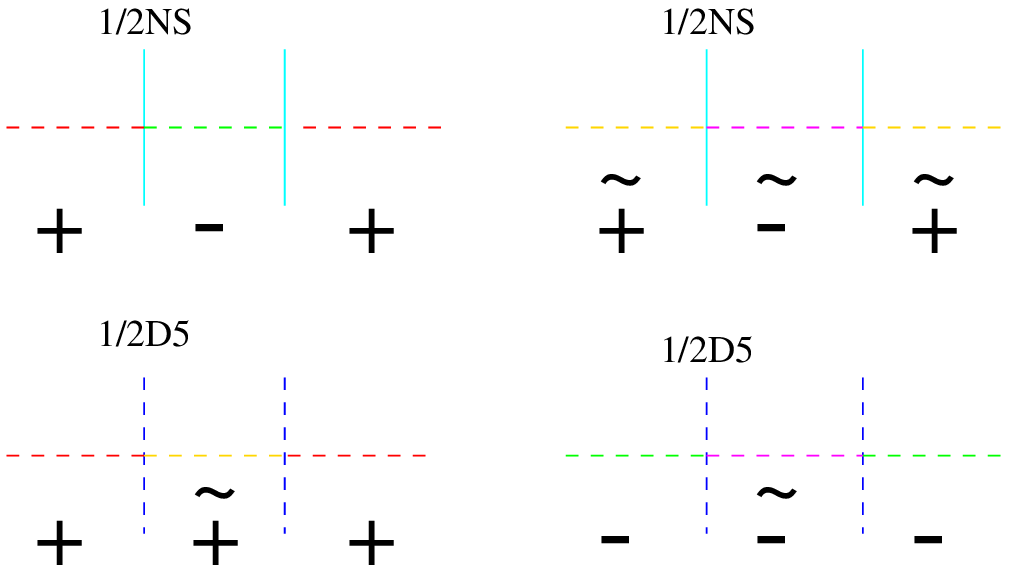}
{The change of four kinds of O3-planes as they cross
1/2NS-branes and 1/2D5-branes. In our brane setup, D3-brane and $O3$-plane
 will extend along $X^{0126}$, D5-brane, $X^{012789}$ and NS-brane,
$X^{012345}$. Henceforth, we use cyan (if the reader
uses colored postscript rendering)  lines to denote the 1/2NS-brane,
blue  lines to denote the 1/2D5-brane,
 dotted horizontal (red)  lines to denote the $O3^+$-plane,  dotted
horizontal (green)  lines to denote the $O3^-$-plane,  dotted
horizontal (yellow)  lines to denote the $\widetilde{O3^+}$  and
finally dotted horizontal (pink)  lines to denote the
$\widetilde{O3^-}$. Furthermore, for simplicity, we use
$-,+,\widetilde{-},\widetilde{+}$ to denote $O3^-,O3^+,
\widetilde{O3^-},\widetilde{O3^+}$ respectively.
\label{fig:four_plane}}

\par
After the discussion of general $Op$-planes, we focus on $O3$-planes which
will be used throughout  this paper.
For $O3$-planes,  the charge of
$O3^-$ is  $-1/4$ while the charges of
$O3^+,\widetilde{O3^-},\widetilde{O3^+}$ are $1/4$. The fact that
the charges for the latter three $O3$-planes are identical  is not
a coincidence and  they are related to each other by the $SL(2,Z)
$ duality symmetry in Type IIB. In particular, under  S-duality
$O3^+$ and $\widetilde{O3^-}$ transform to each other while
$\widetilde{O3^+}$ transforms to itself. $O3^-$ transforms  to itself also
under S-duality because it is the only O3-plane with $-1/4$ charge.
 One immediate application of the
above S-duality property is that the change of O3-planes
crossing the 1/2NS-brane is exactly S-dual to the change of
O3-planes  crossing the 1/2D5-brane. So our rule is consistent.
The above discussions will be useful  later in the study of mirror
symmetry.

\subsection{The supersymmetric configuration}
In the procedures involved in  mirror transformations, we need to
break the D3-branes between the NS-brane and D5-brane to avoid
the so called $s$-rule \cite{Han1}. Furthermore, to read out the
mirror theory from the brane setup it is convenient
 to  move a 1/2NS-brane
along the $X^6$ direction (our notations and conventions for the
brane setups for all kinds of branes is given in the caption of
Figure \ref{fig:four_plane}.)
 to pass through
the 1/2D5-brane
such that the D3-branes ending on the 1/2NS-branes are annihilated
in order to keep the
linking number between 1/2NS-brane and 1/2D5-brane invariant.
All these actions
require the understanding of  supersymmetric configurations in the presence of
O3-planes. We summarize these results
 in this subsection. The tool in our discussion of $s$-configuration is still
 the conservation of linking number between 1/2NS-brane and 1/2D5-brane.
The formula of linking number for 1/2NS-brane and 1/2D5-brane \cite{Han1} is
\begin{equation}
\label{linking}
\begin{array}{lll}
L_{NS} &=&  \frac{1}{2}(R_{D5}-L_{D5}) +(L_{D3}-R_{D3})   \\
L_{D5} &=&  \frac{1}{2}(R_{NS}-L_{NS}) +(L_{D3}-R_{D3})
\end{array}
\end{equation}
where $R_{D5}~ (L_{D5}) $ is the D5-charge to the right (left)  of
NS-brane (1/2D5-brane has 1/2 charge)  and similar definition to
others. Because we have four kinds of O3-planes we  will have four
kinds of  supersymmetric configurations including  one 1/2-NS brane and  one
1/2-D5 brane. These four different cases are:
\begin{equation}
\label{case}
\begin{array}{llll}
(1)  &  O3^-  &  (1/2D5-1/2NS) ~or~(1/2NS-1/2D5)   &
\widetilde{O3^+},\\ (2)  &  O3^+  & (1/2D5-1/2NS)
~or~(1/2NS-1/2D5)   &  \widetilde{O3^-},\\ (3)  &
\widetilde{O3^-} &  (1/2D5-1/2NS) ~or~(1/2NS-1/2D5)   &  O3^+,\\
(4)  & \widetilde{O3^+}  &  (1/2D5-1/2NS) ~or~(1/2NS-1/2D5)   &
O3^-,\\
\end{array}
\end{equation}
where the configuration $O3^-~~  (1/2D5-1/2NS) ~or~(1/2NS-1/2D5)
~~ \widetilde{O3^+}$ means that the $O3^-$ plane is at the left,
$\widetilde{O3^+}$ at the right. In the middle we put 1/2NS-brane
and 1/2D5-brane according to the order  1/2D5-1/2NS from left to
right (see part (a)  of Figure  \ref{fig:4case})
 or  1/2NS-1/2D5 (see part (b)  of Figure  \ref{fig:4case}).

\EPSFIGURE[h]{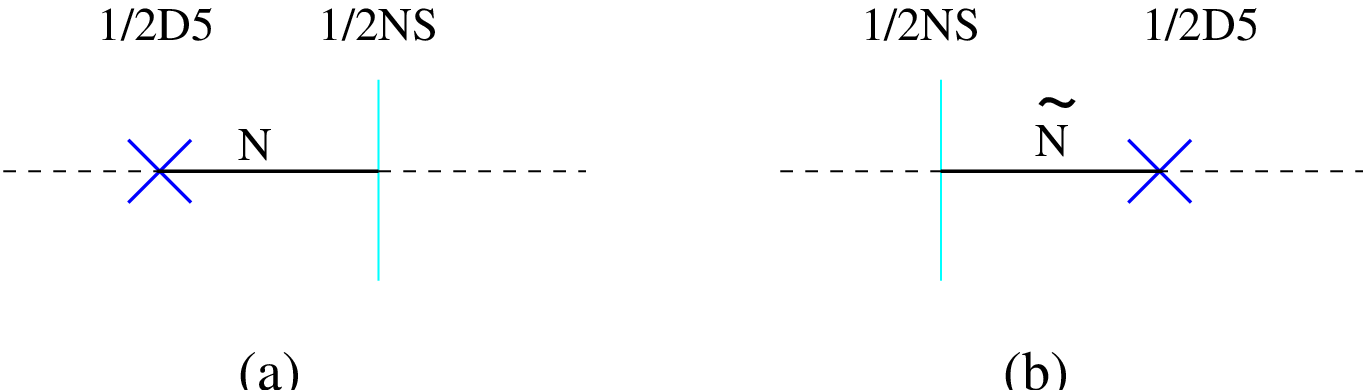}{Starting from any one (left or right figure)
, we move  the $1/2$NS-brane along $X^6$ direction to pass
$1/2$D5-brane and get the other (right or left). To allow
such a process, we must conserve the linking number with the
condition that $N,\tilde{N} \geq 0$. In this figure and
henceforth, when NS-brane and D5-brane show at the same time in
the figure with proper two-dimensional coordinates  (for
example, here  $X=X^6,Y=X^5$), for
clarification we use a line to denote an extended brane in these
coordinates
 and use a
cross to denote  a point-like brane.
\label{fig:4case}}
The general pattern  for the above four supersymmetric configurations
 is shown in Figure  \ref{fig:4case}, where
we assume the number of connected D3-branes (in  physical units)
from the
 left 1/2D5-brane
to the right 1/2NS-brane is $N$ and from the left 1/2NS-brane to the right 1/2
D5-brane is $\widetilde{N}$. So a configuration to be supersymmetric
 is equivalent to the solution of $N,\tilde{N}\geq 0$
such that they conserve the linking number after  crossing.

\subsubsection{The first case: $O3^- --- \widetilde{O3^+}$}
In this case we start from the brane setup (a)  of Figure
\ref{fig:4case} with $O3^-$ plane at the left, $\widetilde{O3^-}$
plane in the middle and $\widetilde{O3^+}$ plane at the right. The
linking numbers are
$
L_{1/2D5}=\frac{1}{2}(\frac{1}{2}-0) +[(-\frac{1}{4})
-(\frac{1}{4}+N) ]=
  -N-\frac{1}{4}$ and $
L_{1/2NS}=\frac{1}{2}(0-\frac{1}{2}) +[(\frac{1}{4}+N)
-(\frac{1}{4}) ]=
  N-\frac{1}{4}$.
Now we move the 1/2D5 along $X^6$ direction to pass through
  1/2NS and get the
(b)  of Figure  \ref{fig:4case} with $O3^-$ plane at the left,
$O3^+$ plane in the middle and $\widetilde{O3^+}$ plane at the
right. For the latter   we have linking numbers as
$
L_{1/2D5}=\frac{1}{2}(0-\frac{1}{2}) +[(\tilde{N}+\frac{1}{4})
-(\frac{1}{4}) ]=
  \tilde{N}-\frac{1}{4}$ and $
L_{1/2NS}=\frac{1}{2}(\frac{1}{2}-0) +[(-\frac{1}{4})
-(\tilde{N}+\frac{1}{4}) ]=
  -\tilde{N}-\frac{1}{4}.$
Comparing these two linking numbers we
get
\begin{equation}
\label{case1}
N=-\tilde{N}
\end{equation}
It is a highly constraining equation. For the supersymmetric configuration,
the only solution is $N=\tilde{N}=0$.
This means that when we break the D3-brane
to go to the Higgs branch, we {\bf can not} put D3-brane between 1/2NS-brane
and 1/2D5-brane in this orientifold configuration.

\subsubsection{The second case: $O3^+ --- \widetilde{O3^-}$}
Starting from brane setup (a)  of  Figure  \ref{fig:4case} with
$O3^+$ at the left, $\widetilde{O3^+}$ in the middle and
$\widetilde{O3^-}$ at the right,  we find the linking numbers as
$
L_{1/2D5}=\frac{1}{2}(\frac{1}{2}-0) +[(\frac{1}{4})
-(\frac{1}{4}+N) ]=
  -N+\frac{1}{4}$ and
$L_{1/2NS}=\frac{1}{2}(0-\frac{1}{2}) +[(\frac{1}{4}+N)
-(\frac{1}{4}) ]=
  N-\frac{1}{4}$. Again by moving  the 1/2D5-brane to pass through
1/2NS-brane we get the brane setup as (b)  with the middle
O3-plane changed from $\widetilde{O3^+}$ in (a)  to $O3^-$ in (b)
(the left and right O3-plane are invariant under the motion). The
linking numbers for the latter are
$
L_{1/2D5}=\frac{1}{2}(0-\frac{1}{2}) +[(\tilde{N}-\frac{1}{4})
-(\frac{1}{4}) ]=
  \tilde{N}-\frac{3}{4}$ and $
L_{1/2NS}=\frac{1}{2}(\frac{1}{2}-0) +[(\frac{1}{4})
-(\tilde{N}-\frac{1}{4}) ]=
  -\tilde{N}+\frac{3}{4}$.
From these relations  we find the equation
\begin{equation}
\label{case2} -N+1= \tilde{N}.
\end{equation}
So for a consistent supersymmetric configuration there are three solutions:
$(N,\tilde{N}) =(0,1) ;(\frac{1}{2},\frac{1}{2}) ;(1,0) $.

\subsubsection{The third  case: $\widetilde{O3^-}--- O3^+$}
For the third case, we start from the brane setup (a)  with
$\widetilde{O3^-}$ at the left, $O3^-$ in the middle and $~O3^+$
at the right. The linking numbers are
$
L_{1/2D5}=\frac{1}{2}(\frac{1}{2}-0) +[(\frac{1}{4})
-(-\frac{1}{4}+N) ]=
  -N+\frac{3}{4}$ and $
L_{1/2NS}=\frac{1}{2}(0-\frac{1}{2}) +[(-\frac{1}{4}+N)
-(\frac{1}{4}) ]=
  N-\frac{3}{4}$.
Now we move the 1/2D5-brane to pass through 1/2NS-brane and get
the brane setup (b)  with $\widetilde{O3^+}$ in the middle. The
linking numbers become
$
L_{1/2D5}=\frac{1}{2}(0-\frac{1}{2}) +[(\tilde{N}+\frac{1}{4})
-(\frac{1}{4}) ]=
  \tilde{N}-\frac{1}{4}$ and $
L_{1/2NS}=\frac{1}{2}(\frac{1}{2}-0) +[(\frac{1}{4})
-(\tilde{N}+\frac{1}{4}) ]=
  -\tilde{N}+\frac{1}{4}$.
By comparing these  relations  we have 
\begin{equation}
\label{case3}
-N+1= \tilde{N}.
\end{equation}
So again there are three solutions: $(N,\tilde{N}) =(0,1)
;(\frac{1}{2},\frac{1}{2}) ;(1,0) $.

\subsubsection{The fourth case: $\widetilde{O3^+}---O3^-$}
For the last case we start from the brane setup (a)  with
$\widetilde{O3^+}$ at the left, $O3^+$ in the middle and $O3^-$ at
the right. The linking numbers are
$
L_{1/2D5}=\frac{1}{2}(\frac{1}{2}-0) +[(\frac{1}{4})
-(\frac{1}{4}+N) ]=
  -N+\frac{1}{4}$ and $
L_{1/2NS}=\frac{1}{2}(0-\frac{1}{2}) +[(\frac{1}{4}+N)
-(-\frac{1}{4}) ]=
  N+\frac{1}{4}$.
Now we move the 1/2D5-brane to pass through 1/2NS-brane and get
the brane setup (b)  with $\widetilde{O3^-}$ in the middle. The
linking numbers change to
$
L_{1/2D5}=\frac{1}{2}(0-\frac{1}{2}) +[(\tilde{N}+\frac{1}{4})
-(-\frac{1}{4}) ]=
  \tilde{N}+\frac{1}{4}$ and $
L_{1/2NS}=\frac{1}{2}(\frac{1}{2}-0) +[(\frac{1}{4})
-(\tilde{N}+\frac{1}{4}) ]=
  -\tilde{N}+\frac{1}{4}$.
From these relations  we have
\begin{equation}
\label{case4}
-N=\tilde{N}.
\end{equation}
The only solution is $(N,\tilde{N}) =(0,0) $ as in the first
case.

\par
Let us summarize the results in the last four subsections. When the
charge of $O3$-planes at the two sides are the same (case two and case three), 
the condition is $N+\tilde{N}=1$, so there is annihilation or creation
of D3-branes in crossing. When the charge of $O3$-planes at the two sides are 
different (case one and case four), the condition is $N=\tilde{N}=0$, 
so there can not be any D3-branes between the 1/2NS-brane and 1/2D5-brane.

\section{The splitting of the physical brane}
To construct the mirror theory by brane setups, we can follow  the
procedure given in the introduction \cite{Han1}. However, in the
presence of the O3-plane, we need one new input: how to {\sl split}
the physical D5-brane into two 1/2D5-branes on the O3-plane. Initially, the
physical D5-brane can be placed off the O3-plane in pairs of
1/2D5-branes (see Figure \ref{fig:split_gen}). We
can move the pair of 1/2D5-branes to touch the O3-plane. After
touching the O3-plane, in principle every 1/2D5-brane can move
freely on the O3-plane. We call such an independent motion of the 
1/2D5-brane as ``{\sl splitting}'' of the  physical D5-brane.
We want to emphasize that the splitting of a physical
D5-brane into  two 1/2D5-branes is a nontrivial dynamical process
in string theory and can be applied to many situations.
Here we need  the splitting  because in the mirror
theory, the gauge theory is given by D3-branes ending on
1/2NS-branes which are the S-dual of 1/2D5-branes in the original
theory. In this paper, we give only a preliminary discussion. We found
some novel results: sometimes there is a creation of one
physical D3-brane between these two 1/2D5-branes; sometimes   
there is an annihilation and sometimes, no
creation and no annihilation. We found these results by matching
the Higgs branch moduli of the $Sp$ or $SO$ theory with the 
correct dimension of moduli space.

\EPSFIGURE[h]{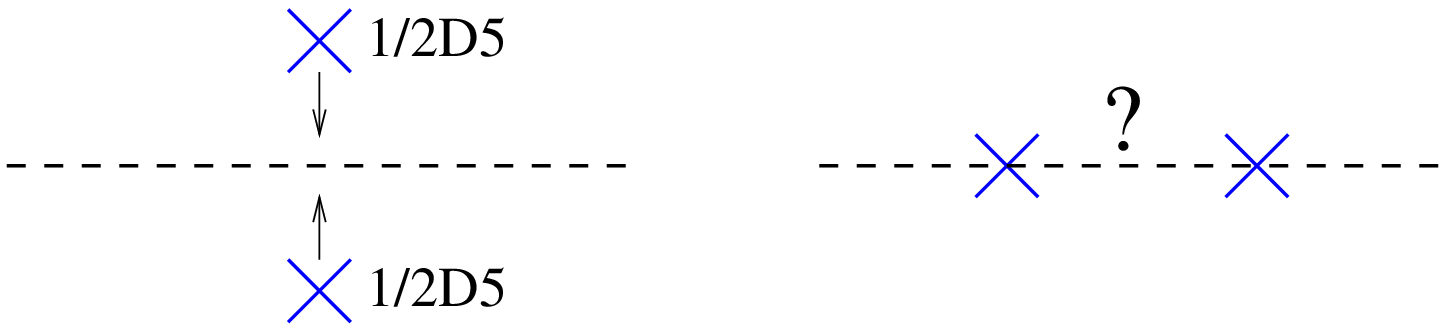,width=15cm} {Splitting of a D5-brane on the
O3-planes. The left figure shows that a pair of 1/2D5-branes moving to
touch the O3-plane. The right figure shows that when they touch the 
O3-plane they can split. The $?$ in the middle of these two 1/2D5-branes
means there is  nontrivial dynamics dependent on different situations.
 \label{fig:split_gen}}

\subsection{The splitting of D5-branes without ending D3-branes}
Before going to the general situation let us discuss the splitting of
D5-branes which do not have any D3-branes ending on them. First
we discuss the case where there is only one physical D5-brane and O3-plane
(see Figure \ref{fig:split_gen}). Before splitting, every 1/2D5-brane
has  linking number {\sl zero}. After splitting, there can be 
$N$ physical D3-branes between these two 1/2D5-branes (to keep 
supersymmetry, there can not be anti-D3-branes between them; furthermore,
because here we do not  have any D3-branes initially, there can not be
annihilation either). Let us calculate the linking number after splitting:

\begin{equation}
\label{linkingchange}
\begin{array}{ccc}
O3~before~splitting~~~~~  & ~~~~~~~ \Delta L_{left}~~~~~~~
&    \Delta L_{right}  \\
O3^+  & -N  &  N  \\
\widetilde{O3^+}  & -N  &  N  \\
O3^-  & -\frac{1}{2}-N  & \frac{1}{2}+N  \\
\widetilde{O3^-}  & \frac{1}{2}-N  & -\frac{1}{2}+N  \\
\end{array}
\end{equation}

In BPS states, we have the tension of D5-branes proportional
to their charge (linking number). 
To have the minimum
tension configuration, it is natural to have $N=0$ for the first three cases.
However, for the last case, $N=0$ and $N=1$ are equally favorable just from the
point view of tension. We will fix the ambiguity in the 
next paragraph. However, 
before we end this paragraph, we want to emphasize that no matter what
case it is, the total change in linking number is always
$$
\Delta L=0 ~~~~~or ~~~~~~\Delta L=\pm \frac{1}{2}~~.
$$

\par
We can fix the ambiguity for the last case by considering  Higgsing.
Starting from the $SO(3)$ gauge theory with one flavor, we can Higgs it
to $SO(2)$ with one singlet (there are $3-1=2$ gauge fields which 
get mass, so
we leave only $3\times 1-2=1$ singlet). In  part (a) of Figure 
\ref{fig:SO3_Higgs} we assume $N=0$ in the splitting process 
and go to the Higgs branch.
By moving 1/2D5-branes outside  we find the final theory is $SO(2)$ without
singlets in part (b). This means that our assumption is wrong. Choosing
the other assumption $N=1$ in part (c), by moving 1/2D5-branes outside we
get the final theory is $SO(2)$ with a
singlet in part (d) which is exactly what we expect from the field theory.
This shows that, for matching the correct moduli dimensions of Higgs branch, 
 in last case of (\ref{linkingchange}) there should be a D3-brane
created in the splitting.

\EPSFIGURE[h]{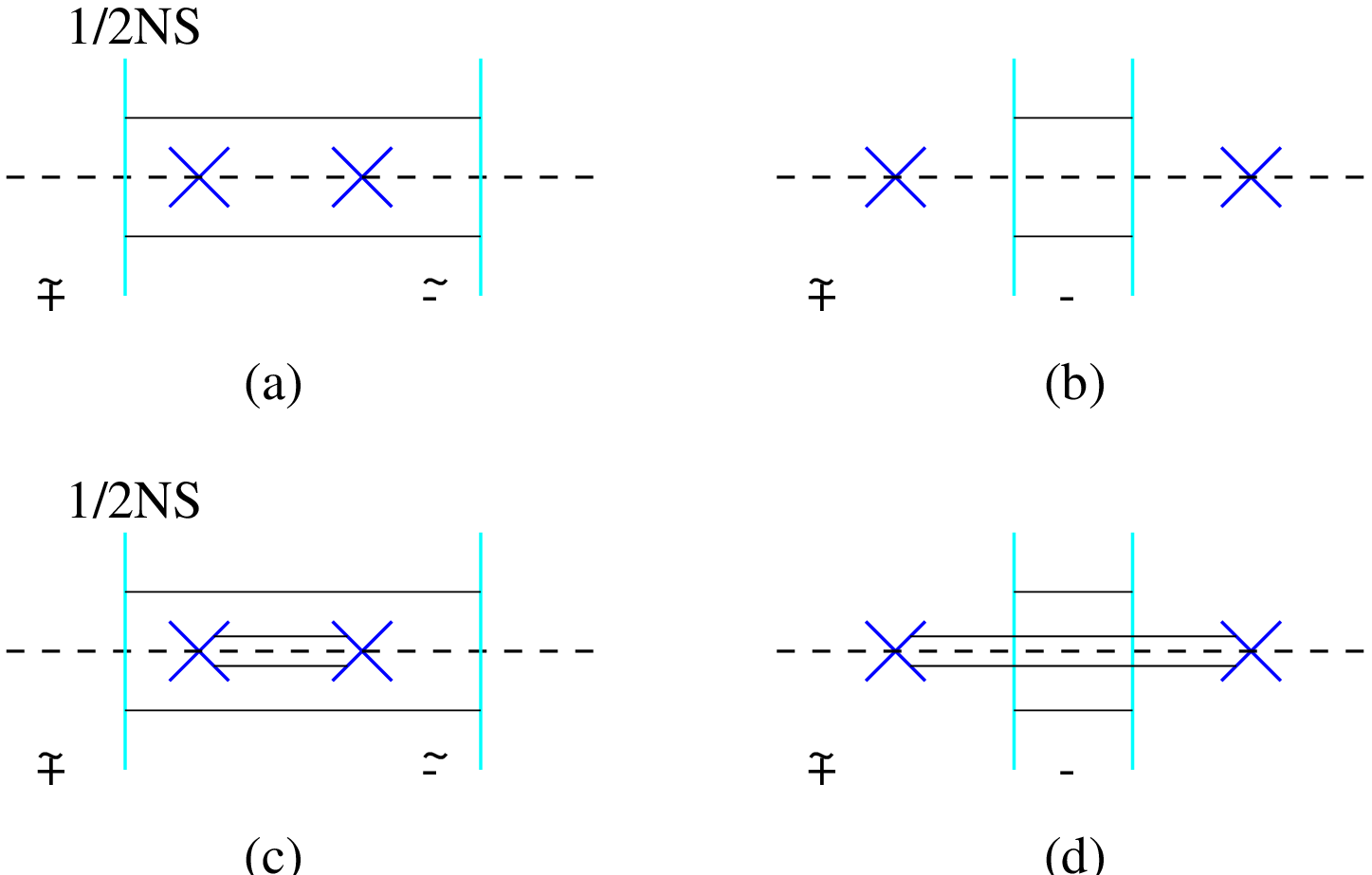,width=15cm} {The Higgs branch of $SO(3)$ with one
flavor. (a) We assume when the splitting, there is no D3-brane generated.
(b) By moving 1/2D5-branes outside, we get $SO(2)$ without singlet.
(c) We assume when the splitting, there is a D3-brane generated.
(d) By moving 1/2D5-branes outside, we get $SO(2)$ with one singlet given
by D3-brane ending on 1/2D5-brane. 
 \label{fig:SO3_Higgs}}

\par
The discussion of the splitting of physical D5-branes becomes more
complex if there are more than one D5-brane to be split. The complexity
manifests in the last two cases in (\ref{linkingchange}) because in these cases
there is a change of linking number ($\Delta L=\pm \frac{1}{2}$)
 for every 1/2D5-brane. Before splitting, we have, for example,
$2n$ 1/2D5-branes with linking number {\sl zero}. After splitting,
we have $n$ 1/2D5-branes with linking number $\frac{1}{2}$ and
$n$, with linking number $-\frac{1}{2}$. The different order of
linking number gives different physical content, i.e.,  the order determines
when there should be  D3-branes created  and when there are no
D3-branes created.

\par
To illustrate our idea, let us see  Figure
\ref{fig:twoD5}. After splitting one D5-brane according to the analysis
in the last paragraph, we continue to split the second D5-brane. However,
in this case, we have two choices. In the first choice,
 the second D5-brane is far away
from the first D5-brane in $X^6$ direction like part (a). So locally
the splitting should be the same as the first D5-brane and we get part (b).
Notice that the order of linking number of 1/2D5-branes is
$-\frac{1}{2},+\frac{1}{2},-\frac{1}{2},+\frac{1}{2}$. In the second
choice, the second D5-brane is in the middle of the pair of first 
1/2D5-branes as part (c). Naively, the second D5-brane will see the
$O3^-$-plane (in fact, D5-brane will see more) 
so the splitting looks like to  go as part (d) with
the order of linking number 
$-\frac{1}{2},-\frac{1}{2},+\frac{1}{2},+\frac{1}{2}$.
However, part (d) is not consistent with the Higgs branch of $SO(3)$
with two flavors. Furthermore, because there are eight supercharges,
the different positions of D5-branes should not effect the physics.
So we argue that from part (c) we should get part (b) too. In part
(c), the second D5-brane sees not only the $O3^-$-plane, 
but also the one created
D3-branes, 1/2D5-branes at left with $\Delta=-\frac{1}{2}$ and
1/2D5-branes at right with $\Delta=+\frac{1}{2}$. This more
complete information determines that the second D5-brane will 
split to part (b).

\EPSFIGURE[h]{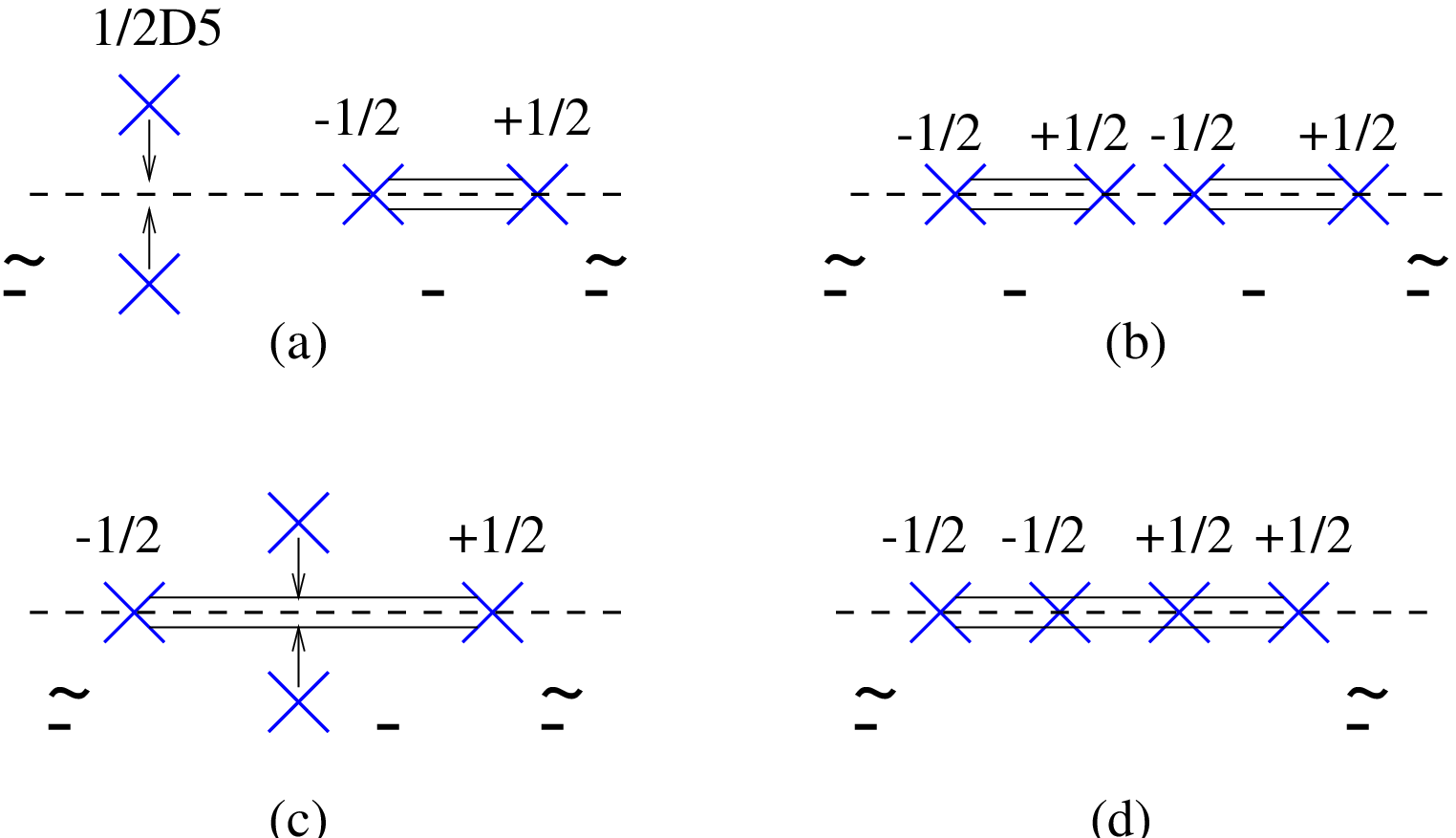} {The splitting of the second D5-brane. (a)
second D5-brane is far away from the first D5-brane. (b) the splitting
of second D5-brane from configuration in part (a). (c). second D5-brane
is in the middle of first D5-brane. (d) the naive splitting of second
D5-brane which turns out to be wrong.
 \label{fig:twoD5}}

\par
From the above observation,
we propose that the correct order of linking number should be
$-\frac{1}{2},+\frac{1}{2},-\frac{1}{2},+\frac{1}{2},....,
-\frac{1}{2},+\frac{1}{2}$ (notice the alternating fashion
 of $-\frac{1}{2}$ and
$+\frac{1}{2}$). We make such a suggestion because it is the only correct
order which can produce the consistent Higgs pattern for $SO(K)$ gauge 
group with $N$ flavors. It will be very interesting if we can 
derive such a rule from string theory. Furthermore, this proposal will
give very interesting predictions which we will discuss later.

\par
Let us pause a moment  to summarize the results we have obtained above.
 Without the D3-brane ending on D5-branes, (1) the change of linking number 
of 1/2D5-branes is $\Delta L=0$ for $O3^+,\widetilde{O3^+}$ and
$\Delta L=\pm \frac{1}{2}$ for $O3^-,\widetilde{O3^-}$;
(2) for the splitting of a bunch of D5-branes, the order of linking
number is $-\frac{1}{2},+\frac{1}{2},-\frac{1}{2},+\frac{1}{2},....,
-\frac{1}{2},+\frac{1}{2}$.

\subsection{The splitting of D5-branes with ending D3-branes}
After the discussion of the splitting of D5-branes without D3-branes ending
on them, we consider the case that there are $N$  D3-branes ending on them.
 The results for this
latter case can be derived from the results in the 
last subsection. For example,
let us discuss the case of 
one D5-brane with one ending D3-brane  
in Figure \ref{fig:oneD3}. We can add one
1/2NS-brane such that the D3-brane ending on it as part (a). Then we can
move D5-brane to the right of 1/2NS-brane and annihilate the D3-brane
as part (b). Now the part (b) is the case we discussed in the last subsection.
We can split the physical D5-brane and move two 1/2D5-branes to left of
1/2NS-brane by using the result in section 2. By this loop, we finally get the
splitting of D5-brane with one ending D3-brane. 
For more D3-branes ending on D5-branes we can add more
1/2NS-brane and repeat the above procedure. 

\EPSFIGURE[h]{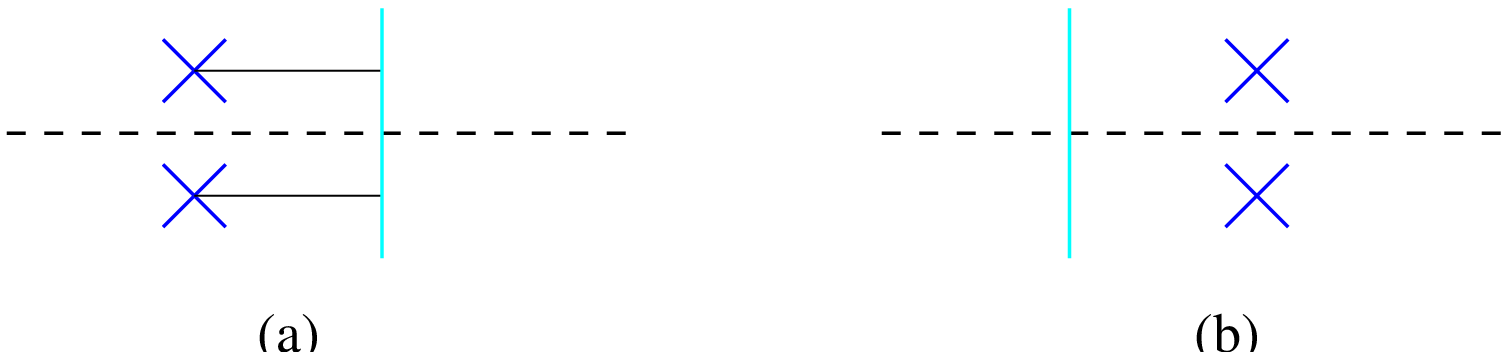} { (a) One D3-brane ends on a physical 
D5-brane. We can add a 1/2NS-brane at the right. It should not affect 
the discussion. (b) By moving D5-brane to right of 1/2NS-brane we annihilate
the ending D3-brane. 
 \label{fig:oneD3}}

\par
Although the above trick solves our problem completely, it is too tedious and
we need  a more direct way to see it. Notice that the change of the linking
number of 1/2D5-branes happens only at splitting. So we can use the changing
of linking number as  the
rule to determine the splitting of D5-brane. In general there will
be
$N_L$ D3-branes ending on D5-brane from the left and $N_R$ D3-branes,
from right. The rule depends only on the absolute difference between
$N_L,N_R$, i.e.,  $N=|N_L-N_R|$.
We summarize
the rule in  Table \ref{tableD5}.
\begin{table}
\label{tableD5}
\caption{The rules of splitting of D5-brane,
where $N=|N_L-N_R|$ is the difference  of D3-branes ending on D5-brane
from the left and the right. }
\begin{tabular}{|c|c|c|}  \hline
O3-plane  &   $N=even$  &  $N=odd$  \\  \hline
$O3^+$    &  $\Delta L=0$  &  $\Delta L=\pm \frac{1}{2}$
in order of $-\frac{1}{2},+\frac{1}{2},-\frac{1}{2},+\frac{1}{2},...$
\\  \hline
$\widetilde{O3^+}$    &  $\Delta L=0$  &  $\Delta L=\pm \frac{1}{2}$
in order of $-\frac{1}{2},+\frac{1}{2},-\frac{1}{2},+\frac{1}{2},...$
\\  \hline
$O3^-$  & $\Delta L=\pm \frac{1}{2}$
in order of $-\frac{1}{2},+\frac{1}{2},-\frac{1}{2},+\frac{1}{2},...$
& $\Delta L=0$  \\  \hline
$\widetilde{O3^-}$  & $\Delta L=\pm \frac{1}{2}$
in order of $-\frac{1}{2},+\frac{1}{2},-\frac{1}{2},+\frac{1}{2},...$
& $\Delta L=0$  \\  \hline
\end{tabular}
\end{table}

\subsection{The splitting of NS-branes and novel predictions of field theory 
in the strong coupling limit}
Making  S-duality, we can get the  rules of  splitting  physical
NS-branes into 1/2NS-branes on O3-plane as Table \ref{table_NS}.
\begin{table}
\label{table_NS}
\caption{The rules of splitting of NS-brane, 
where $N=|N_L-N_R|$ is the difference  of D3-branes ending on NS-brane
from the left and the right.}
\begin{tabular}{|c|c|c|}  \hline
O3-plane  &   $N=even$  &  $N=odd$  \\  \hline
$\widetilde{O3^-}$    &  $\Delta L=0$  &  $\Delta L=\pm \frac{1}{2}$
in order of $-\frac{1}{2},+\frac{1}{2},-\frac{1}{2},+\frac{1}{2},...$
\\  \hline
$\widetilde{O3^+}$    &  $\Delta L=0$  &  $\Delta L=\pm \frac{1}{2}$
in order of $-\frac{1}{2},+\frac{1}{2},-\frac{1}{2},+\frac{1}{2},...$
\\  \hline
$O3^-$  & $\Delta L=\pm \frac{1}{2}$
in order of $-\frac{1}{2},+\frac{1}{2},-\frac{1}{2},+\frac{1}{2},...$
& $\Delta L=0$  \\  \hline
$O3^+$  & $\Delta L=\pm \frac{1}{2}$
in order of $-\frac{1}{2},+\frac{1}{2},-\frac{1}{2},+\frac{1}{2},...$
& $\Delta L=0$  \\  \hline
\end{tabular}
\end{table}

\par
From Table \ref{table_NS}, we get two predictions of $N=4$ three dimensional
field theory 
in the strong coupling limit (see Figure \ref{fig:prediction}). 
In the first case (part (a) of Figure \ref{fig:prediction}), 
the field theory is $SO(2k)\times Sp(k)\times SO(2k)$ with two 
bifundamentals. From the brane setup in part (a), we see that, by
reversing the process of the splitting of the NS-brane, 
 we can move two middle 1/2NS-branes to meet together 
and leave $O3^-$-plane. In field theory, moving two middle 
1/2NS-branes together  corresponds to
the strong coupling limit of  $Sp(2k)$ gauge theory, and moving NS-brane
off the $O3^-$-plane corresponds to turning on  {\sl ``FI-parameters''}
\footnote{In fact, it is a hidden {\sl ``FI-parameters''}. We will discuss
it more in section 4.3}.   So our
brane configuration predicts that, at the strong coupling limit of
$Sp(2k)$ and the turning of FI-parameter, the original theory
$SO(2k)\times Sp(k)\times SO(2k)$ with two 
bifundamentals  will flow
to $SO(2k)$ without any flavor. The second case is given in part (b)
of Figure \ref{fig:prediction}. By the similar  arguments, we  predict
that at the strong coupling limit of
$SO(2k+2)$ and the turning of FI-parameter, the field theory
$Sp(k)\times SO(2k+2)\times Sp(k)$ with two 
bifundamentals  will flow to $Sp(k)$ without any flavor. It will be
interesting to check these two predictions from the field theory point of
view.

\EPSFIGURE[h]{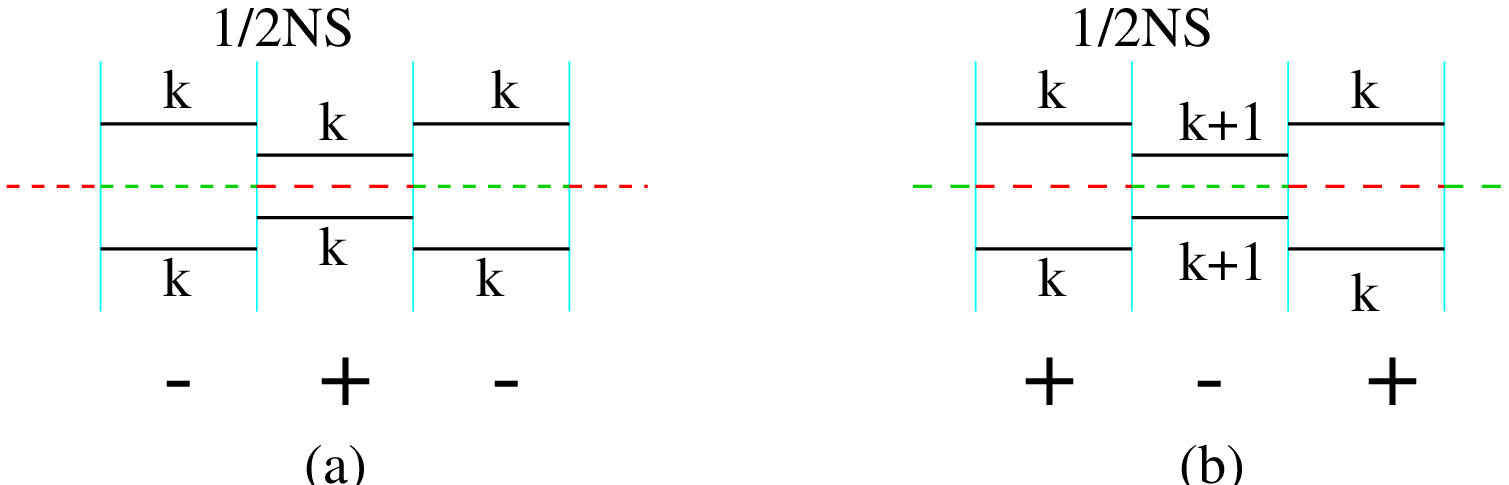} { (a) $SO(2k)\times Sp(k)\times SO(2k)$ gauge 
theory with two 
bifundamentals.  (b) $Sp(k)\times SO(2k+2)\times Sp(k)$ gauge theory with two 
bifundamentals.
 \label{fig:prediction}}

\section{The mirror of  $Sp(k) $ gauge theory}
Now we start to construct mirror pairs  using the above
knowledge. First let us discuss   $Sp(k) $ gauge theory with $N$
fundamental flavors. In this case, the brane setup is  as follows: we
put $2k$  1/2D3-branes, i.e., branes and their images under
the $O$-plane (extended in $X^{0126}$)  ending
 on two 1/2NS branes
(extended in $X^{012345}$)  along $X^6$  while $2N$ 1/2D5-branes
(extended in $X^{012789}$)  are put in the middle (see (a)  of
figure  \ref{fig:Sp1_3}). Then $O3$-planes from the left to right
read as $O3^-,O3^+,O3^-$. This $O3$-plane configuration  reminds
us of the special $s$-configuration discussed in the last section.
Because in the presence of $O3$-planes the $s$-configuration is a
little different from the known $s$-rule in \cite{Han1}, we will
demonstrate the detailed steps for the mirror construction for
$Sp(1) $ with three flavors. Thereafter we  quickly go to the
general $Sp(k) $ case.

\subsection{$Sp(1) $ with the $3$ flavors}
For  the $Sp(1) $ gauge theory with $3$ flavors we have the
following information about the moduli space of the Higgs branch and
the Coulomb branch as well as the FI-parameters and mass parameters:
\begin{equation}
\label{Sp.1}
\begin{array}{lll}
d_v  & = & 1,\\
d_H  & = & 3\times 2-3=3, \\
\#m & = & 3, \\
\# \zeta & = & 0
\end{array}
\end{equation}
After the mirror map, we should have a mirror theory which has
$d_v=3, d_H=1,\#\zeta=3,\#m=0$, i.e. the Coulomb branch and the Higgs branch
are interchanged while the  mass parameters and FI-parameters are
exchanged \cite{Int}. However, when we count these parameters,
sometimes we meet  nontrivial situations, such as
 the ``hidden FI-term'' explained in \cite{Kap}.
We will see later that these ``hidden parameters'' arise in our
 construction and will discuss them in more detail
later.

\par
The details of the mirror construction are given in Figure
\ref{fig:Sp1_3}. Let us go step by step. Part (a)  is just the
brane setup for $Sp(1) $ with three fundamental flavors. By moving
the physical D5-brane to touch the orientifold $O3^+$ plane, i.e.,  
setting the masses to zero, we
can split them into 1/2D5-branes as in part (b). Now we go to the
Higgs branch by splitting the D3-branes between those 1/2NS-branes
and 1/2D5-branes. However, from  (\ref{case1}) and (\ref{case4}),
 we must split these D3-branes as given by part (c). 
The crucial point is that there is no D3-brane connected between
the 1/2NS-brane and its nearest 1/2D5-brane because it is
prohibited by the supersymmetric configuration discussed in section 2. Now we
can use the rules (\ref{case1})  and (\ref{case4})
 to move the
left 1/2NS-brane crossing the neighboring right  1/2D5-brane and
the right 1/2NS-brane crossing the neighboring left  1/2D5-brane.
The result is given by part (d). Notice that  in such a process,
no D3-brane is created or annihilated. Applying (\ref{case2})  and 
(\ref{case3})  to move the 1/2NS-brane across 1/2D5-brane, we reach
part (e). In this process, the physical D3-brane which connects
the 1/2NS-brane and 1/2D5-brane is annihilated. Now we can apply
the mirror transformation to give the result shown in part (f).
However, it is a little hard to read out the final gauge theory
because of the $O3^-$ and $\widetilde{O3^-}$ projections in the
same interval. We can get rid of this ambiguity by applying 
(\ref{case1})  and (\ref{case4})  again to reach the result in part
(g).

\EPSFIGURE[ht]{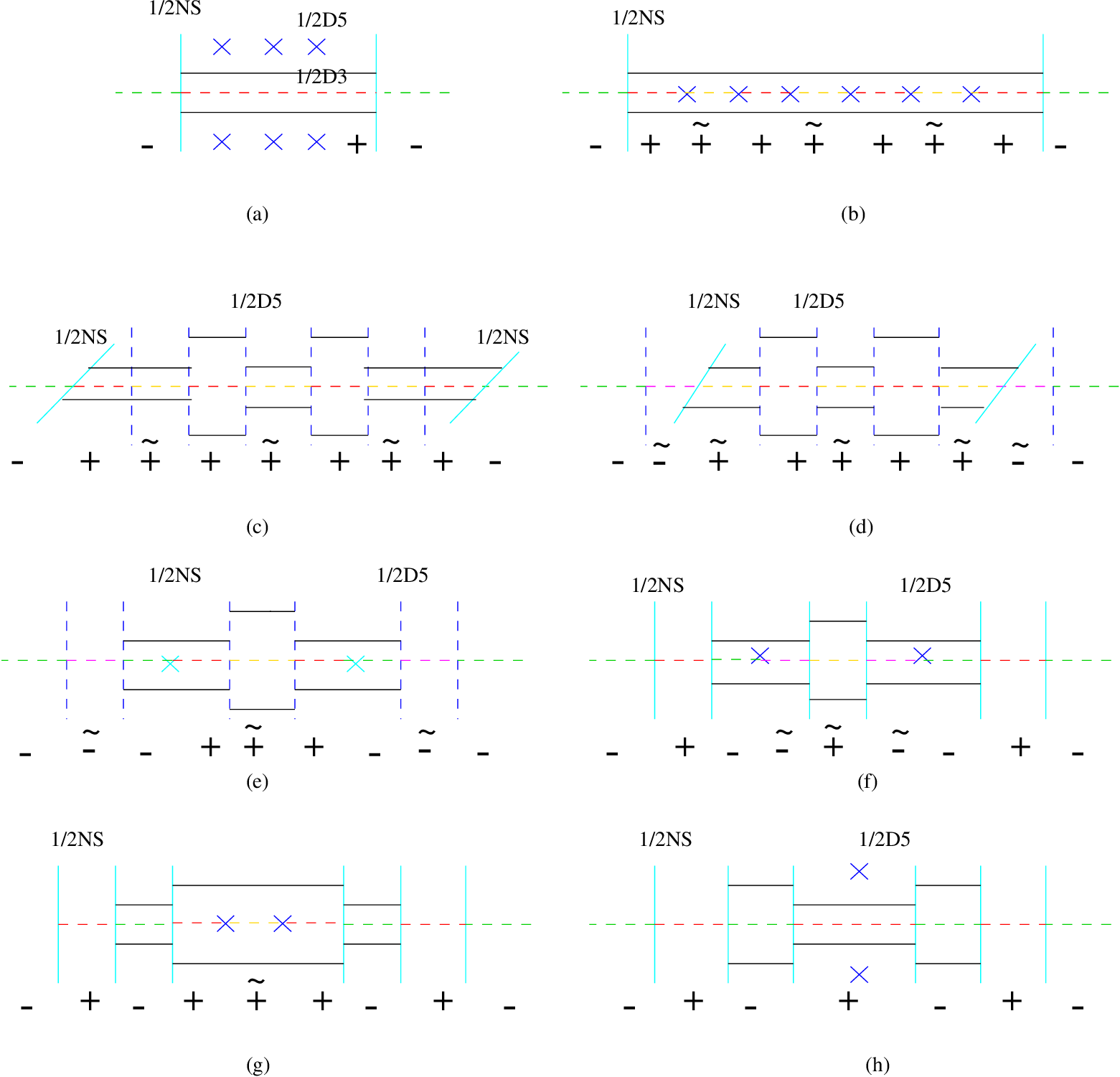,width=15cm}{The detailed steps for getting
the mirror of $Sp(1)$ with three fundamental flavors. (a) The
brane setup. (b)  Splitting of the physical D5-branes. (c) The
Higgs branch obtained by splitting the D3-brane. Notice the
special splitting of these D3-branes. (d)  Using the result of
supersymmetric configuration we can move 1/2NS-brane one step inside. (e)
Using again the rule of supersymmetric  configuration we move the 1/2NS-brane
one further step inside. In this step, the D3-brane ending on the
1/2NS-brane is  annihilated. (f)  S-dual of part (e). (g)  However,
we can not read out the final gauge theory from the brane setup in
(f). For avoiding the ambiguity, we can move 1/2NS-brane one
further step inside. (h)  A special property of our example is
that we can combine two 1/2D5-branes in part (g) together and leave
the $O3^+$ plane. \label{fig:Sp1_3}}


\par
Now we have the brane setup for the mirror theory in part (g)  of
figure
 \ref{fig:Sp1_3}. We can read out the theory directly from the brane setup
according the standard rule: For $2k$ 1/2D3-brane stretching
between two 1/2NS-branes with
$O3^-,\widetilde{O3^-},O3^+,\widetilde{O3^+}$ planes we get
$SO(2k) ,SO(2k+1) ,Sp(k) ,Sp'(k) $ gauge groups respectively. For
one 1/2D5-brane between two 1/2NS-branes it contributes one
fundamental half-hypermultiplet for that gauge group. For one
physical D5-brane between two 1/2NS-branes it contributes one
fundamental hypermultiplet for the gauge group. For two gauge
groups which have a common 1/2NS-brane there is a bifundamental
(in the presence of $O3$-plane, such bifundamental is, more
exactly, half-hypermultiplet). Applying the above rules we
immediately get 
 the mirror theory as $SO(2) \times Sp(1)
\times SO(2) $ with two bi-fundamentals and one fundamental for
$Sp(1) $. Here we want to emphasize that in general we get only
half fundamental hypermultiplets coming from the 1/2D5-brane. The
unusual point for this explicit example is that the two
1/2D5-branes are in the same interval such that they can combine
together and leave the orientifold (see section 3).
Now let us calculate the moduli spaces and parameters to see if
they are really mirror to each other. For the mirror theory in the
part(h)  of Figure  \ref{fig:Sp1_3}, it is easy to get the
dimensions of moduli spaces as $d_v=1+1+1=3$ and $d_H=(2\times
2+2\times 2) /2+1\times 2-(1+1+3) =6-5=1$, so we see the results
match when comparing to  \ref{Sp.1}. However, when we turn to
calculate the mass parameters and FI-parameters, a mismatch
occurs. In the mirror theory, we have two bifundamentals and one
fundamental. For the two bifundamentals we do not know how to turn
the mass parameters so we get the $\# m=1$. Because there are no
 $U(1) $ factors in the mirror theory, it seems that we should get
$\# \zeta=0$. Now comparing with the original theory, we find a
mismatch in the mass parameters and FI-parameters. The solution of
the above mismatch is given by the concept of ``hidden FI-term''
which we will discuss later \cite{Kap}.

\subsection{Another method to go to the Higgs branch}
In the above procedure, we split D5-branes first, then went
 to the Higgs branch
by splitting the D3-branes. However, we can go to the Higgs branch
in another way by splitting the D3-brane first on the physical
D5-brane and then splitting the D5-brane on the O3-plane. The
procedure of this second method is drawn in Figure
\ref{fig:second}. In  part (a) , we keep the D5-branes off the
O3-plane and split the D3-branes to go the Higgs branch (such
splitting is very familiar to us already, see \cite{Han1}). By
moving the physical D5-brane to cross the 1/2NS-brane, we can get
rid of the D3-brane ending one D5-brane and 1/2NS-brane. The
result is shown in part (b). Now we move the D5-brane to the
O3-plane and split them. For consistency with the first method in
the last subsection we must require the splitting of D5-brane with
one D3-brane ending on it as the rule given in section 3.2. In fact,
as we discussed above, we find all rules in section 3.2 in this way. It is
easy to check that in this example we should get the same result
as part (e)  of Figure  \ref{fig:Sp1_3}.

\EPSFIGURE[h]{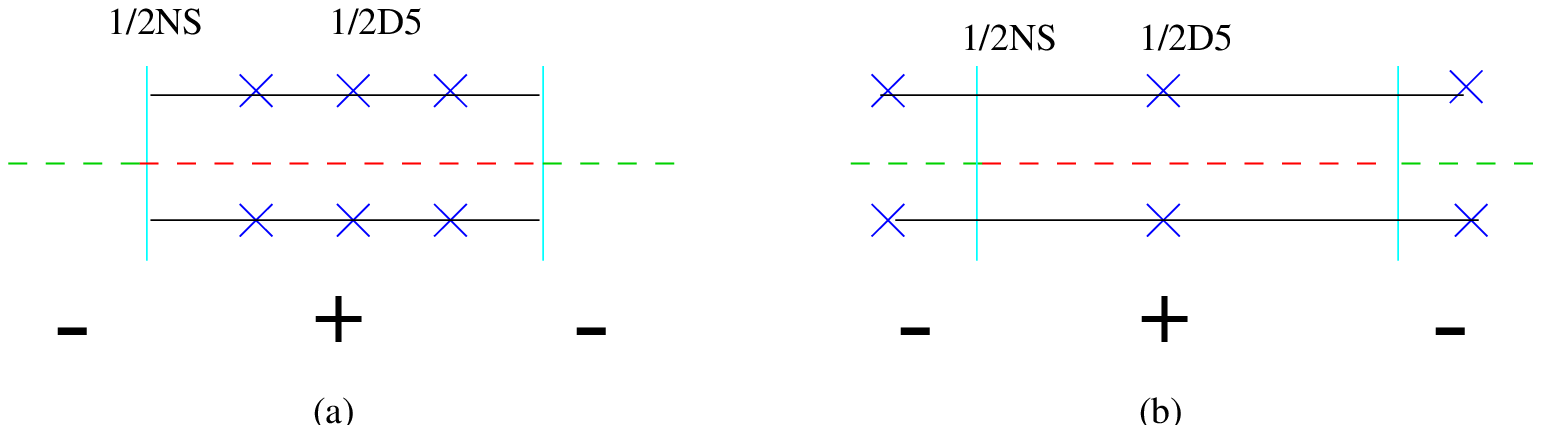,width=14cm} {The other method to go to
the Higgs branch: (a)  The incomplete Higgs branch when D5-branes
are off the O3-plane. (b)  By moving 1/2NS-brane one step inside,
we get rid of the D3-brane ending on 1/2NS-brane.
\label{fig:second}}

\subsection{The ``hidden FI-term''}
We have met the mismatch of mass and FI parameters in the above
mirror pair. It is  time for  us to talk more about it in this
subsection. In fact, such a mismatch of mass and FI parameters in
mirror pair is not new to us. Kapustin found this problem in
\cite{Kap}. In that paper, he considers the mirror of $Sp(k) $
with an antisymmetric tensor and $n$ fundamental flavors. He found
that when $n=2,3$ the quivers of the mirror theory are in fact
affine $A_1$ for $n=2$ and affine $A_3$ for $n=3$. However, it is
a well-known fact that a gauge theory given by an affine $A_n$
quiver has one mass parameter. On the other hand, classically
 the original $Sp(k) $ theory does not have any FI parameters. Kapustin
suggests the concept of {\sl {\bf ``hidden FI term''}} to resolve the
conflict. Such a term arises as the deformation in the infrared
limit and has the same quantum number as a FI-term. Because it is a
quantum effect, these deformations need not have a Lagrangian
description in the ultraviolet. 
To count the number of hidden FI
deformation we simply count the mass parameters in the mirror
theory. Now applying Kapustin's explanation to our example, we
find there is one ``hidden FI-term'' for the original theory and
three ``hidden FI-terms'' for the mirror theory. This result is
consistent with Kapustin's result. Notice that for $k=1$ the
antisymmetric tensor of $Sp(k) $ does not exist, so his theory is
in  exact agreement with our original theory and we both find
one ``hidden FI-term''. Hereafter we do not discuss the matching
of the mass and FI parameters anymore, but we will mention the
case when there exists a ``hidden FI-term'' for the original
theory.

\par

The appearance of  the {\sl ``hidden FI term''} indicates another important
aspect of  the possible
enhanced hidden global symmetry. In \cite{Int}, the authors observed
that the fixed point can have global symmetries which are manifest in one
description but hidden in another  (i.e., can be seen only  
quantum mechanically). For example, the $U(1)$ with two flavors is a 
self-mirror theory. On a classical level we have $SU(2)\times U(1)$ global
symmetry, where $SU(2)$ is the flavor symmetry and $U(1)$ is the global 
symmetry connecting  one FI-parameter (FI-parameter can be considered
as a component in the background vector supermultiplet of $U(1)$). However,
at the fixed point, the $U(1)$ global symmetry is enhanced to $SU(2)$.
This enhanced symmetry can be easily seen in the brane setup of the
mirror theory because in this special case ($U(1)$ with two flavors), the
two D5-branes (the S-dual of two NS-branes in original symmetry) meet in
same interval. This is another advantage of brane setup because we can see
a lot of nontrivial phenomena pictorially. In  later sections, when we
find the case where there is a {\sl ``hidden FI term''}, we will also 
discuss the
enhanced global symmetry.

\par
There is another interesting aspect which is worth mentioning.
 If our construction is right, it seems that we have two
different theories which are mirror to the same one because in
\cite{Kap,Han2} we can construct the mirror of $Sp(k) $ gauge
theory by using the $O5^-$ plane. This is also met by Kapustin in
\cite{Kap}. He noticed that two theories, (1) the $Sp(k) $ gauge
theory
 with an antisymmetric tensor plus two or three fundamental and
(2) the  $U(k) $ gauge theory with an adjoint plus two or four
fundamental flavors, are mirror to the same affine $A_1$ or $A_3$
quiver theory. Because  mirror symmetry is a property in the
infrared limit of gauge theory, such a non-uniqueness is allowed.
Actually the brane picture provides a definition of the theory
beyond the  infrared limit and the non-uniqueness can be seen in
nature by having two different brane representations of the same
field theory.

\subsection{$Sp(2) $ with $6$ flavors}
With the experience of $Sp(1) $ gauge theory, we can deal with the
$Sp(2) $ with $6$ flavors very quickly. The moduli spaces for the
original theory  have
 $d_v=2$ and $d_H=6\times 4-10=14$
(as mentioned above, in the following discussion we do not discuss
the issue of mass parameters and FI-parameters). The steps for
getting the mirror theory is in Figure  \ref{fig:Sp2_6}. From the
brane setup (d)  in Figure  \ref{fig:Sp2_6} we read out that the
mirror theory as $SO(2)  \times Sp(1)  \times SO(4)  \times Sp(2)
\times SO(5)  \times Sp(2) \times SO(4) \times Sp(1) \times SO(2)
$ with $8$ bifundamentals and two fundamental half-hypermultiplets
one for each $Sp(2) $ gauge theory. By an easy calculation, we can
check the moduli spaces as: $d_v=4\times 1+ 5\times 2=14$ and
$d_H=(2\times 4+2\times 8+ 2\times 16+2\times 20+2\times 4)
/2-(2+2\times 3+2\times 6+3\times 10)
 =52-50=2$.
\EPSFIGURE[h]{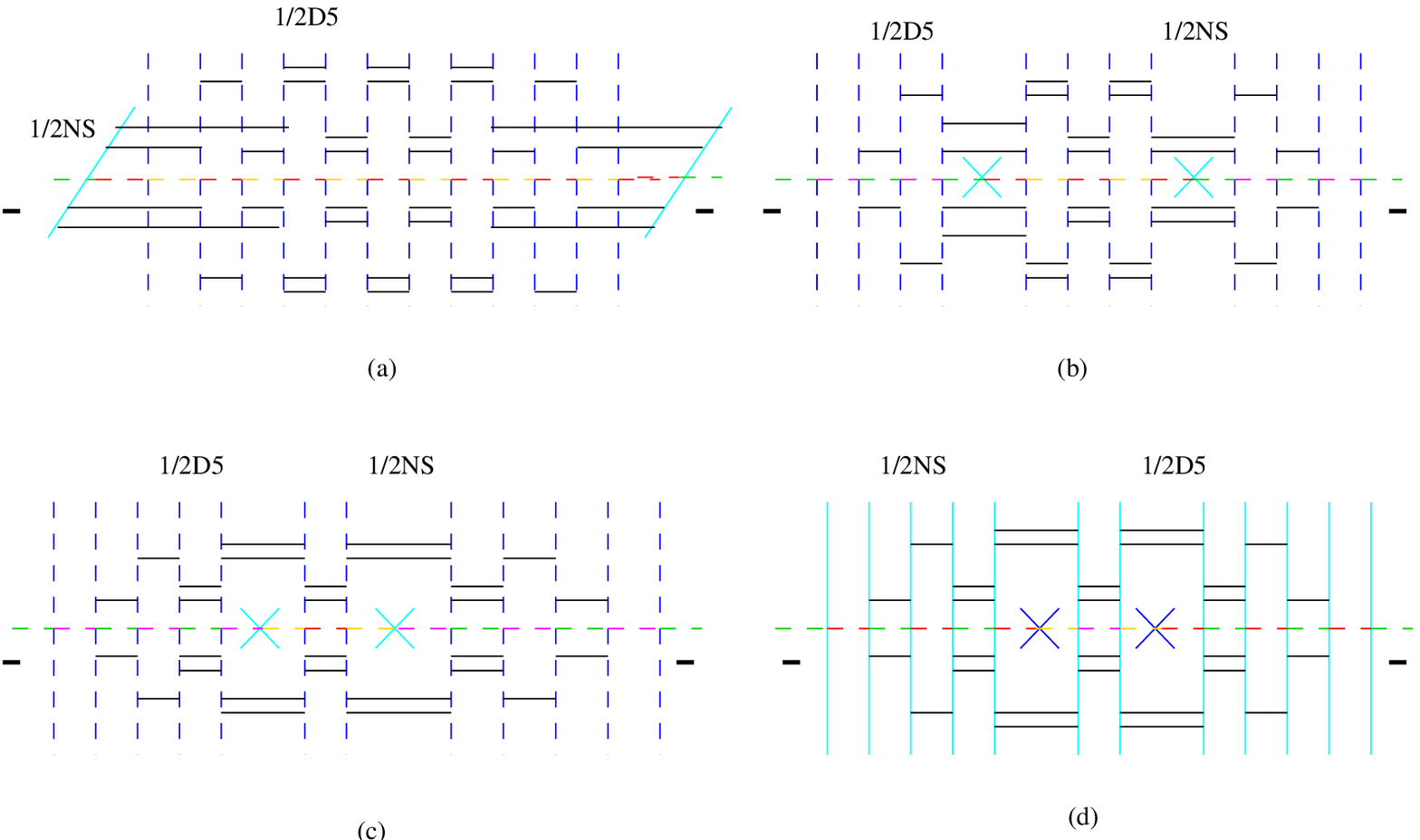,width=15cm} {(a)  The Higgs branch of $Sp(2) $ with $6$
flavors. Notice how we split the D3-branes according the
supersymmetric configuration. (b)  By moving 1/2NS-brane across the
1/2D5-brane, we get rid of the D3-brane ending on 1/2NS-brane. (c)
However, to read out the correct mirror theory, we need to move
the 1/2NS-brane one step further inside. (d)  By S-duality of part
(c)   we get the brane setup of the mirror theory.
\label{fig:Sp2_6}}

\subsection{The general case}
Now we discuss the general case, i.e.,  $Sp(k) $ with $N$ fundamental
flavors (to get the complete Higgsing, we have to assume that $N
\geq 2k$). The moduli space has $d_v=k$ and $d_H=2kN-k(2k+1) $.
 The steps for getting the
mirror theory are  shown in Figure  \ref{fig:Sp_gen}. 
\EPSFIGURE[ht]{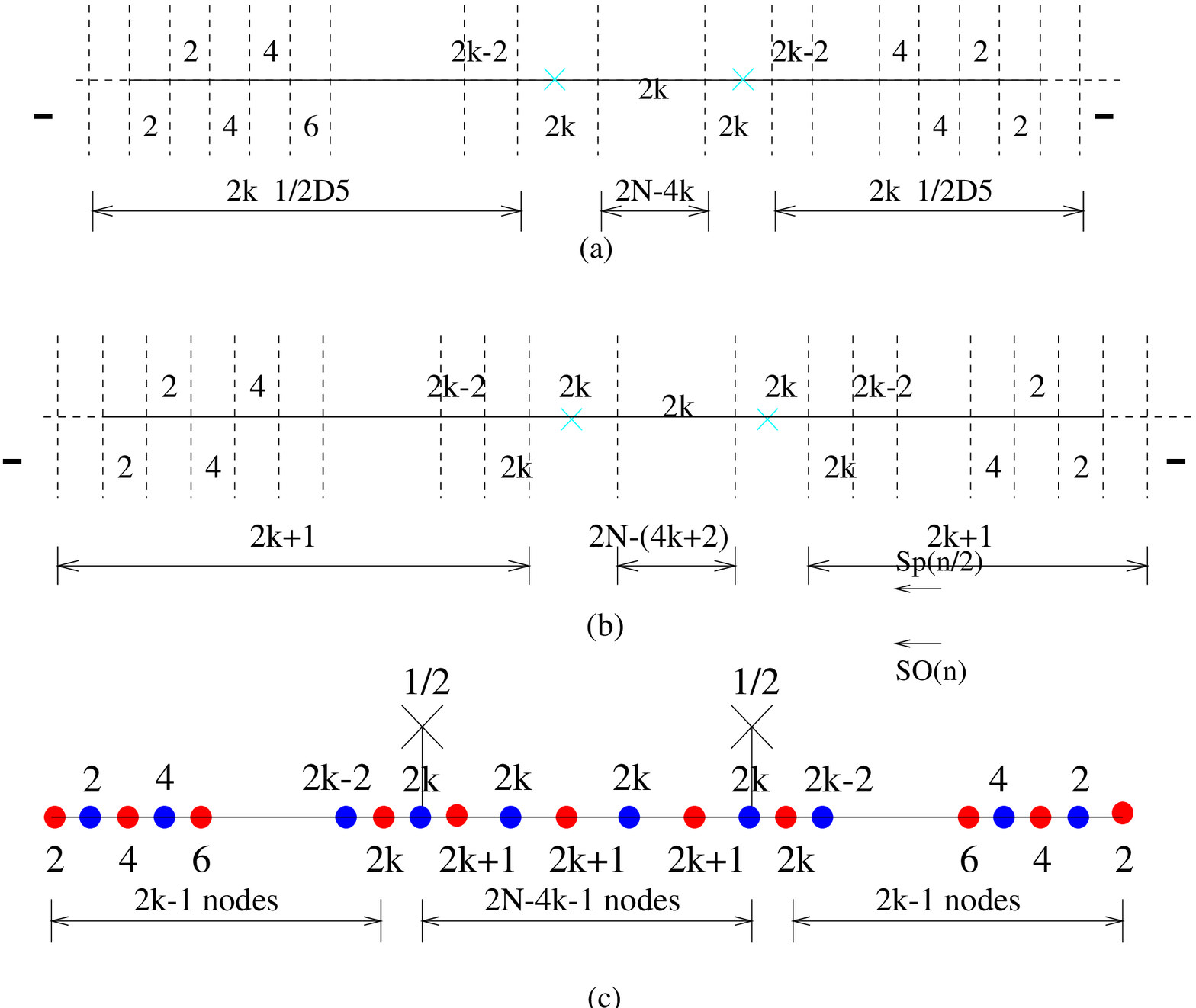,width=15cm} {The mirror of  $Sp(k) $ gauge theory with
$N$ flavors. Because of the complexity, in this figure we do not
keep
 track of the
change of O3-plane anymore and  use dotted horizontal black line to
express all O3-planes. However, we do keep track of
the intervals which give the
$Sp$ or $SO$ group in the final mirror theory by using the number above the
O3-plane to denote the $Sp$ group and below to denote $SO$ group.
 (a) The Higgs branch of $Sp(k) $ with $N$ flavors. Notice that the
pattern of the number of $1/2$-D3 branes between two nearby 1/2D5-branes
 is, from left to right,
$0,2,2,4,4,6,...2k-2,2k^{2N-4k+1},2k-2,2k-2,...,4,4,2,2,0$. (b)
To read out the mirror theory in general we need to  move the
 1/2NS-brane one step further inside. However, we
can consider (b)  as the brane setup of the mirror theory too by
just thinking of the dotted  vertical  line as 1/2NS brane and the
cross as 1/2D5-brane. (c) For convenience, we draw the quiver
diagram.
 We use  red dots for $SO$ groups
and blue dots for $Sp$ groups.
We also write the number above for an $Sp$ group and
under for an $SO$ group.
\label{fig:Sp_gen}}
From it we
can read out that the mirror theory  are $SO(2) \times Sp(1)
\times SO(4)  \times Sp(2) \cdots \times Sp(k-1)  \times SO(2k)
\times (Sp(k) \times SO(2k+1) ) ^{n-2k-1} \times Sp(k)  \times
SO(2k)  \times Sp(k-1)  \cdots \times Sp(1)  \times SO(2) $ with
bifundamentals and one fundamental half-hypermultiplet for each
the first and the last $Sp(k) $ gauge groups. For clarity,
the corresponding quiver diagram of the above mirror theory is also
drawn in part (c)  of this figure. Now we can calculate the moduli
spaces of the mirror  as
\begin{equation}
\label{Spquiver}
\begin{array}{lll}
d_v & = & 4\sum_{n=1}^{k-1}n  + (2N-4k+1) k=2Nk-k(2k+1) , \\ d_H &
= & [\frac{1}{2}\times 2 \sum_{n=1}^{k-1} ( (2n) ^2+2n(2n+2) )
+\frac{1}{2}( 2(2k) ^2+ (2N-4k-2) 2k(2k+1) ) + 2k]  \\ & - & [
2\sum_{n=1}^{k-1}(n(2n-1) +n(2n+1) )  + (2N-4k-1) k(2k+1)
+2k(2k-1) ]
\\ & = & k.
\end{array}
\end{equation}
As mentioned above, for general $N,k$ we get only half-hypermultiplets
coming from the 1/2D5-branes in the mirror theory.
However, there are two degenerate cases where one fundamental hypermultiplet
 does exist
instead of two half-hypermultiplets. The first case is when
$N=2k$. In this case, we do not need to move the 1/2NS brane
further from part (a)  to part (b)  in Figure  \ref{fig:Sp_gen}.
Instead, we can make the mirror transformation directly from
part(a). In the mirror theory, we get only one $SO(2k) $ gauge
group but with one flavor for this $SO(2k) $. As explained above,
such a flavor  hints a ``hidden FI-term'' in the original theory.
The second case is when $N=2k+1$, where we get only one $Sp(k) $
gauge group in mirror theory, but also with one flavor of the
$Sp(k) $ which  also suggests a ``hidden FI-term'' in the original
theory. 
For $k=1$, the two cases where  a ``hidden FI-term'' shows
is given in \cite{Kap}. For $k \geq 2$ it is a new result.

\par
As we mentioned in section 4.3, in the case where a ``hidden FI-term''
shows we should consider the possible enhancement of global symmetry.
In general the theory has global $SO(2N)$ flavor symmetry. When
$N=2k$, the global symmetry will be enhanced to $SO(2N)\times  Sp(1)$.
The factor $Sp(1)$ can be seen from the mirror theory, where two 1/2D5-branes
meet and give one flavor to the $SO(2k)$ gauge group (notice the flavor symmetry
for $Sp(k)$ gauge groups is $SO(2N)$, for $Sp'(k)$ gauge groups, $SO(2N+1)$, 
for $SO(2k)$ gauge groups,
$Sp(N)$ and for $SO(2k+1)$ gauge groups, $Sp'(N)$). When 
$N=2k+1$, the global symmetry will be enhanced to $SO(2N) \times SO(2)$
because in this case, the one extra flavor in mirror theory belongs to the
$Sp(k)$ gauge group.

\section{The mirror of $Sp'(k) $ gauge theory}
We know that the $O3^+$ and $\widetilde{O3^+}$ projections both
give  $Sp(k) $ gauge theory. To distinguish them, we denote the
gauge theory given by $O3^+$ projection as $Sp(k) $ and that by
$\widetilde{O3^+}$ as $Sp'(k) $. After the discussion of the
mirror of $Sp(k) $ gauge group in the last section, we now address
the $Sp'(k) $ case in this section. The brane setup of $Sp'$  is
just to replace the $O3^{\pm}$ in $Sp(k) $ by
$\widetilde{O3^{\pm}}$ (for example, see figure
\ref{fig:Sp1_prime_3}). However, by such a replacement, the
theory becomes $Sp(k) $ gauge theory with $n$ flavors plus  two
half-hypermultiplets contributed from the $\widetilde{O3^-}$ at
the two sides (notice that $\widetilde{O3^-}$ can be considered as
 $O3^-$ plus a 1/2D3-brane).
We will start the discussion also from a simple example, then go
to the general case. Furthermore, we will compare the mirror of
$Sp(k) $ and $Sp'(k) $ and show that in fact they give the same mirror
theory.

\subsection{$Sp'(1) $ with 3 fundamental flavors}
In this example, the theory is $Sp(1) $ gauge group with three
 hypermultiplets and two half-hypermultiplets. The moduli space has
 $d_v=1$ and
$d_H=3\times 2 +2\times 2/2-3=5$. The steps for finding the mirror
theory are drawn in Figure  \ref{fig:Sp1_prime_3}. First we go to the
Higgs branch. Now equations (\ref{case2}) and (\ref{case3})  allow us the
break the D3-branes between the 1/2NS-branes and the neighboring
1/2D5-brane as part (a). Form part (a)  we move 1/2NS-brane
inside to pass one 1/2D5-brane and get part (b). In this step,
the 1/2NS-branes get rid of the D3-brane ending on them already.
However, this brane setup does not readily give the correct mirror
theory and we need go to the next step, i.e., moving 1/2NS-brane one
step further inside as in part (c).
 Finally, we make the S-duality transformation
 and get the mirror theory in part (d). The mirror theory
is $SO(2)  \times Sp(1)  \times SO(3)  \times Sp(1)  \times SO(2)
$ with four bifundamentals and two half-hypermultiplets one for
each $Sp(1) $ gauge group. We can check the moduli spaces of the
mirror theory as having $d_v=5$ and $d_H=(2\times 4+2\times
6+2\times 2) /2-(2+3\times 3)  =12-11=1$.

\EPSFIGURE[h]{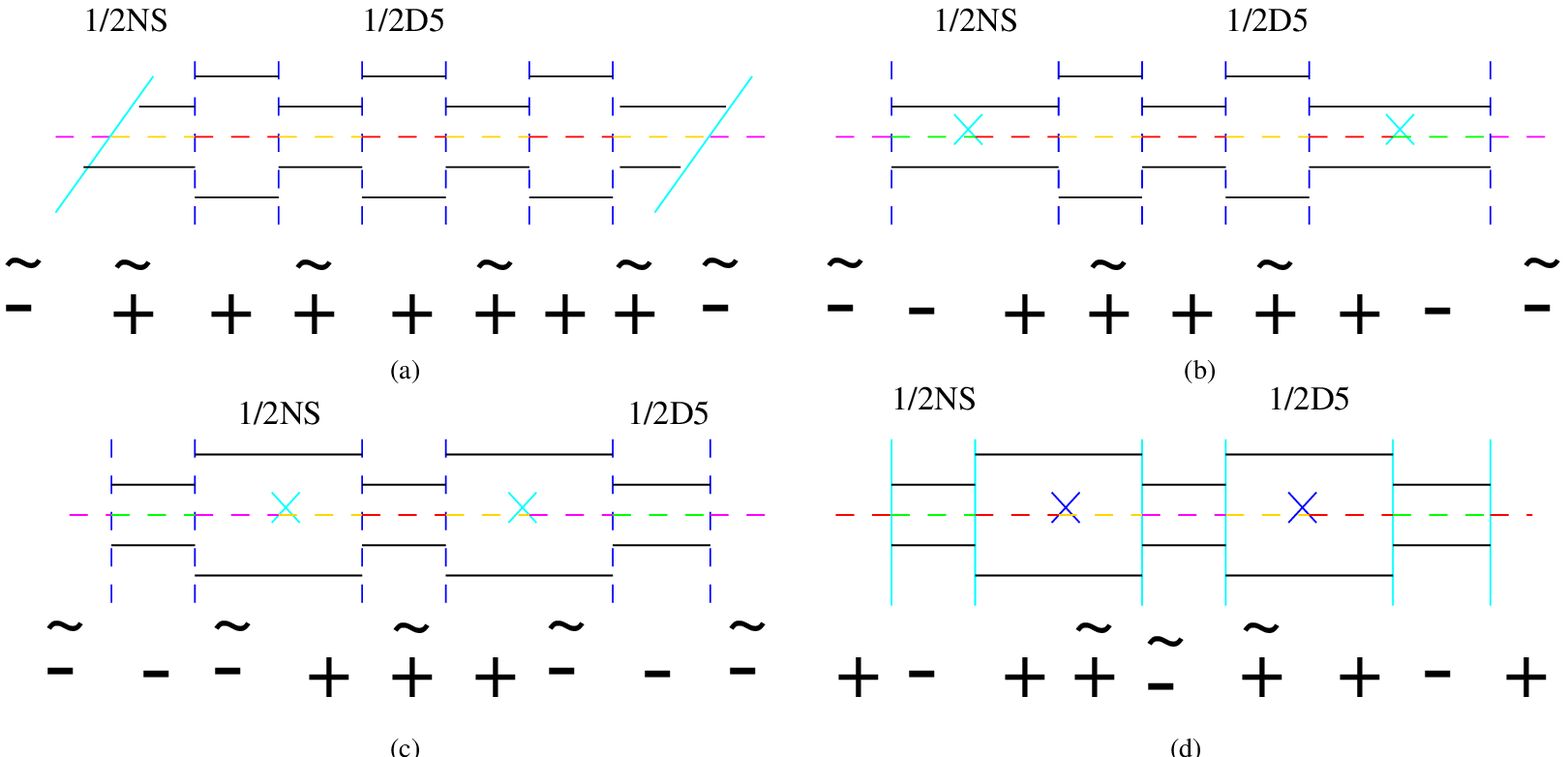,width=15cm} {(a)  The Higgs branch of $Sp'(1) $
with three flavors. Notice how we split D3-branes to satisfy the
supersymmetric configuration. (b) (c)  Using the rule of supersymmetric
configuration,
we reach the brane setup which is good for the mirror
transformation. (d)  The brane setup of the mirror theory.
\label{fig:Sp1_prime_3}}

\subsection{The general case}
Now we discuss the general case, i.e., $Sp'(k) $ with $n$
hypermultiplets and two half-hypermultiplets. The moduli spaces
have $d_v=k$ and $d_H=2nk+2k-k(2k+1) =2nk-k(2k-1) $. The main
steps to get the mirror theory are in Figure  \ref{fig:Sp_prime}.
\EPSFIGURE[ht]{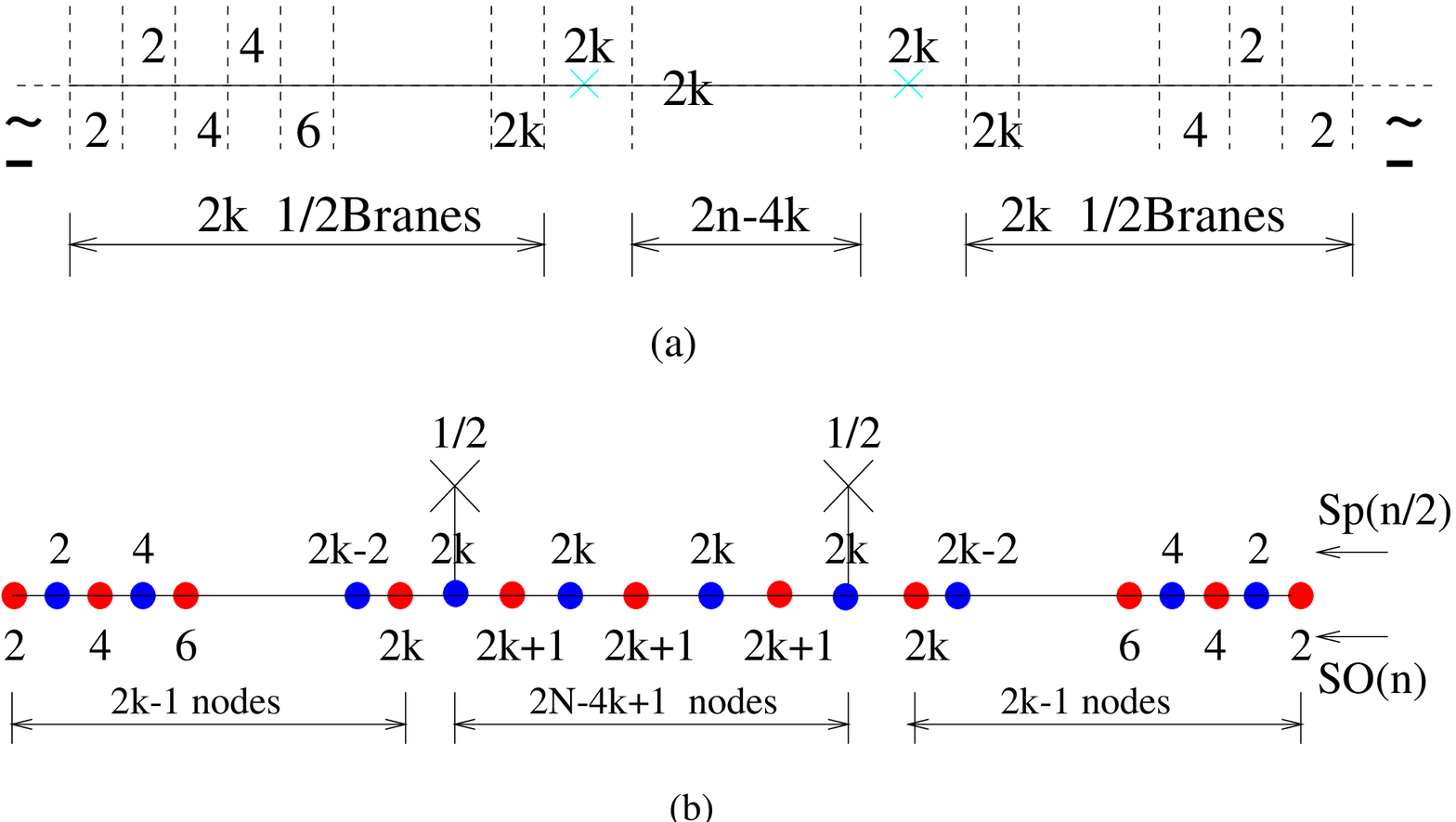,width=15cm} {(a)  The Higgs branch of $Sp'(k) $ with
$N$ flavors after  moving the two 1/2NS-branes inside. As before,
the numbers above and below  mean the number of 1/2D3-branes which
connect  two neighboring 1/2D5-branes. We can also consider it as
the brane setup of the mirror theory just by considering  the
vertical line as 1/2NS-brane instead of 1/2D5-brane and the crosses as
1/2D5-branes instead of 1/2NS-branes. (b)  The quiver diagram of the
mirror theory. The numbers written above the (blue)  node denote
the $Sp$ gauge group and the numbers written under the (red)  node
denote the $SO$ gauge group. \label{fig:Sp_prime}}
We can read out the mirror theory from the quiver diagram in
part(b)  and check the  moduli space as having
\begin{equation}
\label{Sp_prime}
\begin{array}{lll}
d_v &  =  &  4\sum_{t=1}^{k} t + k(2n-4k-1) =2nk-k(2k-1) ,  \\ d_H
&  = & [2\sum_{t=1}^{k-1}( (2t) ^2+2t(2t+2) ) + 2(2k) ^2+ (2n-4k)
2k(2k+1) +2(2k) ]/2    \\ & - & [2\sum_{t=1}^{k-1}(t(2t-1)
+t(2t+1) ) + 2k(2k-1) +(2n-4k+1) k(2k+1) ] \\ & = & k
\end{array}
\end{equation}
As the case of $Sp$ gauge group
 when $n=2k$, the mirror theory has only one
$Sp(k) $ gauge group and the two half-hypermultiplets combine
together to give one flavor for $Sp(k) $. It means that we have a
``hidden FI-term'' in the original theory.
However, it is not the
end of the story. By careful observation, we find that when
$n=2k-1$, the mirror theory has only one $SO(2k) $ gauge group and
two half-hypermultiplets also combine together to give one flavor
for $SO(2k) $ (this happens  because  in this case, we do not need
move 1/2NS-brane one further step as we did from part (b)  to part
(c)  in Figure
 \ref{fig:Sp1_prime_3}). So we get a ``hidden FI-term''  
in this case also. This
is not expected  initially because it seems that for $n=2k-1$ we
can not  get the complete Higgs branch, but this is not true. By
studying the part (a)  of Figure  \ref{fig:Sp1_prime_3}, we find
that for $n=1$ in $Sp'(1) $ we indeed get  complete Higgsing.
Furthermore by the discussion in the next subsection we will see more
clearly  the reason why $n=2k-1$ gives a ``hidden FI-term''.

\par

Now let us discuss the global symmetry. The results are very similar
to those at the end of section 4. In the general case we have
global $SO(2N+1)$ flavor symmetry\footnote{From the discussion in the next
subsection, the mirrors of  
 single $Sp'(k)$ with $N$ flavors and single  
$Sp(k)$ with $N+1$ flavors are identical. In the latter case, the
flavor symmetry is $SO(2N+2)$, but in the former case, we see only an obvious
$SO(2N+1)$ flavor symmetry. 
However, in current situation of product gauge theories 
the argument of section 5.3 can not be applied directly. There is true
distinguishing between $Sp(k)$ and $Sp'(k)$ gauge theories}. 
When $N=2k-1$, the global symmetry
goes to $SO(2N+1)\times Sp(1)$. When $N=2k$, the global symmetry
goes to $SO(2N+1)\times SO(2)$.

\subsection{Comparing the mirror of $Sp(k) $ and $Sp'(k) $}
In the above, we have discussed the mirror of two kinds of $Sp$
gauge groups, i.e.,  $Sp(k) $ and $Sp'(k) $. We want to ask ourselves
whether   there is any relation between the mirrors of these two $Sp$ gauge
groups? By checking the two quivers in Figure  \ref{fig:Sp_gen}
and Figure  \ref{fig:Sp_prime}, we find that these two quivers are
exactly the same, except that $N$ flavors in $Sp'(k) $ should
correspond to $N+1$ flavors in $Sp(k) $. This is reasonable
because for $Sp'(k) $ with $N$ flavors there are two
half-hypermultiplets which give the same degrees of freedom as one flavor.
However, in principle there is a difference between one flavor and
two half-hypermultiplets: for the former we can involve one mass
parameter, but for the latter there is no such mass parameter. We will
show, in the case of $Sp'(k) $, that the two half-hypermultiplets do
combine to give one flavor with the mass parameter. To see
this, we move one 1/2D5-brane from infinity at each side to pass
the 1/2NS-brane. By using the $s$-configuration in section 2, we
get the brane setup of $Sp(k) $ with an additional flavor. The
whole discussion is shown in figure  \ref{fig:equi}. Furthermore,
it is easily to show that the two cases where a ``hidden
FI-parameter'' shows in $Sp(k) $ and $Sp'(k) $  exactly match
 each other.

\EPSFIGURE[ht]{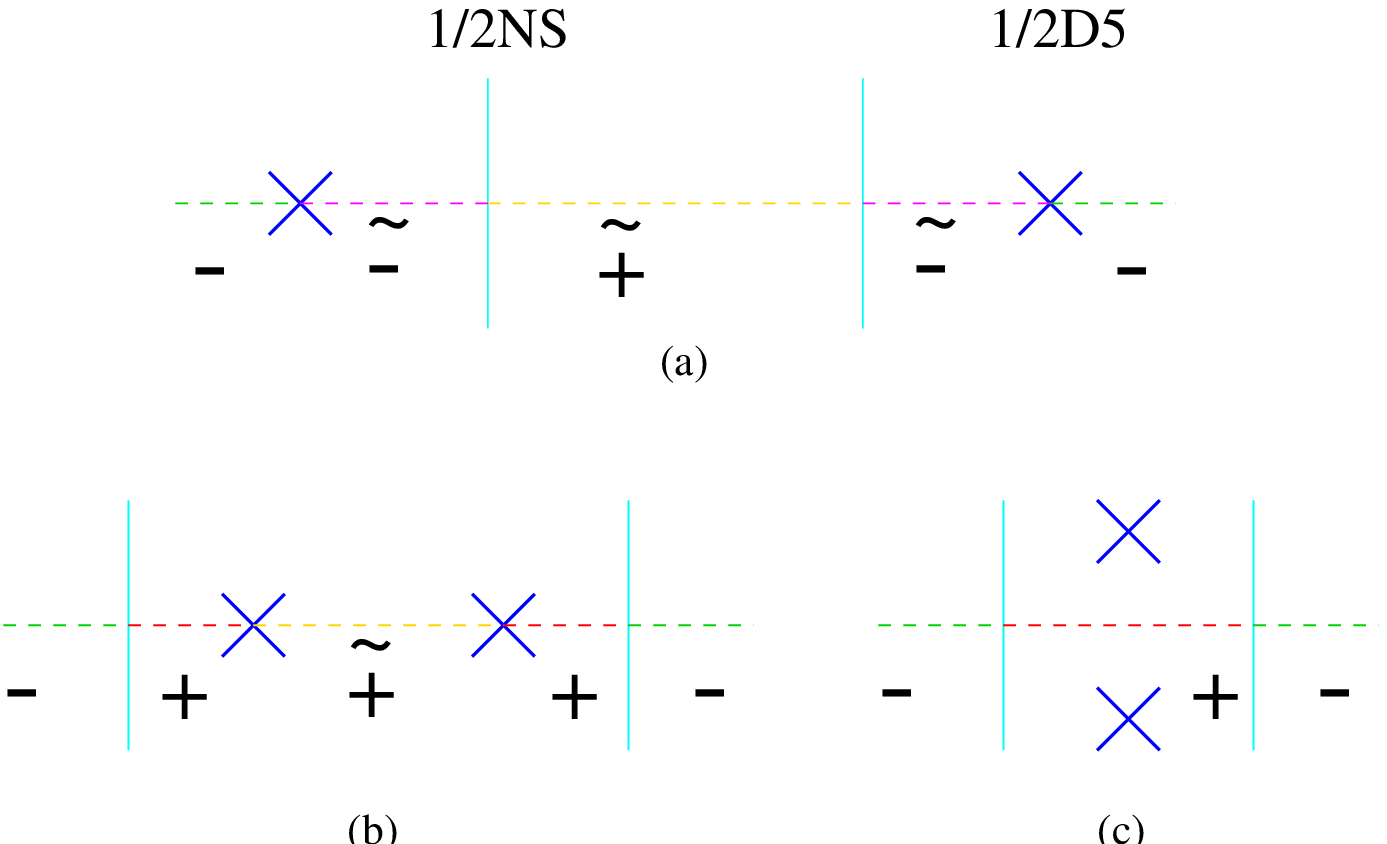,height=7cm} {(a)  We move one 1/2D5-brane from the left
and right infinity. (b)  By using the $s$-configuration in section
2, we change the position of 1/2D5-branes inside. (c)  By combining
the two 1/2D5-brane we get one physical D5-brane which can be
moved off the $O3^+$-plane. From it we see that we change $Sp'(k) $ 
to $Sp(k) $ with one additional flavor. \label{fig:equi}}

\section{The mirror of   $SO(2k) $ gauge theory }
After the discussion of the mirror theories for $Sp(k) $ gauge
groups, we now discuss $SO(2k) $. There are no known results for
the  mirror of $SO(2k) $ gauge groups and it is the main
motivation of  this paper to calculate it  using the $O3$ plane.
As in the last two sections, we first present the simple case of
$SO(2) $ with three flavors, then give the general results for
$SO(2k) $ with $N$ flavors.

\subsection{$SO(2) $  with 3 flavors}
For $SO(2) $ gauge theory with three  flavors, the moduli spaces
 have  $d_v=1$ and $d_H=3\times 2-1=5$.
The steps for the mirror transformation are given in Figure
\ref{fig:SO23}. In part (a) , we break the D3-branes by preserving
the supersymmetric configurations, then use  (\ref{case2}) and  (\ref{case3})
to move the 1/2NS-brane passing the 1/2D5-brane to get  part (b).
Unlike the $Sp$ case, part (b)  is already convenient  for the mirror
transformation, so we can make S-duality directly and get part (c)
. From the brane setup in part (c)  we read out  the mirror
theory to be $Sp(1)  \times SO(2)  \times Sp(1)  \times SO(2)
\times Sp(1) $ with four bifundamentals and two
half-hypermultiplets for the leftmost $Sp(1) $ and two
half-hypermultiplets for the rightmost $Sp(1) $. Here we want to
emphasize that in the two half-hypermultiplets for the leftmost
$Sp(1) $, one comes from the 1/2D5-brane and the other from
the $\widetilde{O3^-}$ projection (same for the rightmost $Sp(1)
$). That the half-hypermultiplets come from different sources is a
general phenomenon in $SO(2k)$. However, for our simple example,
we can combine these two half-hypermultiplets together by moving
the 1/2D5-brane in part (c)  to go part (d). Now we have one
flavor of $Sp(1) $ given by one physical D3-brane stuck between
the 1/2NS-brane and the 1/2D5-brane. We need to emphasize that
because the physical D3-brane is stuck between the 1/2NS-brane and
the 1/2D5-brane, it does not contribute to the mass parameter.  It
will be interesting to compare it with the discussion in section
4.3, where we find that the two half-hypermultiplets of $Sp'(k) $
can combine to give a flavor with free mass parameter. Finally,
we calculate the dimension of the moduli spaces of mirror
theory as $d_v=5$ and $d_H=(4\times 4/2+2\times 2) -(2+3\times 3)
=12-11=1$.

\EPSFIGURE[ht]{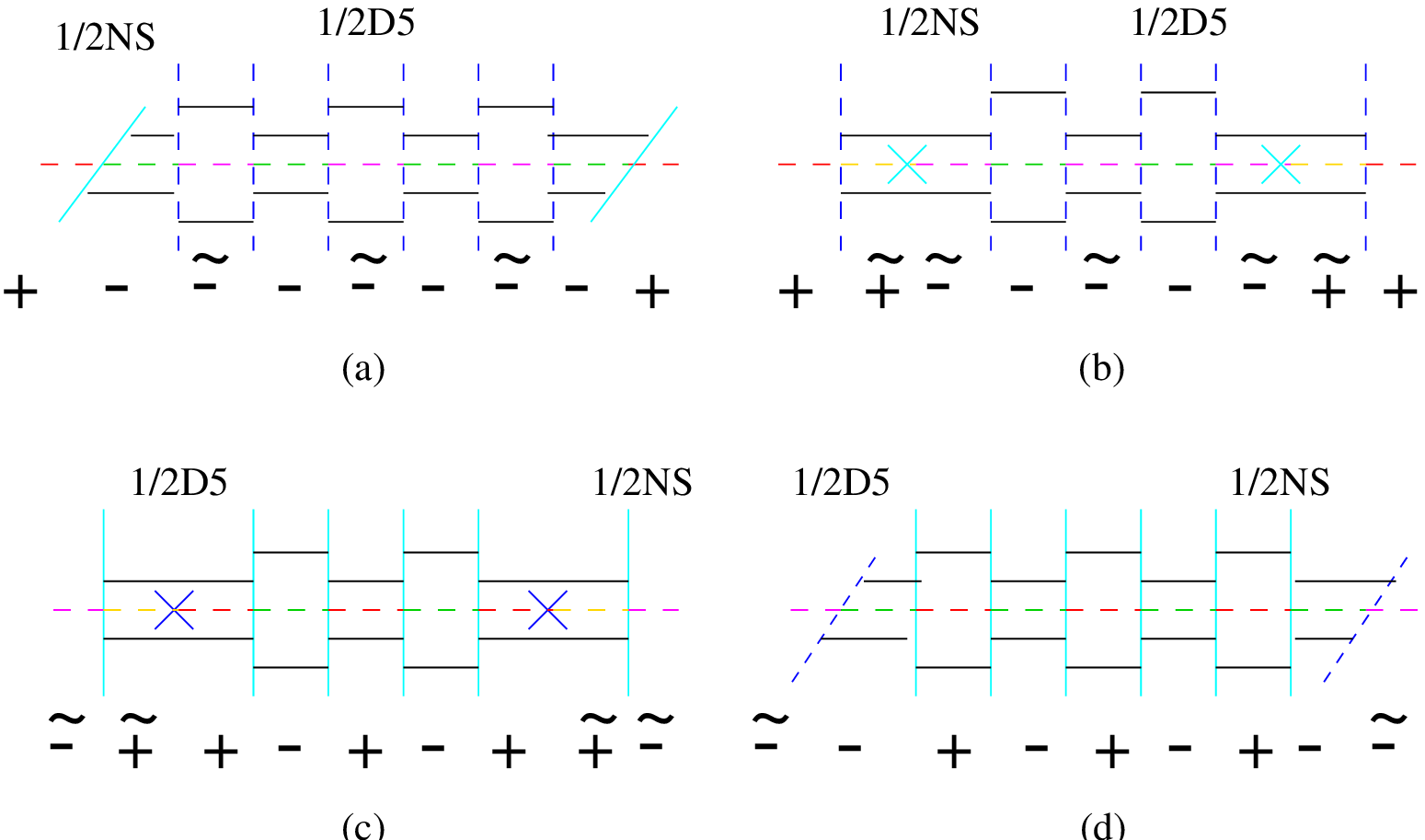,width=15cm} {(a)  The Higgs branch of $SO(2) $ with three
flavors. Notice how we split the D5-brane to satisfy the
supersymmetric configurations. (b)  Using the rule of supersymmetric
configuration, we
move the 1/2NS-brane one step to reach the brane setup which is
convenient for the mirror transformation. (c)  The brane setup of the
mirror theory. (d)  By moving the 1/2D5-brane one step outside, we
combine the two half-hypermultiplets into one hypermultiplet.
\label{fig:SO23}}

\subsection{An exotic example:  $SO(2) $ with 2 flavors}
In this subsection, we discuss the mirror of $SO(2) $ with two flavors.
This theory will show one nontrivial phenomenon. 
 The moduli are $d_v=1$ and $d_H=2\times 2-1=3$.
According to the standard procedure introduced in the last subsection
we get the Higgs branch as part (a) in Figure  \ref{fig:SO22}
and the mirror theory in part (b). The dimensions of 
moduli spaces of the mirror theory in part (b)  are $d_v=1+1+1=3$
and $d_H=2\times 4/2+4\times 2/2-(3+3+1) =9-8=1$.
\EPSFIGURE[h]{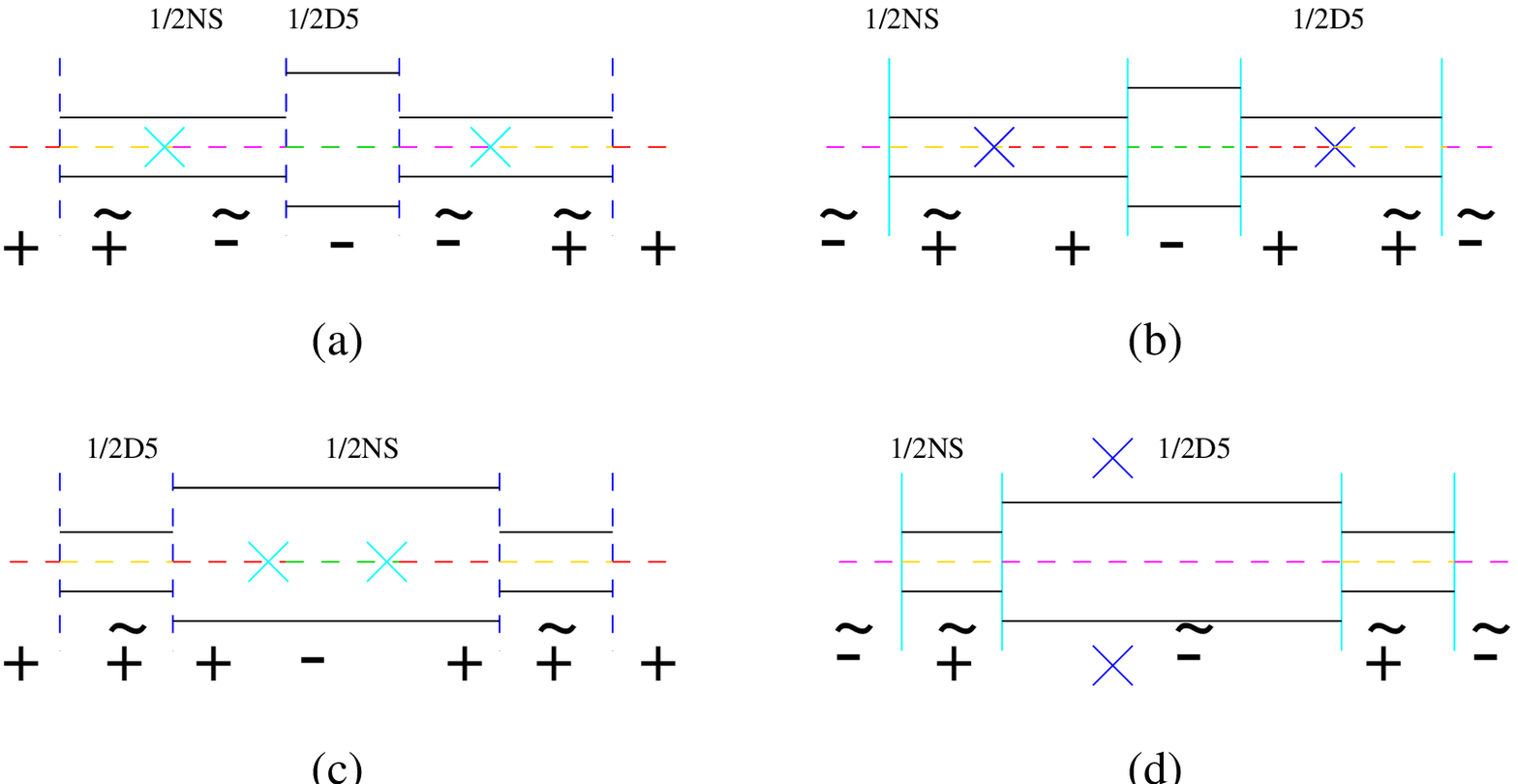,width=15cm} {(a)  The Higgs branch of $SO(2) $ with two
flavors.
 (b) By S-duality, we get the mirror theory as $Sp(1) 
 \times SO(2)  \times Sp(1) $
with two bifundamentals  and four half-hypermultiplets for the two
$Sp(1) $ gauge theories.
 (c) However, for this special case, it seems we can get another
 mirror theory by
 moving the 1/2NS-brane one further step
inside from part (a)  to part (c). In our case, now two
1/2NS-branes are in same interval. If they can  combine together
and leave the $O3^+$ plane, we can make the S-duality to get part
(d). (d)  The mirror theory got from part (c)  is $Sp(1) \times
SO(3)  \times Sp(1) $ with two bifundamentals , two
half-hypermultiplets for two $Sp(1) $ and
 one fundamental for $SO(3)$.
\label{fig:SO22}}

\par
However it seems we can get another
 possible Higgs branch in part (c) by
 moving the 1/2NS-brane one further step
inside from part (a). If these two
1/2NS-branes do not meet together, the brane setup is not  convenient
to perform  S-duality to get the mirror theory and we must go
back to part (a). But in this special example, these two
1/2NS-branes do meet together. Now if these two 1/2NS-brane can
combine to leave the  $O3^+$ plane, we do get another mirror theory
like part (d). Let us assume it is correct first and calculate the
moduli spaces. In the part (d), the mirror theory is $Sp(1) \times
SO(3)  \times Sp(1) $ with two bifundamentals, two
half-hypermultiplets for the two $Sp(1) $ and
 one fundamental for $SO(3) $, so the moduli are $d_v=3$ and
$d_H=2\times 6/2+ 2\times 2/2+3-(3+3+3) =11-9=2$. Therefore the
results do not match. There is
another inconsistent result because in the mirror theory of
part(d)  we get one ``hidden FI-term'' which does not exist in the
mirror theory of part (b).

\par
What is the resolution for the above inconsistency?
Notice the combination of two 1/2NS-branes
on the $O3^+$ is S-dual to  the combination of two 1/2D5-branes on
the $\widetilde{O3^-}$. We have discussed this configuration in section 3.1,
where we showed, only when there is an extra physical D3-brane between these
two 1/2D5-branes (1/2NS-branes)  can they combine and leave the 
O3-plane. So the
conclusion is that the two 1/2NS-branes in part (c) can not combine and
leave the 
O3-plane. We are left with only one correct mirror theory in part (b).

\subsection{The general $SO(2k) $ with $N$ flavors}
With the experience of the  $SO(2) $ case, we can now work on the
general $SO(2k) $ with $N$ flavors. The moduli  for this theory
are $d_v=k$ and $d_H=2kN-k(2k-1) $. The steps for the mirror theory
are given in Figure  \ref{fig:SO_gen}. 
\EPSFIGURE[ht]{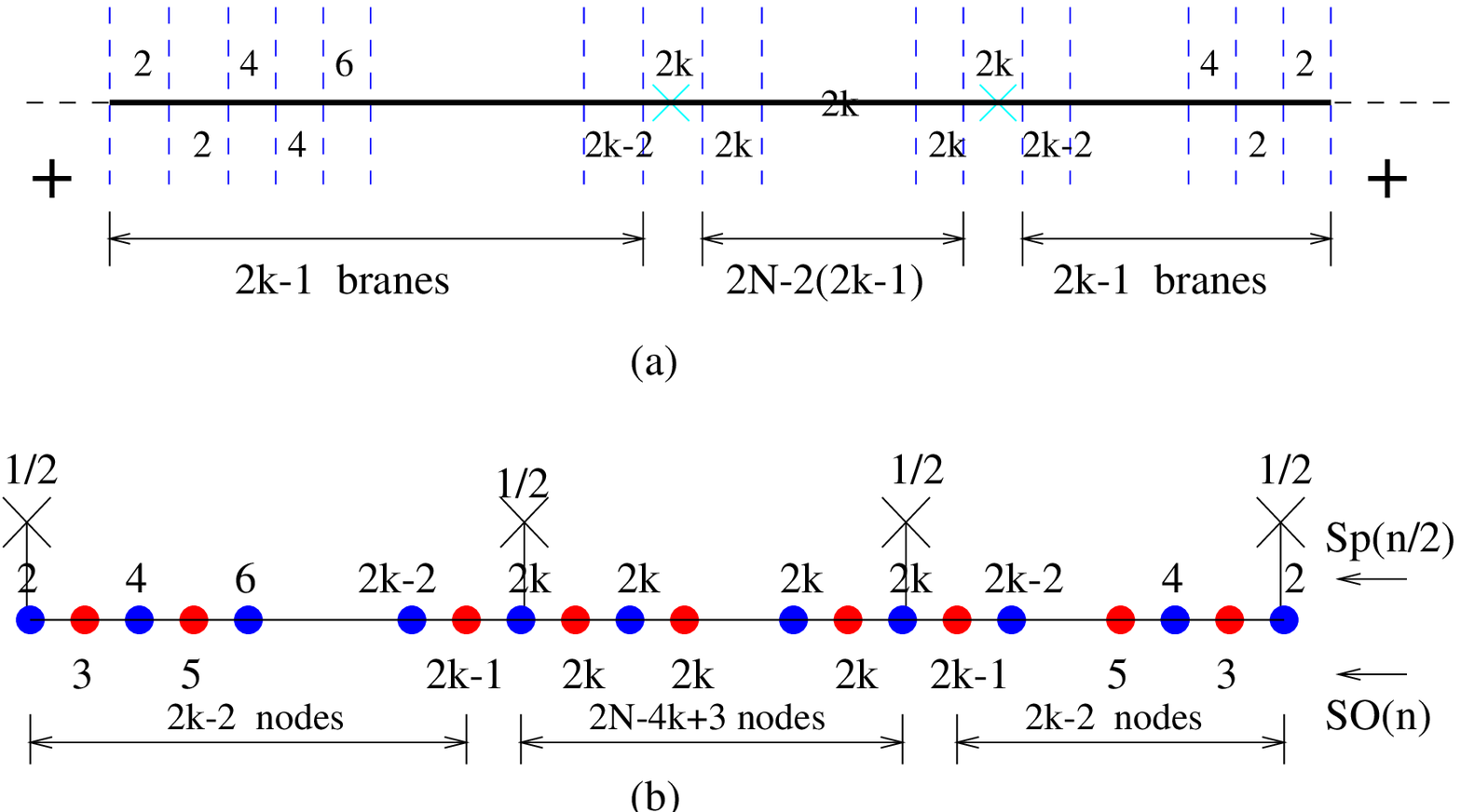,width=15cm} {(a) The Higgs branch of $SO(2k) $ with $N$
flavors in the setup of the D5-brane splitting.  The numbers in
the interval denote the number of 1/2D3-branes connecting the two
neighboring 1/2D5-branes. (b) The quiver diagram of the mirror
theory of $SO(2k) $ with $N$ flavors. Notice that the index above
the node means $Sp(n/2) $ and index below the node means $SO(n) $.
The $1/2$ means the half-fundamental. \label{fig:SO_gen}}
Again, we first break the
D3-branes according to the supersymmetric configuration, then move the
1/2NS-branes inside to go to part(a). The brane setup in part(a)
can be considered as the brane setup of S-duality just by
exchanging the roles of the 1/2NS-brane and the 1/2D5-brane  and
putting in a proper O3-plane. For clarity, we draw the
quiver diagram of the mirror theory in part(b). Let us check the
result again by calculating the moduli of the  mirror theory as
\begin{equation}
\label{SOquiver}
\begin{array}{lll}
d_v & = & 4\sum_{n=1}^{k-1} n + k(2N-4k+3) =2kN-k(2k-1) ,\\ d_H &
= & [\frac{1}{2}\times 2 \sum_{n=1}^{2k-2} (n+1) (n+2)  +
  \frac{1}{2} 4k^2(2N-4k+2) + \frac{1}{2}(2k+2k+2+2) ]  \\
&- & [ 4\sum_{n=1}^{k-1} n(2n+1)  + k(2k+1) (N-2k+2) +k(2k-1)
(N-2k+1) ]  \\ & = & k.
\end{array}
\end{equation}
By checking  part (a)  in Figure  \ref{fig:SO_gen}, we find that
there is a ``hidden FI-parameter'' in the original theory when
$N=2k-1$ because two 1/2NS-branes will meet in same interval of
$\widetilde{O3^+}$ plane.
For general $N,k$, 
 the global symmetry is an $Sp(N)$ flavor symmetry, but in the case
$N=2k-1$ it is enhanced to $Sp(N)\times SO(3)$. We want to point out
that  there is only one case where {\sl ``hidden FI-parameters''} show 
in $SO(2k)$ while 
for $Sp(k)$ and $Sp'(k)$ there are two cases. This difference can be 
seen very clearly in part (c) of Figure \ref{fig:SO22}. In that case
two 1/2NS-branes do meet in same interval, but they can not combine and
leave $O3^+$-plane. So there is no {\sl ``hidden FI-parameters''}.


\section{The mirror  of $SO(2k+1)$ gauge theory}
In this section, we discuss the mirror theories of $SO(2k+1)$ to complete
our study of single gauge groups. We first present the simple example  of
$SO(3)$ with two flavors, then give the general results for
$SO(2k+1)$ with $N$ flavors.

\subsection{$SO(3)$ with 2 flavors}
For $SO(3)$ with two flavors, the dimensions of moduli space are
$d_v=1$ and $d_H=2\times 3-3=3$. The steps to get the mirror theory are
shown in Figure \ref{fig:SO3_2}. In part (a) we split the physical D5-branes
into the 1/2D5-branes according the rules given in section two. In such a 
process
we see the generation
of two physical D3-branes which is necessary to account for the correct Higgs
branch. In part (b) we split the D3-brane between the 1/2D5-branes and
1/2NS-branes to go to the Higgs branch. Notice that there is no D3-branes 
connecting 1/2NS-brane and the nearest 1/2D5-brane which is required by
$s$-rule. In part (c) we move the 1/2NS-branes inside to get rid of the
D3-branes ending on them. Now we can make S-duality to give the
mirror theory in part (d). However, in our example, there is a special
property: two 1/2D5-branes can combine together and leave the 
$\widetilde{O3^+}$-plane to give one flavor.

\EPSFIGURE[h]{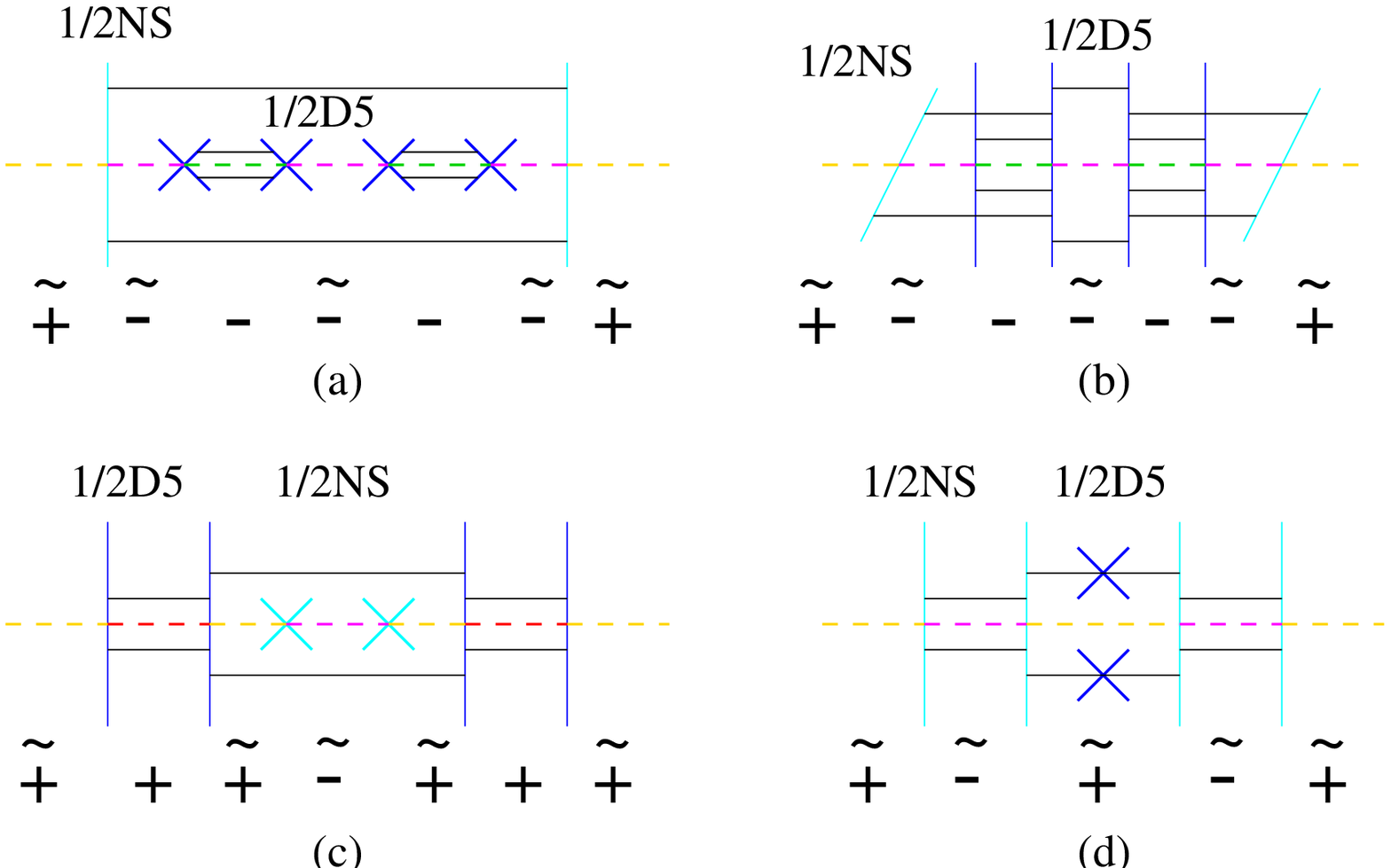,width=15cm} { The mirror of $SO(3)$ with two flavors. 
(a) Splitting of D5-branes according the rules given above. Notice the
generation of D3-branes between 1/2D5-branes. (b) The Higgs branch of
$SO(3)$ theory. (c) By moving 1/2NS-branes inside we get rid of D3-brane
ending on 1/2NS-branes and ready to go to the mirror theory. 
(d) The mirror theory. However, here we combine two 1/2D5-branes to give
one physical D5-brane. 
 \label{fig:SO3_2}}

Now we can read out the mirror theory as 
$SO(3) \times Sp(1) \times SO(3)$ with two bifundamentals and one
flavor for $Sp(1)$. Let us calculate the dimension of moduli space.
For the Coulomb branch, we have $d_v=1+1+1=3$ which matches the Higgs branch
of the original theory. For the Higgs branch, naively we should have
$d_H=[\frac{1}{2} (6+6)+2]-[3+3+3]=-1$. However, the dimension can never 
be negative. The negative result means that our naive calculation is
wrong. The reason is that in our naive calculation we assumed that
there is  complete Higgsing. However, in our example, there is no 
complete Higgsing in the mirror theory. After Higgsing, we still keep 
two $SO(2)$ gauge groups which give the correct $d_H=[8]-[9-2]=1$
and match the Coulomb branch in the original theory. Furthermore, in our 
example, we have one flavor in the mirror theory which means that there is
a ``hidden FI-term'' in the original theory.

\subsection{The general case: $SO(2k+1)$ with $N$ flavors}
Now let us discuss the mirror of $SO(2k+1)$ with $N$ flavors. The 
dimensions of moduli spaces are $d_v=k$ and $d_H=(2k+1)N-k(2k+1)$.
The steps to get the mirror theory are given in Figure \ref{fig:SO_odd}.
In part (a), we give the brane setup of the Higgs branch. In fact, we can
consider it as well as the brane setup of the mirror theory by just changing the
role of the vertical line and cross line (in Higgs branch, vertical lines denote
1/2D5-branes and cross lines, 1/2NS-branes; in the mirror theory, vertical
lines denote 1/2NS-branes and cross lines, 1/2D5-branes). For convenience,
we give the quiver diagram of the mirror theory in part (b).  

\EPSFIGURE[h]{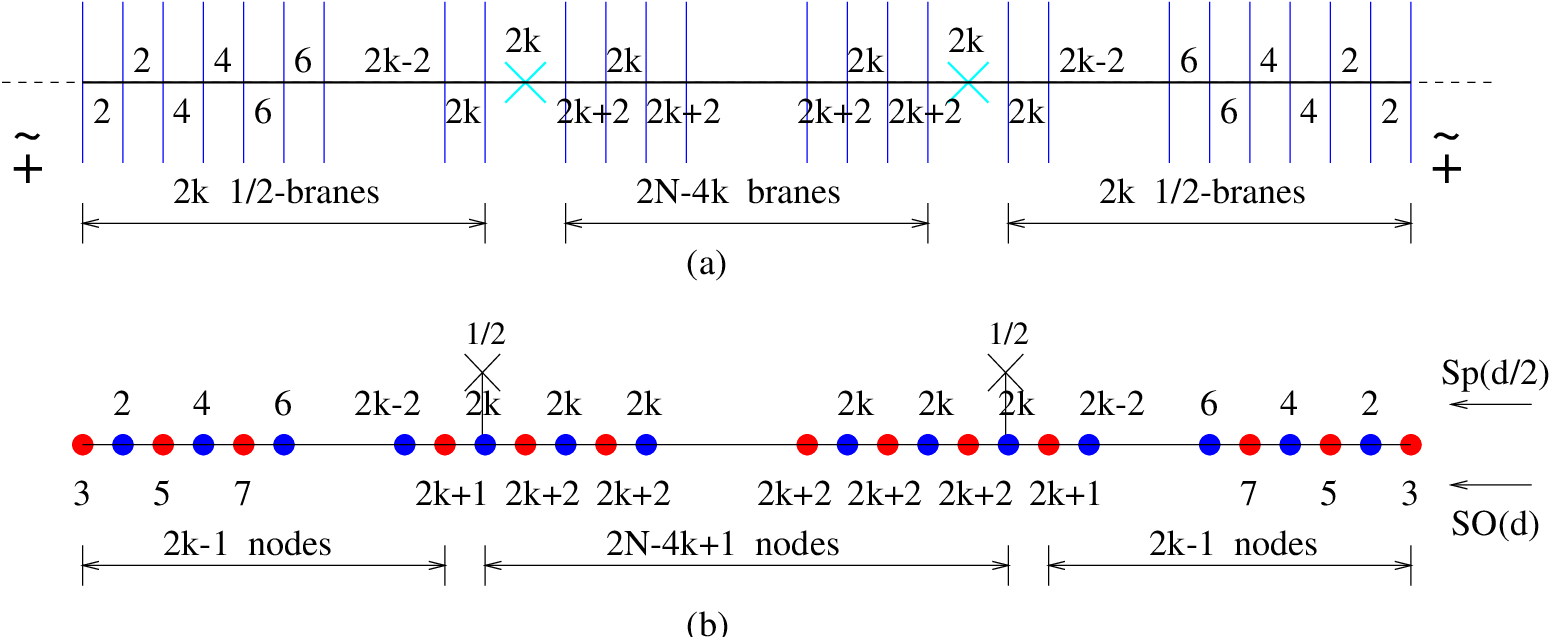,width=15cm} { The mirror of $SO(2k+1)$ with $N$ flavors.
(a) The Higgs branch of the original theory or the Coulomb branch of the mirror
theory. (b) The quiver diagram of the mirror theory.
 \label{fig:SO_odd}}

Let us calculate the dimensions of the moduli spaces of the mirror theory to see
if they match the dimensions of the moduli spaces of the original theory. The
calculations are given as

\begin{equation}
\label{SO_odd}
\begin{array}{lll}
d_v  &  =  & [\sum_{i=1}^{k-1} 4i]+k(N-2k+3)+(k+1)(N-2k)  \\
& = & (2k+1)N-k(2k+1)  \\
d_H & = & 2\sum_{i=1}^{k-1} [\frac{(2i+1)2i}{2}+\frac{2i(2i+3)}{2}-2i(2i+1)]\\
&+ & (2N-4k)\frac{2k(2k+2)}{2}-(N-2k+1)k(2k+1)-(N-2k)(k+1)(2(k+1)-1) \\
&+ & 2 \frac{2k}{2}+2\frac{2k(2k+1)}{2}-2k(2k+1)  \\
&+ & N \\
&= & [2k(k-1)]+[-(N-2k)-k(2k+1)]+[2k]+[N]  \\
& = & k
\end{array}
\end{equation}
Notice that we add  $N$ when we calculate $d_H$ because after the Higgsing,
the mirror theory still keep $N$ $SO(2)$ gauge group. Furthermore,
from the part (a) in Figure \ref{fig:SO_odd} we see when $N=2k$, two
1/2-branes can combine together and leave the orientifold plane. This means
that when $N=2k$ there is a {\sl ``hidden FI-term''} in the original theory.
 This also means that in the special case, the original theory has an enhanced global 
$Sp'(N) \times SO(3)$ symmetry instead of $Sp'(N)$ flavor symmetry in 
general.

\subsection{Comparing  the mirrors of $SO(2k)$ and $SO(2k+1)$}
At the end of this section, let us compare the mirror theories of
$SO(2k)$ and $SO(2k+1)$. First we can start from the $SO(2k+1)$ with
$N+1$ flavors to go to $SO(2k)$ with $N$ flavors by Higgsing one flavor.
At the other side, by comparing the quivers in Figure \ref{fig:SO_gen}
and Figure \ref{fig:SO_odd}, it is obvious that if we change the 
$SO(d)$ gauge group in Figure \ref{fig:SO_odd} to $SO(d-2)$ while
keeping the $Sp(d/2)$ gauge group we get exactly the quiver in 
Figure \ref{fig:SO_gen}. In particular, the two $SO(3)$ gauge group in 
Figure \ref{fig:SO_odd} go to $SO(1)$ and disappear as a gauge group but 
add two half-hypermultiplets to two $Sp(1)$ at the two ends of 
quiver in Figure \ref{fig:SO_gen}. This pattern can also be found if
we higgs $SO(2k)$ with $N$ flavors to $SO(2k-1)$ with $N-1$ flavors.
In the latter case, we change the $Sp(d/2)$ gauge group in 
Figure \ref{fig:SO_gen} to $Sp(d/2-1)$ gauge group while keeping 
$SO(d)$ gauge group. After such a change, the quiver in 
Figure \ref{fig:SO_gen} becomes exactly the quiver in 
Figure \ref{fig:SO_odd} (the two nodes at the ends in Figure \ref{fig:SO_gen}
disappear). Notice that the Higgsing in the original theory should
correspond to the reduction of the Coulomb branch in the mirror theory. The 
change of gauge group is exactly the required reduction of 
the Coulomb branch in the mirror theory.

\par
The above pattern passes another consistency check. Notice that
for $SO(2k+1)$ gauge theory with $N+1$ flavors, it has an enhanced 
$SO(3)$ global symmetry when $N+1=2k$. After Higgsing, we get
$SO(2k)$ with $N$ flavors. For the latter, it has an enhanced 
$SO(3)$ global symmetry exactly when $N=2k-1$. We see such 
hidden global symmetry is not broken by the Higgs mechanism as it
should be.

\section{The mirror of  $Sp(k)\times SO(2m)$}
We have discussed the mirror for a single  $Sp$ or $SO$ group
above. In this section, we generalize the above construction to
the case of the product of $Sp$ and $SO$ gauge groups. Because
after crossing the 1/2NS-brane $O3^{\pm}$($\widetilde{O3^{\pm}}$)
change to $O3^{\mp}$($\widetilde{O3^{\mp}}$)
 and vise versa,
we get two series of  products  $SO(2n_1)  \times Sp(k_1)  \times
SO(2n_2)  \times Sp(k_2)  ..$  and  $SO(2n_1+1)  \times Sp'(k_1)
\times SO(2n_2+1)  \times Sp'(k_2)  ..$. In this section, we 
discuss the first  series and leave the second series to next section.
For simplicity, we
will discuss only the product of two gauge groups, i.e.,  $ Sp(k)
\times SO(2m) $ (the case of more product groups can be directly
generalized). For this simple case, we still have two choices,
the so called ``elliptic model'' \cite{Wit2} ($X^6$ direction is
compactified) , or the ``non-elliptic model'' ($X^6$ direction is not
compactified). We discuss these two models one by one.

\subsection{The non-elliptic model}
For the non-elliptic model, there are $N$ fundamentals for $SO(2m)
$, $H$ fundamentals for $Sp(k) $ and one bifundamental (for
simplicity we assume that $N,H$ are sufficiently large. For $N,H$
too small, there are a lot of special cases which  need to be
discussed individually and are tedious without providing too much
new insight). The moduli are $d_v=m+k$ and
$d_H=2mN+2kH+2mk-m(2m-1) -k(2k+1) $. In 
constructing the mirror theory, we need to study three
cases: $m>k$, $m=k$ and $m<k$. Let us start with the case of
$m>k$. The mirror theory is given in Figure  \ref{fig:SPOm}. 
\EPSFIGURE[h]{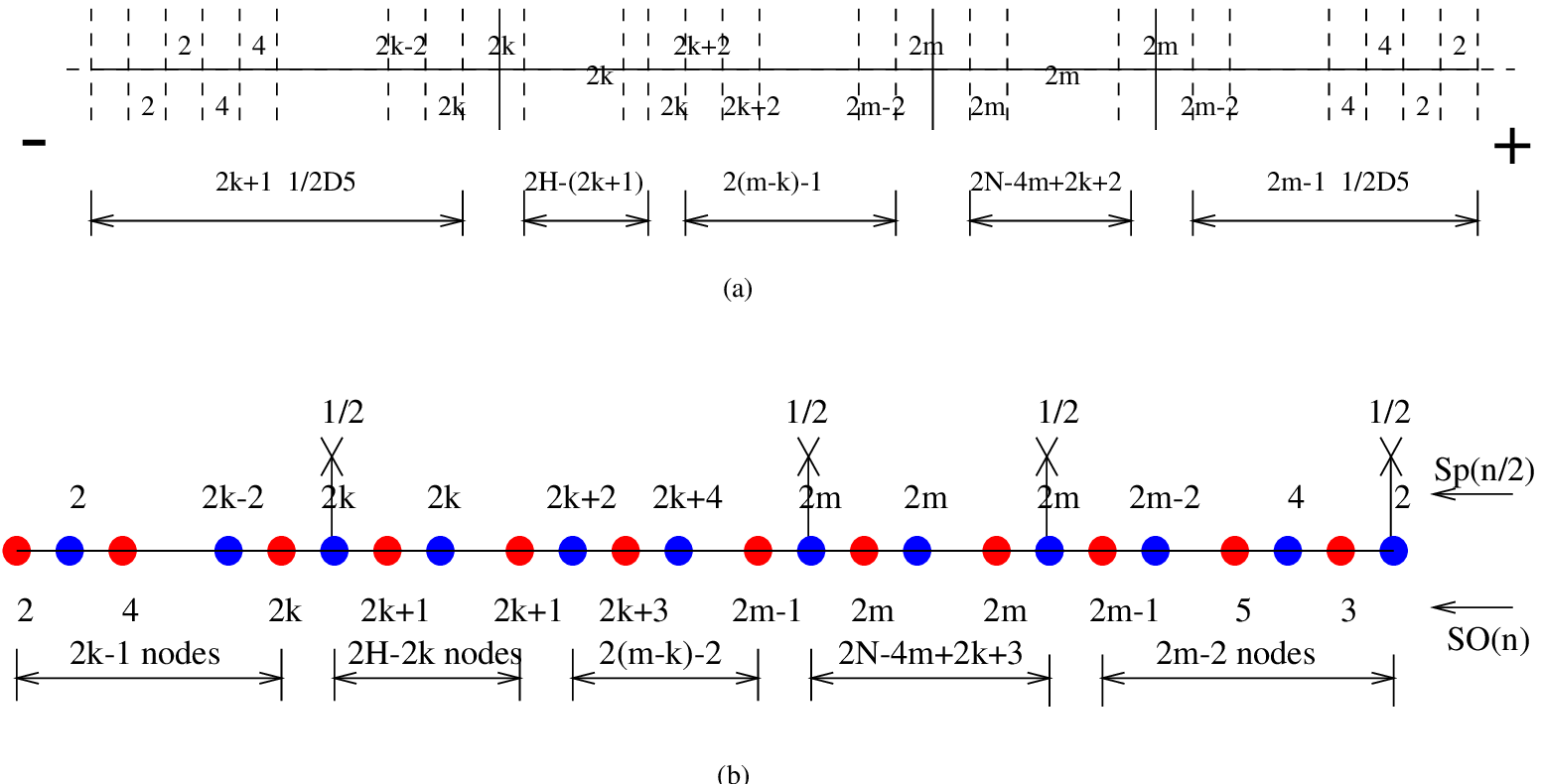,width=15cm} {(a) The Higgs branch of $Sp(k) \times SO(2m)
$ with $N$ fundamentals for $SO(2m) $, $H$ fundamentals for $Sp(k)
$ and one bifundamental in the case of $m>k$. (b)  The quiver
diagram of the mirror theory of part(a). Notice that the index
$n$ above the node denotes $Sp(n/2) $ and index $n$ below the node
denotes $SO(n) $. The $1/2$ means half-hypermultiplets.
\label{fig:SPOm}}
When
we go to the Higgs branch, we can connect the D3-branes at the two
sides of middle 1/2NS-brane. Because  $m>k$, we can connect only
$k$ D3-branes such that they end on the left and the right
1/2NS-branes. There are still $m-k$ D3-branes ending on the middle
1/2NS-brane
 from the right. To get rid of those D3-branes, we must move the middle
1/2NS-brane to the right. The final Higgs branch after such a motion
is given in part (a)  of Figure  \ref{fig:SPOm} and the quiver
diagram of the mirror, in part (b). The moduli of the mirror can
be calculated as
\begin{equation}
\label{SPOmquiver}
\begin{array}{lll}
d_v & = & [\sum_{p=1}^{p=k-1} 2p]+(2H-2k+1) k
    +[2\sum_{p=k+1}^{m-1} p] \\
    & + & (2N-4m+2k+3) m+[2\sum_{p=1}^{p=m-1} p] \\
    & = & 2mN+2kH+2mk-m(2m-1) -k(2k+1) , \\
d_H & = & [\sum_{i=1}^{i=k-1} (2i\times 2i+2i\times(2i+2) )
/2-i(2i-1)
    -i(2i+1) ]  \\
    & + & [4k^2/2+(2H-2k-1) 2k(2k+1) /2-(H-k) 2k(2k+1) -k(2k-1) ]  \\
    & + & [\sum_{i=1}^{m-k-1}(2k+2i-1) (2k+2i) /2+(2k+2i) (2k+2i+1) /2
    -2(k+i) (2k+2i+1) ]  \\
    & + & [2m(2m-1) /2+4m^2(2N-4m+2k+2) /2 \\
    &  & -(N-2m+k+1) (m(2m-1) +m(2m+1) )
    -m(2m+1) ]\\
    & + & [\sum_{i=1}^{m-1} 2i(2i+1) /2+(2i+1) (2i+2) /2
    -2i(2i+1) ] \\
    & + & [2k/2+2m/2+2/2+2m/2] \\
& = & [k(k-1) ]+[-2k^2]+[(-k-m) (m-k-1) ]+[-2m]+[m^2-1]+[k+2m+1]
\\ & = & k+m
\end{array}
\end{equation}
From  part (a)  of Figure  \ref{fig:SPOm}, we see that when
$2N-4m+2k+2=0$, the two 1/2NS-branes meet together which indicates
a ``hidden FI-parameter'' in the original theory.

\par
After the discussion of the $m>k$ case, we go to the $m=k$ case. 
Here, by connecting the D3-branes between the two sides of
the middle 1/2NS-brane, we get the Higgs branch looking like
part (a)  in Figure  \ref{fig:SPOmk}. 
\EPSFIGURE[h]{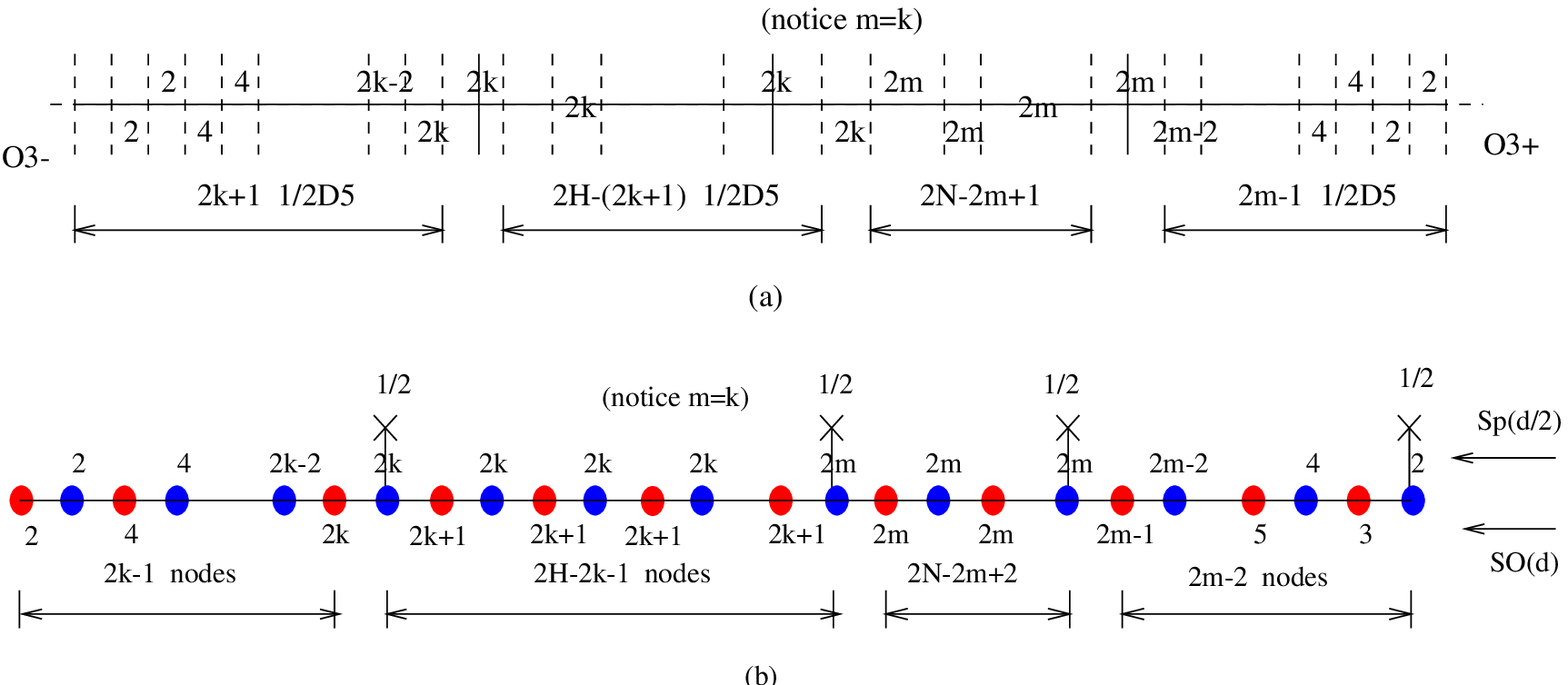,width=15cm} {(a) The Higgs branch of $Sp(k) \times
SO(2m) $ with $N$ fundamentals for $SO(2m) $, $H$ fundamentals for
$Sp(k) $  and one bifundamental in the case of $m=k$. (b)  The
quiver diagram of the mirror theory of part(a). Notice that the
index $n$ above the node denotes $Sp(n/2) $ and index $n$ below
the node denotes $SO(n) $. The $1/2$ means the
half-hypermultiplet. \label{fig:SPOmk}}
From the quiver diagram
part (b)  we recalculate the moduli space as:
\begin{equation}
\label{SPOmkquiver}
\begin{array}{lll}
d_v & = & [\sum_{p=1}^{p=k-1} 2p]+(2H-2k-1+2N-2m+2+1) k
    +[2\sum_{p=k+1}^{m-1} p] \\
    &=& 2kN+2kH-2k^2  \\
    & = & 2mN+2kH+2mk-m(2m-1) -k(2k+1) ~~~~when~~m=k, \\
d_H & = & [\sum_{i=1}^{i=k-1} (2i\times 2i+2i\times(2i+2) )
/2-i(2i-1)
    -i(2i+1) ]  \\
    & + & [4k^2/2+(2H-2k-2) 2k(2k+1) /2-(2H-2k-1) k(2k+1) -k(2k-1) ]  \\
    & + & [(2N-2m+2) 4m^2/2-(N-m+1) (m(2m+1) +m(2m-1) ) ]  \\
    & + & [\sum_{i=1}^{m-1} 2i(2i+1) /2+(2i+1) (2i+2) /2
    -2i(2i+1) ] \\
    & + & [2k/2+2m/2+2/2+2m/2] \\
& = & [k(k-1) ]+[-2k^2]+[0]+[m^2-1]+[k+2m+1] \\ & = & k+m.
\end{array}
\end{equation}
From the figure again, when $2H-2k=0$ or $2N-2m+2=0$, 
the two 1/2NS-branes meet
together to give a ``hidden
FI-term'' in the original theory.

\par
Now we are left with only one case, i.e., $m<k$. In this last case,
to get rid of the D3-branes, the middle 1/2NS brane should move to
the left direction. The result is shown in Figure  \ref{fig:SPOk}.

The moduli of the mirror theory are
\begin{equation}
\label{SPOkquiver}
\begin{array}{lll}
d_v & = & [\sum_{p=1}^{p=k-1} 2p]+(2H-4k+2m+1) k
    +[2\sum_{p=m+1}^{k-1} p] \\
    & + & (2N-2m+3) m+[2\sum_{p=1}^{p=m-1} p] \\
    & = & 2mN+2kH+2mk-m(2m-1) -k(2k+1) , \\
d_H & = & [\sum_{i=1}^{i=k-1} (2i\times 2i+2i\times(2i+2) )
/2-i(2i-1)
    -i(2i+1) ]  \\
    & + & [4k^2/2+(2H-4k+2m-2) 2k(2k+1) /2+4k^2/2 \\
    & &  -(2H-4k+2m-1) k(2k+1) -2k(2k-1) ]  \\
    & + & [\sum_{i=1}^{k-m-1}2(m+i) 2(m+i) /2+2(m+i) 2(m+i+1) /2 \\
    && -(m+i) (2(m+i) -1) -(m+i) (2(m+i) +1) ]  \\
    & + & [2m(2m+2) /2+4m^2(2N-2m+2) /2 \\
    &  & -(N-m+1) (m(2m-1) +m(2m+1) ) -m(2m+1) ]\\
    & + & [\sum_{i=1}^{m-1} 2i(2i+1) /2+(2i+1) (2i+2) /2
    -2i(2i+1) ] \\
    & + & [2k/2+2k/2+2/2+2m/2] \\
& = & [k(k-1) ]+[-2k^2+k]+[(k+m) (k-m-1) ]+[m]+[m^2-1]+[2k+m+1] \\
& = & k+m
\end{array}
\end{equation}
There is also a possible ``hidden FI-term'' in original theory when
$2H-4k+2m-2=0$ which can be explicitly seen in part(a)  in Figure
\ref{fig:SPOk}.

\EPSFIGURE[ht]{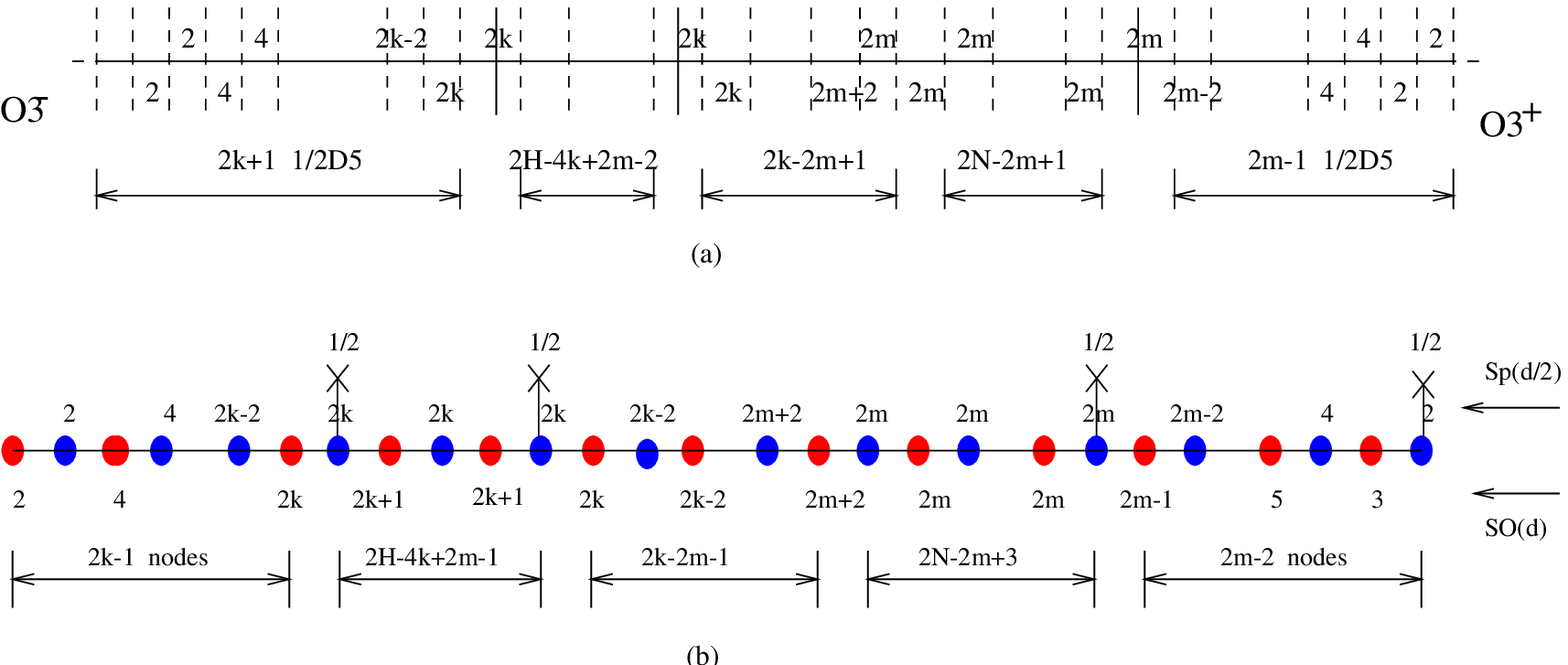,width=15cm} {(a) The Higgs branch of $Sp(k) \times SO(2m)
$ with $N$ fundamentals for $SO(2m) $, $H$ fundamentals for $Sp(k)
$ and one bifundamental in the case of $m<k$. (b)  The quiver
diagram of the mirror theory of part(a). Notice that the index
$n$ above the node denotes $Sp(n/2) $ and index $n$ below the node
denotes $SO(n) $. The $1/2$ denotes the half-fundamental.
\label{fig:SPOk}}

\subsection{The elliptic model}
In the elliptic model, the $X^6$ direction is compactified such
that for consistency, we must have an even number of 1/2NS-branes
and an even number of gauge groups where half of them are $Sp$
gauge groups  and the other half, $SO$ gauge groups.  We discuss the
case of only two gauge groups, i.e., $Sp(k) \times SO(2m) $ with $H$
fundamentals for $Sp(k) $, $N$ fundamentals for $SO(2m) $ and two
bifundamentals. The moduli for this theory are $d_v=k+m$ and
$d_H=2mN+2kH+4mk-m(2m-1) -k(2k+1) $. The mirror theory for the
elliptic model is similar to the non-elliptic model. The only
difference is that in the non-elliptic model we can connect the
D3-branes only at the middle 1/2NS-brane, but here in the elliptic
model we can connect the D3-branes to all 1/2NS-branes (here two
1/2NS-branes). Again we divide into three cases to
discuss. The simple one is the case $m=k$. In this case, we can
connect all D3-branes such that no
 D3-brane is left to end on the 1/2NS-branes. The Higgs branch and the
quiver of the mirror are given in parts (a)  (b)  of Figure
\ref{fig:Ell_mk} and the moduli are:
\begin{equation}
\label{Ell_mk}
\begin{array}{lll}
d_v & = & (2H+2N) k  \\
    & = & 2mN+2kH+4mk-m(2m-1) -k(2k+1) ~~~when~~m=k, \\
d_H & = & [(2H-2) 2k(2k+1) /2-(H-1) 2k(2k+1) -k(2k+1) ]  \\
    & + & [(2N+2) 4k^2/2-N(k(2k+1) +k(2k-1) ) -k(2k-1) ]  \\
    & + & [2k/2+2k/2]  \\
    & = & [-k(2k+1) ]+[2k^2+k]+[2k] = k+m
\end{array}
\end{equation}
\EPSFIGURE[h]{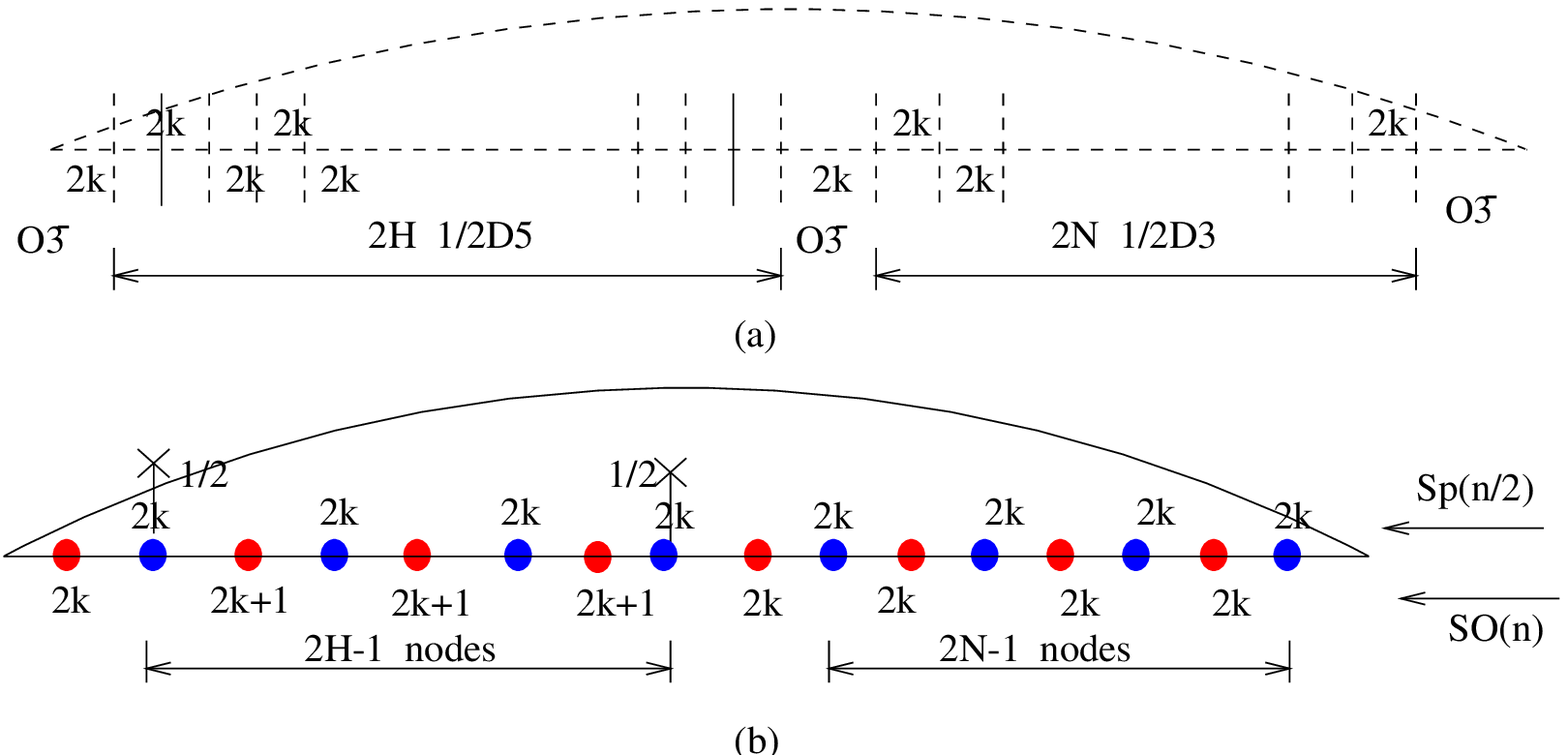,width=15cm} {(a) The Higgs branch of elliptic $Sp(k)
\times SO(2m) $ with $N$ fundamentals for $SO(2m) $, $H$
fundamentals for $Sp(k) $  and two bifundamentals in the case of
$m=k$. The number here denotes how many 1/2D3-branes are connected
to neighboring 1/2D5-branes. (b)  The quiver diagram of the mirror
theory of part(a). Notice that the index $n$ above the node
denotes $Sp(n/2) $ and index $n$ below the node denotes $SO(n) $.
The $1/2$ denotes the half hypermultiplets. \label{fig:Ell_mk}}

There is still one case where the ``hidden FI-term'' appears, namely when
$H=1$. In this case, two 1/2NS-branes in part(a)  of Figure
\ref{fig:Ell_mk} meet together.

\par
Now we move to the case of $k>m$. The Higgs branch looks like the
superposition of the Higgs branch of the $m=k$ case together with that of
 a single $Sp(k-m) $ gauge
theory. We give the result in Figure  \ref{fig:Ell_k}. To check
the result, we calculate the moduli as
\begin{equation}
\label{Ell_k}
\begin{array}{lll}
d_v & = & [4\sum_{i=1}^{i=k-m-1} (m+i) ]+k(2H-4(k-m) +1) +m(2N+3)
\\
    & = & 2mN+2kH+4mk-m(2m-1) -k(2k+1) , \\
d_H & = & [2\sum_{i=1}^{i=k-m-1} 2(m+i) 2(m+i) /2+2(m+i) 2(m+i+1)
/2
\\
    &  &  -(m+i) (2(m+i) -1) -(m+i) (2(m+i) +1) ]  \\
    & + & [2\times 4k^2/2+(2H-4(k-m) -2) 2k(2k+1) /2 \\
        & & -(2H-4(k-m) -1) k(2k+1) -2k(2k-1) ]  \\
    & + & [(2N+2) 4m^2/2+2\times 2m(2m+2) /2-(N+1) (m(2m+1) +m(2m-1) ) -m(2m+1) ] \\
    & + & [2\times 2k/2]   \\
    & = & [2(k^2-m^2-k-m) ]+[-2k^2+k]+[2m^2+3m]+[2k]  \\
    & = & k+m.
\end{array}
\end{equation}
\EPSFIGURE[ht]{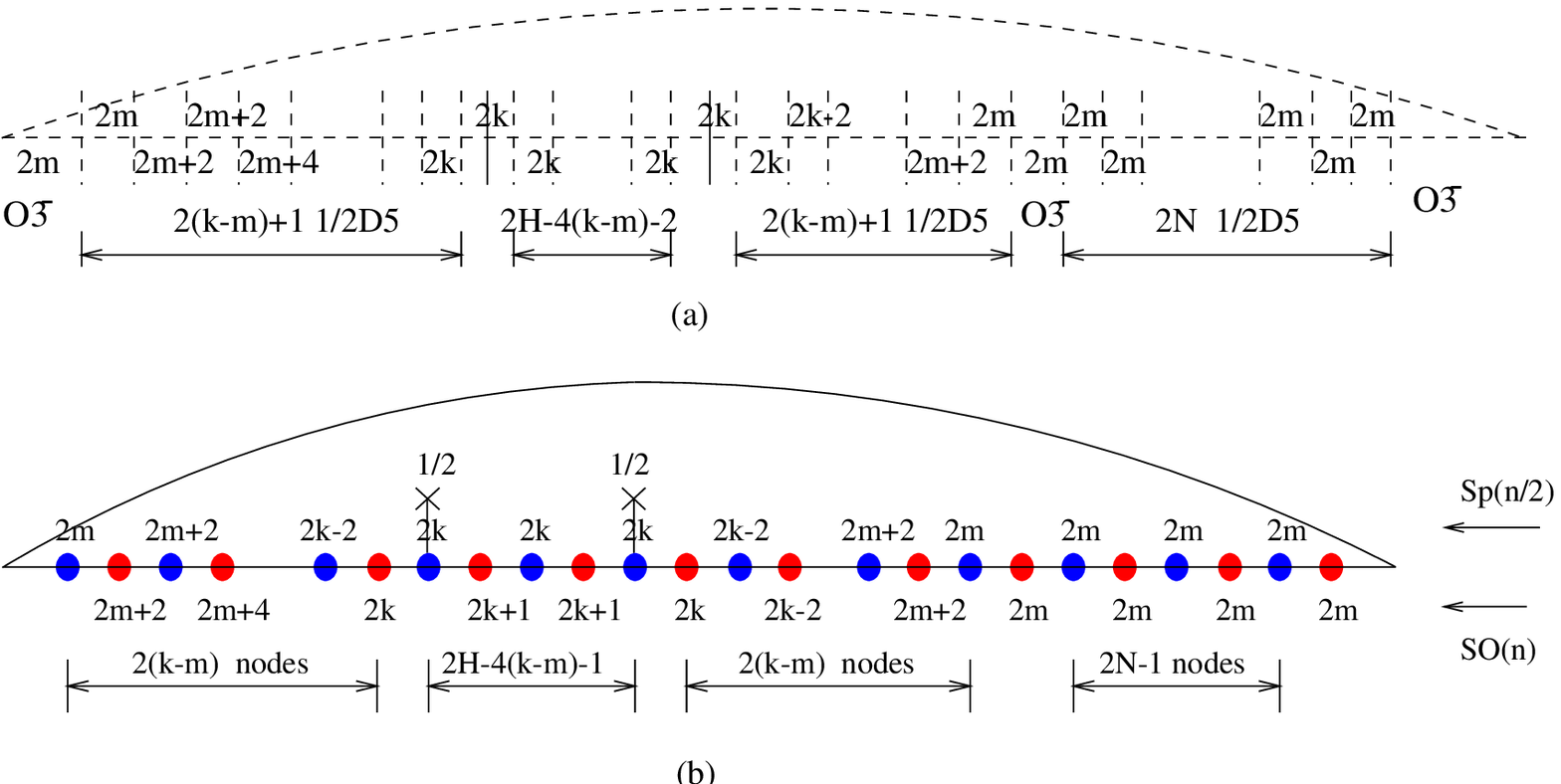,width=15cm} {(a) The Higgs branch of elliptic $Sp(k)
\times SO(2m) $ with $N$ fundamentals for $SO(2m) $, $H$
fundamentals for $Sp(k) $  and two bifundamentals in the case of
$m<k$. The number here denotes how many 1/2D3-branes are connected
to neighboring 1/2D5-branes. (b). The quiver diagram of the
mirror theory of part(a). Notice that the index $n$ above the
node denotes $Sp(n/2) $ and index $n$ below the node denotes
$SO(n) $. The $1/2$ denotes the half hypermultiplets.
\label{fig:Ell_k}}
When $2H-4(k-m) -2=0$, two 1/2NS-branes will meet together in
part(a)  of Figure  \ref{fig:Ell_k}. This is the condition that a
``hidden FI-term'' exists.

\par
Now the remainder case is $m>k$. In this case, the Higgs branch
looks like the superposition of two Higgs branches: that of the
$m=k$ case and that of  a single $SO(2(m-k) ) $ theory. The result
can be found in Figure  \ref{fig:Ell_m}. The moduli of the mirror
theory are
\begin{equation}
\label{Ell_m}
\begin{array}{lll}
d_v & = & [4\sum_{i=1}^{i=m-k-1}(k+i) ]+k(2H+1) +m(2N-4(m-k) +3)
\\
    & = & 2mN+2kH+4mk-m(2m-1) -k(2k+1) , \\
d_H & = & [2\sum_{i=1}^{i=m-k-1} 2(k+i) (2k+2i+1) /2+(2k+2i+1)
(2k+2i+2) /2  \\
    &  &  -2(k+i) (2k+2i+1) ]  \\
    & + & [(2N-4(m-k) +2) 4m^2/2-(N-2(m-k) +1) (m(2m+1) +m(2m-1) ) \\
& & -m(2m+1) ]  \\
& + & [2h2k(2k+1) /2+2(2k+1) (2k+2) /2-(2H+1) k(2k+1) ] \\
    & + & [2\times 2m/2]   \\
    & = & [2m^2-2(k+1) ^2]+[-2m^2-m]+[2k^2+5k+2]+[2k]  \\
    & = & k+m
\end{array}
\end{equation}
When $2N-4(m-k) +2=0$, there is a ``hidden FI-term'' in the
original theory.

\EPSFIGURE[h]{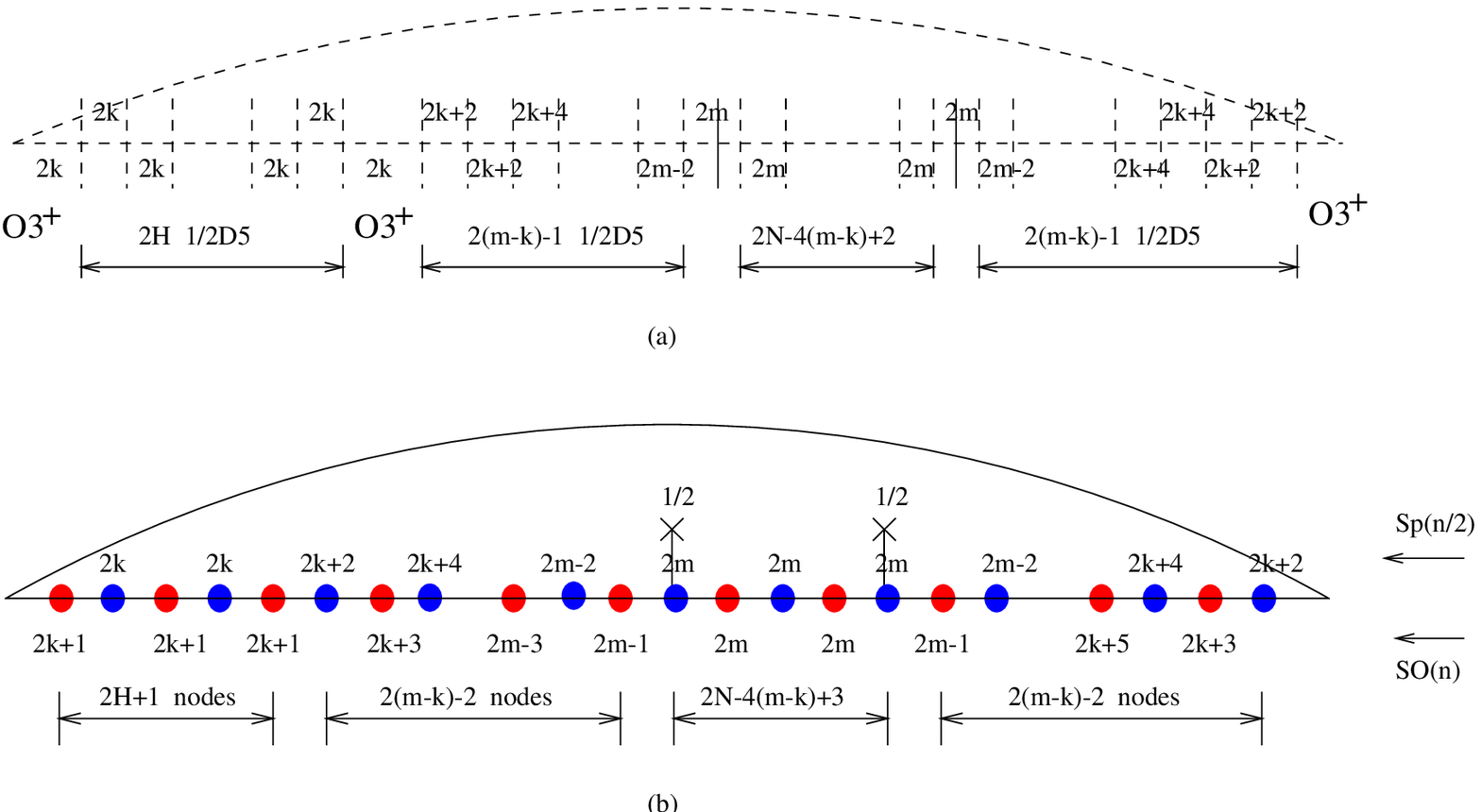,width=15cm} {(a).The Higgs branch of elliptic $Sp(k)
\times SO(2m) $ with $N$ fundamentals for $SO(2m) $, $H$
fundamentals for $Sp(k) $  and two bifundamentals in the case of
$m>k$. The number here means how many 1/2D3-branes are connected
to neighboring  1/2D5-branes. (b). The quiver diagram of the
mirror theory of part(a). Notice that the index $n$ above the
node denotes $Sp(n/2) $ and index $n$ below the node denotes
$SO(n) $. The $1/2$ denotes the half hypermultiplets.
\label{fig:Ell_m}}

\section{The mirror  of $Sp'(k)\times SO(2m+1)$}
For completion, we give one more example: the mirror theory of
$Sp'(k)\times SO(2m+1)$. We assume that there are $H$ flavors for 
$Sp'(k)$ gauge theory and $N$ flavors for $SO(2m+1)$ gauge theory. Besides,
there are one or two bifundamentals and half-hypermultiplet for $Sp'(k)$
depend on different situations.
Again we divide our discussion into two parts: non-elliptic model and
elliptic  model.

\subsection{The non-elliptic model}
Let us start from the non-elliptic model. In this case, the dimensions of
moduli spaces are $d_v=k+m$ and $d_H=2kH+(2m+1)N+k(2m+1)+k
-k(2k+1)-m(2m+1)=2kH+(2m+1)N-2k^2+k-2m^2-m+2km$ (here
again, for simplicity we assume $N,H$ are sufficiently large to avoid
special cases). The mirror theory depends on whether $m>k$ , $m=k$ 
or $m<k$. We first give the mirror of the case $m=k$ because in this
particular case, we can combine the D3-branes at the two sides of
middle 1/2NS-brane such that there is no D3-branes ending on the middle 
1/2NS-brane anymore. The mirror theory is given in Figure \ref{fig:pSPO_mk}.
\EPSFIGURE[h]{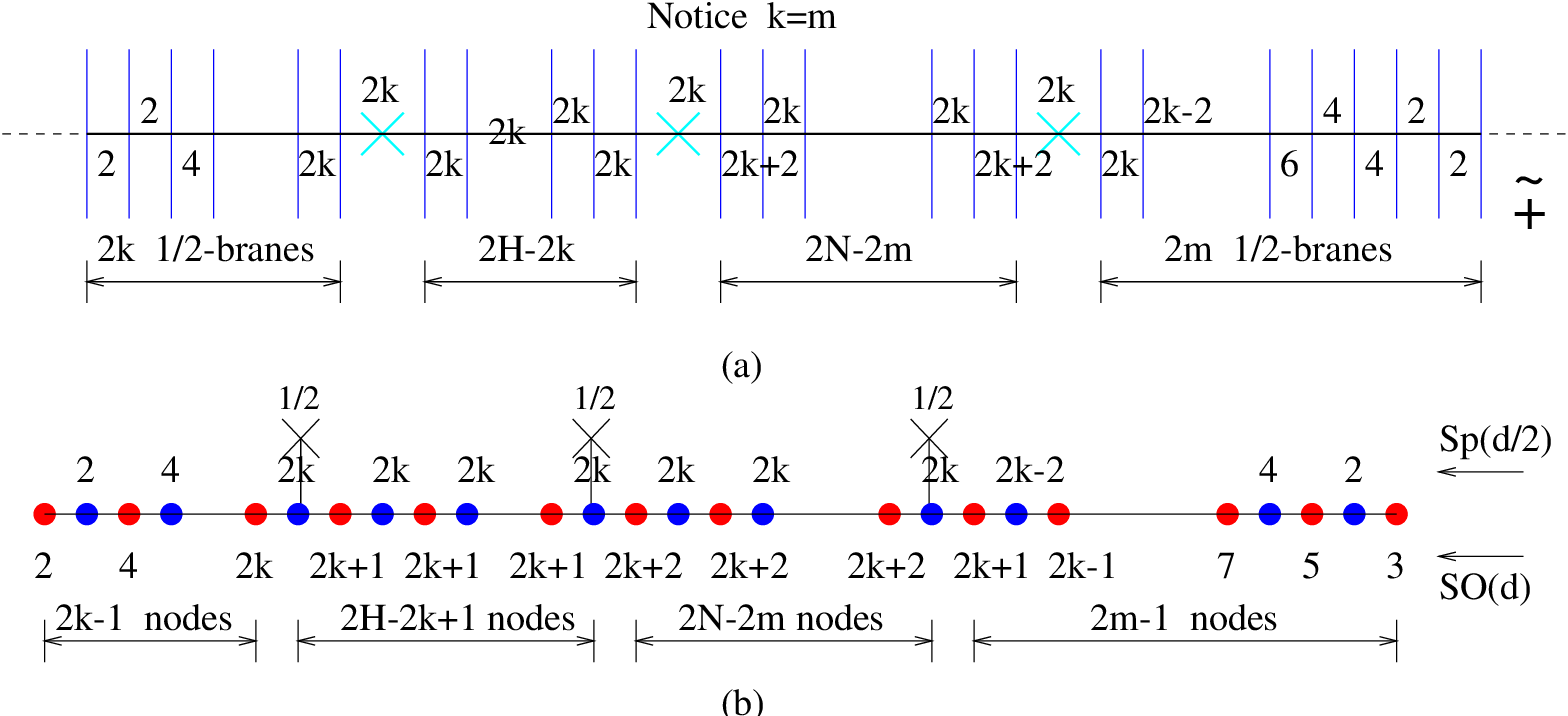,width=15cm} { The mirror of 
$Sp'(k) \times SO(2m+1)$ with $H$ flavors for $Sp'(k)$ , $N$ flavors 
for $SO(2m+1)$, a half-hypermultiplet for $Sp'(k)$ and one bifundamental
 in case of $m=k$. 
(a) The Higgs branch of original theory or the Coulomb branch of the mirror
theory. (b) The quiver diagram of mirror theory.
 \label{fig:pSPO_mk}}
Let us check it by calculating the dimensions of moduli spaces of the
mirror theory:
\begin{equation}
\label{pSPO_mk}
\begin{array}{lll}
d_v & = & 2\sum_{i=1}^{k-1} i + (2H-2k+2)k+ (N-k)(k+k+1)+k+2\sum_{i=1}^{k-1} i
\\
& = &  2kH+(2k+1)N-2k^2 \\
d_H & = & \sum_{i=1}^{k-1} [ \frac{2i2i}{2}+\frac{2i(2i+2)}{2}-
i(2i-1)-i(2i+1)] \\
& + &\sum_{i=1}^{k-1} [\frac{(2i+1)2i}{2}+\frac{2i(2i+3)}{2}-2i(2i+1)]\\
& + & \frac{2k(2k+1)}{2}(2H-2k)-(H-k)2k(2k+1) \\
& + & \frac{2k(2k+2)}{2}(2N_2k)-(N-k)(k(2k+1)+(k+1)(2k+1))  \\
& + & \frac{2k2k}{2}-k(2k+1)-k(2k-1)+\frac{2k(2k+1)}{2}-k(2k+1)\\
& + & 3\frac{2k}{2}+N  \\
& = & [k^2-k]+[k^2-k]+[0]+[-(N-k)]+[-2k^2]+[3k+N] \\
& = & 2k, \\
\end{array}
\end{equation}
where when we calculate the $d_H$ we add the term $N$ to account for  the 
remaining $H$ $SO(2)$ gauge groups after Higgsing (this happens for latter
examples so we will not mention it every time). When $2H-2m=0$ there
is a {\sl ``hidden FI-term''} in the original theory.

\par
Now we go to the case that $k>m$. In this case, after connecting the 
D3-branes at the two sides of the middle 1/2NS-brane, we still have
$k-m$ D3-brane ending on the middle 1/2NS-brane from the left. The
mirror theory is given in Figure \ref{fig:pSPO_k}.
\EPSFIGURE[h]{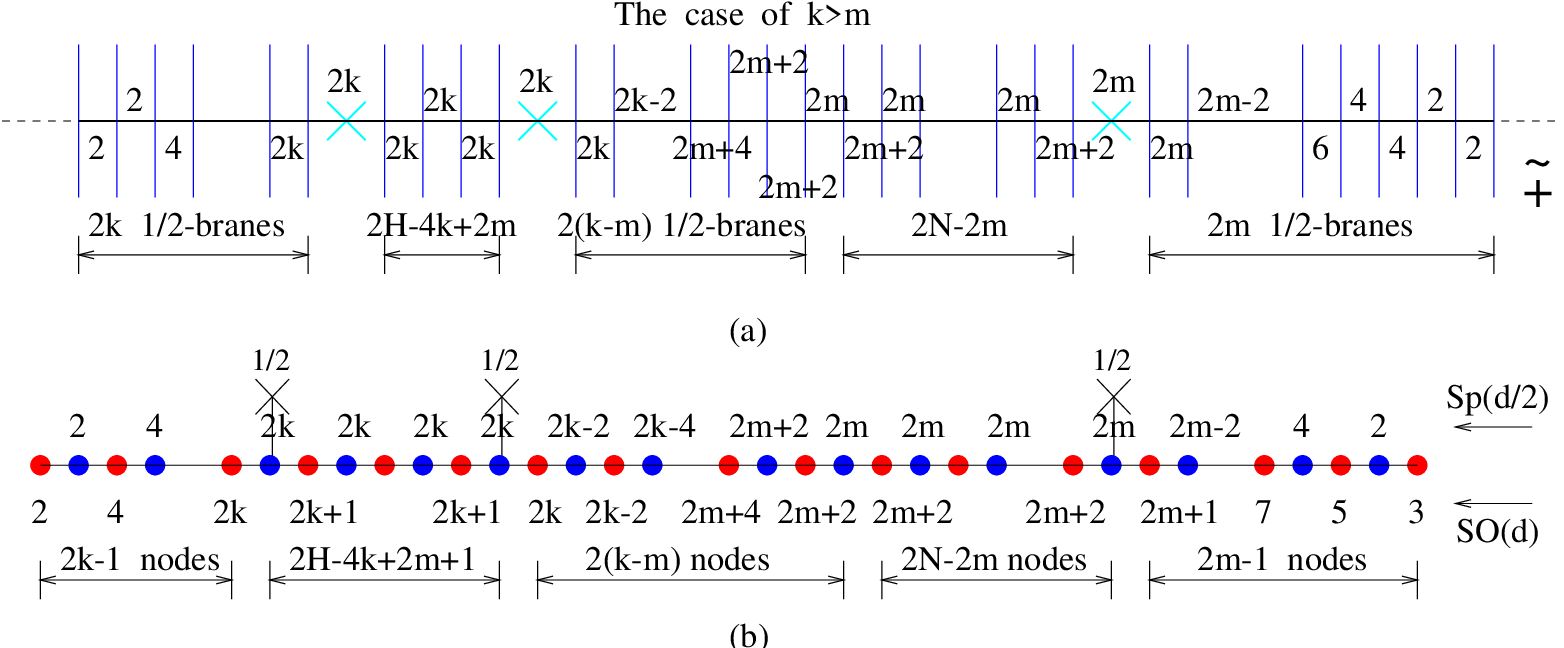,width=15cm} { The mirror of 
$Sp'(k) \times SO(2m+1)$ with $H$ flavors for $Sp'(k)$ , $N$ flavors 
for $SO(2m+1)$, a half-hypermultiplet for $Sp'(k)$ and one bifundamental
 in case of $k>m$. 
(a) The Higgs branch of original theory or the Coulomb branch of the mirror
theory. (b) The quiver diagram of mirror theory.
 \label{fig:pSPO_k}}
The dimensions of the moduli space of the mirror theory are
\begin{equation}
\label{pSPO_k}
\begin{array}{lll}
d_v  &= & [2\sum_{i=1}^{k-1} i] +[2\sum_{i=1}^{m-1} i]+(2H-4k+2m+3)k \\
& + & [2\sum_{i=1}^{k-m-1} (m+i)]+[(N-m+2)m+(N-m)(m+1)]  \\
& = & [k^2-k]+[m^2-m]+[2kH-4k^2+2km+3k]  \\
& + & [k^2-m^2-k-m]+[(2m+1)N-2m^2+m]  \\
& = & 2kH+(2m+1)N-2k^2-2m^2+k-m+2km  \\
d_H & = & \sum_{i=1}^{k-1} [ \frac{2i2i}{2}+\frac{2i(2i+2)}{2}-
i(2i-1)-i(2i+1)] \\
& + & 2\frac{2k2k}{2}+\frac{2k(2k+1)}{2}(2H-4k+2m)-
(2H-4k+2m+1)k(2k+1)-2k(2k-1) \\
& + & \sum_{i=1}^{k-m-1}[ \frac{(2(m+i))^2}{2}+\frac{2(m+i)2(m+i+1)}{2}
-(m+i)(2(m+i)-1)-(m+i)(2(m+i)+1)] \\
& + & \frac{2m(2m+2)}{2}(2N-2m+1)-(N-m+1)m(2m+1)-(N-m)m(2(m+1)-1)  \\
& + & \sum_{i=1}^{m-1} [\frac{(2i+1)2i}{2}+\frac{2i(2i+3)}{2}-2i(2i+1)]\\
& + & \frac{2m(2m+1)}{2}-m(2m+1)+2\frac{2k}{2}+\frac{2m}{2}+N  \\
& = & [k^2-k]+[-2k^2+k]+[k^2-m^2-k-m]+[-N+2m]+[m^2-m]+[N+2k+m] \\
& = & k+m.
\end{array}
\end{equation}

After the discussion of above two cases, we go to the last case: $k<m$.
In this case, because $k<m$, after the combination of D3-branes 
at the two sides of middle 1/2NS-brane, we still leave $m-k$ D3-brane 
ending on it from the right. The mirror theory is given in 
Figure \ref{fig:pSPO_m}. 
\EPSFIGURE[h]{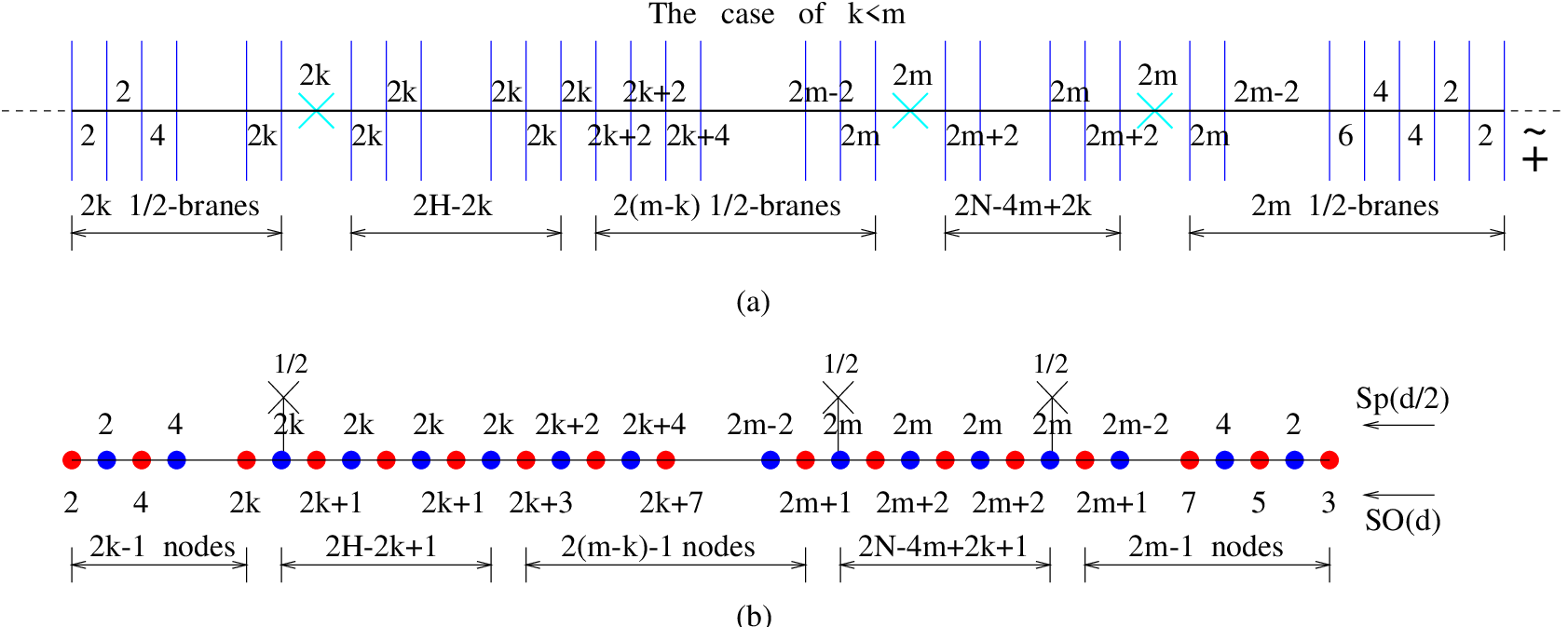,width=15cm} { The mirror of 
$Sp'(k) \times SO(2m+1)$ with $H$ flavors for $Sp'(k)$ , $N$ flavors 
for $SO(2m+1)$, a half-hypermultiplet for $Sp'(k)$ and one bifundamental
 in case of $k<m$. 
(a) The Higgs branch of the original theory or the Coulomb branch of the mirror
theory. (b) The quiver diagram of the mirror theory.
 \label{fig:pSPO_m}}
Let us calculate the dimensions of moduli spaces:
\begin{equation}
\label{pSPO_m}
\begin{array}{lll}
d_v & = & [2\sum_{i=1}^{k-1} i] +[2\sum_{i=1}^{m-1} i]+(2H-2k+2)k \\
& + & 2\sum_{i=1}^{m-k-1} (k+i)+m(N-2m+k+3)+(m+1)(N-2m+k)  \\
& = & [k^2-k]+[m^2-m]+[2kH-2k^2+2k]+[m^2-k^2-k-m] \\
& + & [(2m+1)N-4m^2+m+k+2km] \\
& = & 2kH+(2m+1)N+2km-2k^2-2m^2-m+k  \\
d_H & = & \sum_{i=1}^{k-1} [ \frac{2i2i}{2}+\frac{2i(2i+2)}{2}-
i(2i-1)-i(2i+1)] \\
& + & \sum_{i=1}^{m-1} [\frac{(2i+1)2i}{2}+\frac{2i(2i+3)}{2}-2i(2i+1)]\\
& + & \frac{(2k)^2}{2}+\frac{2k(2k+1)}{2}(2H-2k)-k(2k-1)-(2H-2k+1)k(2k+1) \\
& + & \sum_{i=1}^{m-k-1}[\frac{(2k+2i+1)2(k+i)}{2}+\frac{2(k+i)(2k+2i+3)}{2}
-2(k+i)(2(k+i)+1)]\\
& + &[ 2\frac{2m(2m+1)}{2}+\frac{2m(2m+2)}{2}(2N-4m+2k)\\
& & -(N-2m+k+3)m(2m+1)-(N-2m+k)(m+1)(2m+1) ] \\
& + & \frac{2k(2k+3)}{2}+k+2m+N  \\
& = & [k^2-k]+[m^2-m]+[-2k^2]+[m^2-k^2-k-m]\\
& + & [-N-2m^2+m-k]+[2k^2+4k+2m+N]\\
& = & m+k.  \\
\end{array}
\end{equation}

\subsection{The elliptic model}
In this section, we discuss the mirror theory of $Sp'(k)\times SO(2m+1)$
in the elliptic model. Now because  $X^6$ is compact, the matter contents
are $H$ flavors for $Sp'(k)$, $N$ flavors for $SO(2m+1)$ and two 
bifundamentals. The dimensions of the moduli spaces are 
$d_v=k+m$ and $d_H=2kH+(2m+1)N+2k(2m+1)-k(2k+1)-m(2m+1)=
2kH+(2m+1)N+4km-2k^2-2m^2-m+k$. Again, our investigation will be divided into
three cases $k=m$, $k>m$ and $k<m$. 

\par
Let us start from the case $k=m$. In this case, because we can combine all
D3-branes at the two sides of 1/2NS-branes, it makes the mirror theory 
very simple as shown in Figure \ref{fig:pEll_mk}. 
\EPSFIGURE[h]{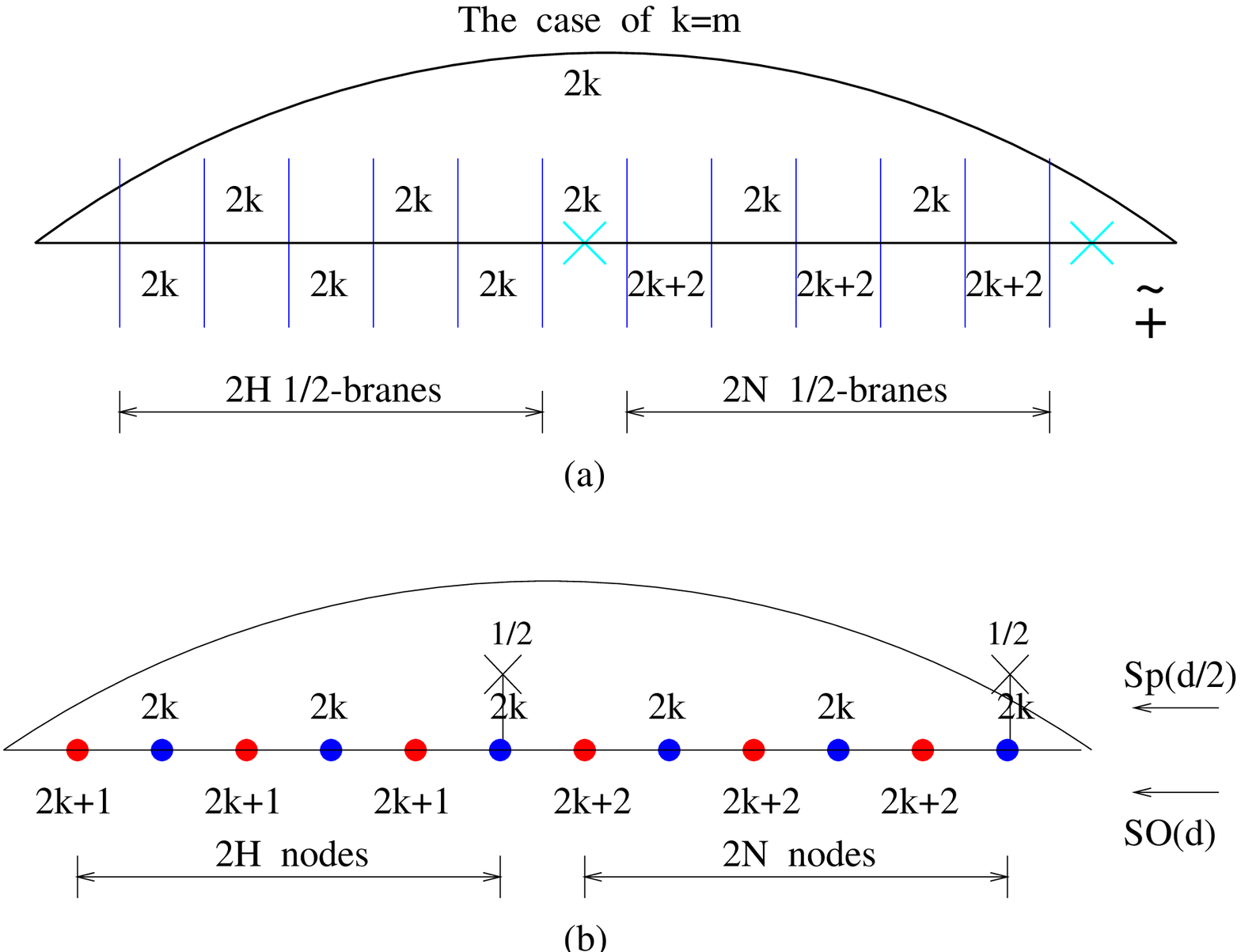,width=15cm} { The mirror of 
$Sp'(k) \times SO(2m+1)$ with $H$ flavors for $Sp'(k)$ , $N$ flavors 
for $SO(2m+1)$ and  two bifundamentals
 in case of $k=m$. 
(a) The Higgs branch of the original theory or the Coulomb branch of the mirror
theory. (b) The quiver diagram of the mirror theory.
 \label{fig:pEll_mk}}
Let us check the dimensions of moduli spaces:
\begin{equation}
\label{pEll_mk}
\begin{array}{lll}
d_v & = & 2kH+Nk+N(k+1)=2kH+(2k+1)N  \\
d_H & = & [\frac{2k(2k+1)}{2}2H-2H k(2k+1)]+ [N]+[2\frac{2k}{2}] \\
& + & [\frac{2k(2k+2)}{2}2N-Nk(2k+1)-N(k+1)(2k+1)] \\
& = & 2k. \\
\end{array}
\end{equation}

\par
Now we go to the case of $k>m$. In this case, After combining the D3-branes,
we still leave $k-m$ D3-branes in the interval of $\widetilde{O3^+}$-plane.
The mirror theory is given in Figure \ref{fig:pEll_k}.
\EPSFIGURE[h]{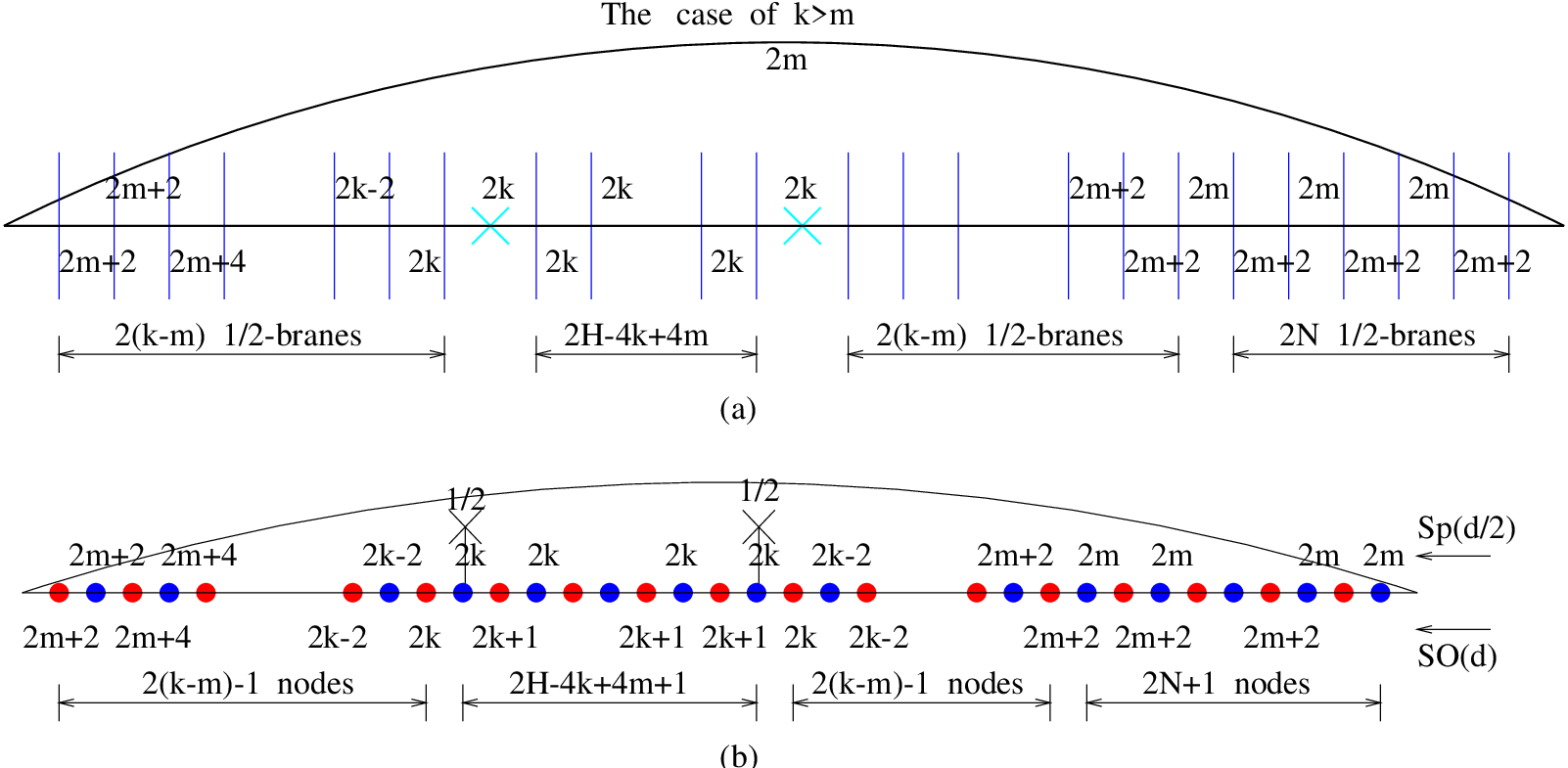,width=15cm} { The mirror of 
$Sp'(k) \times SO(2m+1)$ with $H$ flavors for $Sp'(k)$ , $N$ flavors 
for $SO(2m+1)$ and  two bifundamentals
 in case of $k>m$. 
(a) The Higgs branch of the original theory or the Coulomb branch of the mirror
theory. (b) The quiver diagram of the mirror theory.
 \label{fig:pEll_k}}
The dimensions of moduli spaces are 
\begin{equation}
\label{pEll_k}
\begin{array}{lll}
d_v & = & 2\sum_{i=1}^{k-m-1} 2(m+i) + k(2H-4k+4m+3)+m(N+1)+(m+1)N \\
& = & [2k^2-2m^2-2k-2m]+[2kH+(2m+1)N+4km-4k^2+3k+m] \\
& = & 2kH+(2m+1)N+4km-2k^2-2m^2+k-m  \\
d_H & = &2 \sum_{i=1}^{k-m-1}[\frac{(2(m+i))^2}{2}+\frac{2(m+i)2(m+i+1)}{2}
-(m+i)(2m+2i-1)-(m+i)(2m+2i+1)]  \\
& + & 2\frac{(2k)^2}{2}+\frac{2k(2k+1)}{2}(2H-4k+4m)-(2H-4k+4m+1)k(2k+1)
-2k(2k-1) \\ 
& + & 2k+N \\
& = &[2k^2-2m^2-2k-2m]+[-2k^2+k]+[-N+2m^2+3m]+[N+2k] \\
& = & k+m \\
\end{array}
\end{equation}

\par
We are left only one more example, i.e., the case of $k<m$. For this case,
after the combination, we still have $m-k$ D3-branes in the 
interval of $\widetilde{O3^-}$-plane. The mirror theory is given in
Figure \ref{fig:pEll_m}.
\EPSFIGURE[h]{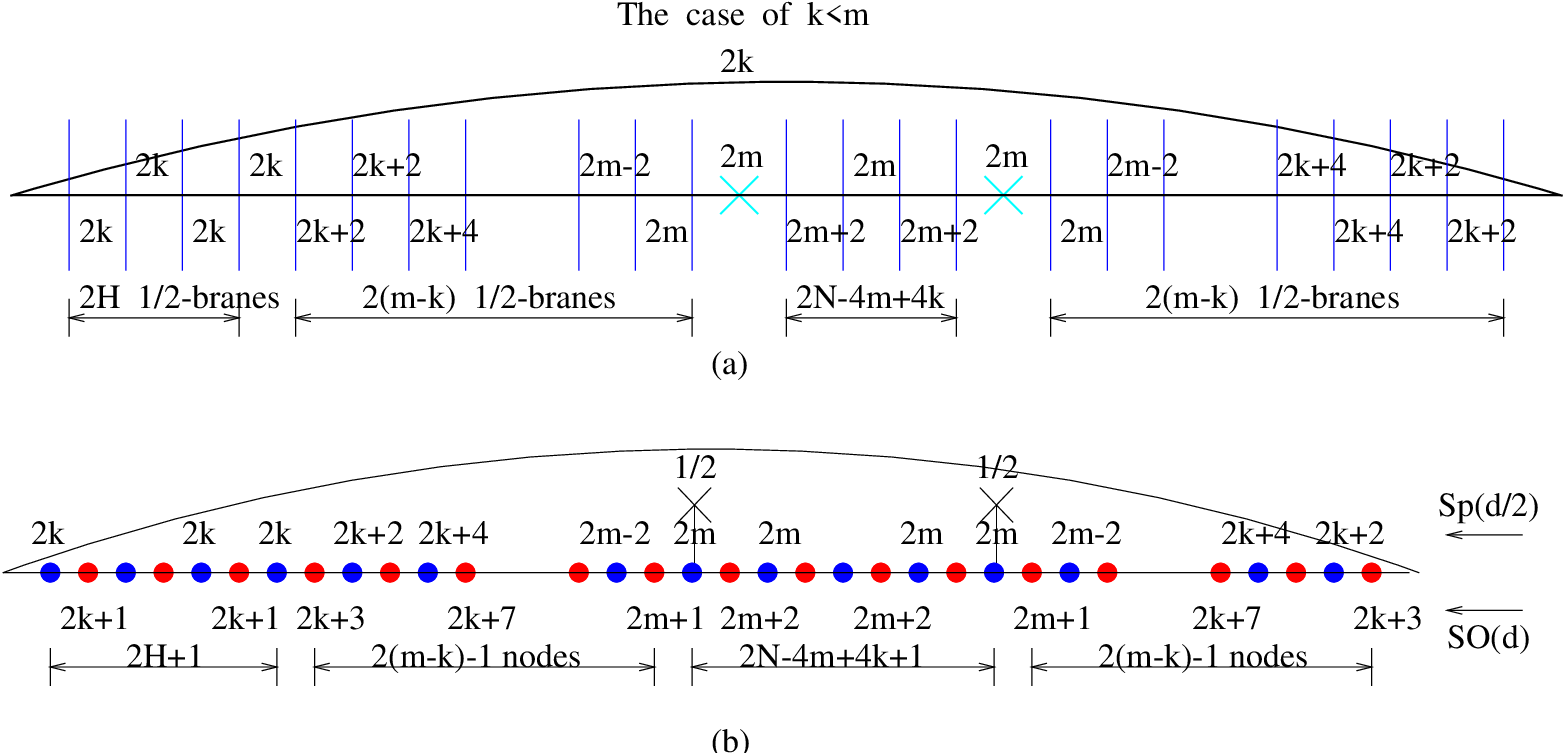,width=15cm} { The mirror of 
$Sp'(k) \times SO(2m+1)$ with $H$ flavors for $Sp'(k)$ , $N$ flavors 
for $SO(2m+1)$ and  two bifundamentals
 in case of $k<m$. 
(a) The Higgs branch of the original theory or the Coulomb branch of the mirror
theory. (b) The quiver diagram of the mirror theory.
 \label{fig:pEll_m}}
The dimensions of moduli spaces are
\begin{equation}
\label{pEll_m}
\begin{array}{lll}
d_v & = &2 \sum_{i=1}^{m-k-1} 2(k+i)+k(2H+1)+ +m(N-2m+2k+3)+(m+1)(N-2m+2k) \\
& = & [2m^2-2k^2-2k-2m]+[2kH+(2m+1)N+4km+m+3k] \\
& = & 2kH+(2m+1)N-2m^2-2k^2-m+k  \\
d_h & = & 2\sum_{i=1}^{m-k-1}[\frac{(2k+2i)(2k+2i+1)}{2}+
\frac{(2k+2i)(2k+2i+3)}{2}-2(k+i)(2k+2i+1)]  \\
& + &[ 2\frac{2m(2m+1)}{2}+\frac{2m(2m+2)}{2}(2n-4m+4k) \\
& & -(N-2m+2k+3)m(2m+1)-(N-2m+2k)(m+1)(2m+1)] \\
& + & \frac{2k(2k+1)}{2}(2H)-(2H+1)k(2k+1)+2\frac{2k(2k+3)}{2} \\
& + & 2m+N \\
& = & [2m^2-2k^2-2k-2m]+[-N-2m^2-2k+m]+[2k^2+5k]+[2m+N] \\
& = & k+m. \\
\end{array}
\end{equation}

\section{Conclusion}
In this paper, we give the mirror theories of $Sp(k)$ and $SO(n)
$ gauge theories. In particular, for
 the first time  the mirror of $SO(n) $ gauge theory is given. 
In the
construction of the mirror, we have made an assumption about 
the splitting of  D5-branes  on O3-planes in the {\sl brane-plane}
system\footnote{For more details about the brane-plane system see
\cite{Han4}}.
We want to emphasize that  because
the splitting of D5-brane on O3-plane is a nontrivial
dynamical process and we do not  fully understand it at this moment,
 we can not
really prove our assumption by calculation. However, although 
our discussions in
this paper indicate that our assumption is consistent, the other independent
checks are favorable.  This gives one
direction of further work as to prove our observation.

\par
Furthermore, as we discussed in section three, our rules observed in
this paper about the splitting of physical brane predict some 
nontrivial strong coupling limit of a particular field theory. It will
be very interesting to use the Seiberg-Witten curve 
\cite{Witten,Land} to show whether it is true.

\par
There is another direction to pursue our investigation. By
rotating one of the 1/2NS-branes \cite{Eli1,Eli2,Ahar,Boer3}  we break the
$N=4$ theory in three dimensions to an $N=2$ theory. Then we can
discuss the mirror of $N=2$ in three dimension as we have done in
this paper. However, because there is less supersymmetry in the
$N=2$ case, things become more complex (for a detailed explanation
of new features in $N=2$, see \cite{Boer3}). Indeed, we can even
break the supersymmetry further to discuss the mirror symmetry in
the  $N=1$ case \cite{Gre}.

\section*{Acknowledgements}
We would like to extend our sincere gratitude to A. Kapustin, K. Intriligator
and  A. Uranga
 for fruitful discussions. Furthermore A. Hanany would like to thank
A. Zaffaroni for discussion and B.Feng
would like to thank Yang-Hui He, A. Naqvi,
and J. S. Song for their suggestions and help.

\par
This project is supported by the DOE under grant no. DE-FC02-94ER40818.
A.H is also partially supported by the National Science Foundation 
under grant
no. PHY94-07194,  by
an A.P.Sloan Foundation Fellowship and by a DOE OJI award.

\bibliographystyle{JHEP}

\end{document}